\documentclass{aa}
\usepackage{amssymb}
\usepackage{booktabs, multirow, tabularx}
\usepackage{graphicx}
\usepackage{longtable}
\usepackage{lscape}
\usepackage{txfonts}
\usepackage[hyperfootnotes=false]{hyperref}
\usepackage{subcaption}
\usepackage{textgreek}
\usepackage{bm}
\usepackage{float}
\usepackage{tikz}
\usepackage{comment}
\usetikzlibrary{arrows.meta, positioning, shapes.geometric}
\usetikzlibrary{positioning,decorations.pathreplacing}

\usepackage{microtype}

\begin{document}

\title{The burgeoning rise of the 21-cm forest}
\subtitle{I: constraints on the optical depth of the intergalactic medium at $5.38 < z < 5.84$}

\author{C.~Kongprachaya\inst{1,2}
          \and
          G.~Bernardi\inst{2,3,4}
          \and
          E.~Ceccotti\inst{2} 
          \and
          A.~Mesinger\inst{5} 
          \and
          B.~Ciardi\inst{6} 
          \and
          O.M.~Smirnov\inst{3,4,2} 
          }

\institute{Department of Physics and Astronomy, University of Bologna, Via Gobetti 93/2 – 40129 Bologna, Italy
        \and 
        INAF -- Istituto di Radioastronomia, Via P.~Gobetti 101, 40129 Bologna, Italy \\
        \email{chanasorn.kongprachaya@inaf.it}
        \and
            Centre for Radio Astronomy Techniques and Technologies (RATT), Department of Physics and Electronics, Rhodes University, Makhanda 6140, South Africa
        \and
            South African Radio Astronomy Observatory, Cape Town 7700, South Africa
        \and
            Department of Physics and Astronomy ``Ettore Majorana'', University of Catania, Via Santa Sofia 64, 95123 Catania, Italy
        \and
            Max-Planck-Institut für Astrophysik, Karl-Schwarzschild-Str. 1, D-85748 Garching b. München, Germany \\
             }
\date{Received ---; accepted ---}

\abstract{The redshifted 21-cm line is a promising probe of the Epoch of Reionization (EoR), during which the first generation of stars ionized the InterGalactic Medium (IGM).}{We aimed to constrain the IGM neutral Hydrogen fraction and spin temperature via redshifted 21-cm absorption against high-redshift radio sources.}
{We analysed an 18-hour observation of the radio-loud quasar PSO~J352.4034-15.3373 ($z = 5.84$) in the $203{-}222.5$~MHz band. We obtained a continuum image with a 0.66~mJy~beam$^{-1}$ rms noise and a spectrum with 3.6~mJy~beam$^{-1}$ rms noise per 390~kHz-wide channel. We fit the quasar spectrum with a power-law continuum modified by an intervening 21-cm absorption, testing two scenarios: an \textit{island model}, representing a residual cold neutral Hydrogen patch surviving at the end of reionization, and a \textit{global model}, approximating the overall decline of the IGM HI fraction and spin temperature with redshift.}
{Combining archival data spanning 150~MHz to 3.0~GHz with our measurements, we determined a quasar flux density of $86.8 \pm 1.1$~mJy at 200~MHz and a spectral index of $-0.88 \pm 0.02$ across the full frequency range. We found no evidence for the 21-cm absorption from intervening neutral Hydrogen at $5.38<z<5.84$. Assuming the \textit{island model}, we set a 95\% confidence lower limit on the IGM spin temperature of 1.73~K in residual HI regions. Assuming the \textit{global model}, we constrained the 21-cm optical depth to $\tau_{21}< 0.02$ (95\% C.L.).}
{These results provide constraints on the 21-cm optical depth near the end of reionization, over the $5.38 < z < 5.84$ range, and confirm that the IGM was heated above the adiabatic cooling limit ($\sim 0.8$~K at $z = 5.68$), consistent with theoretical predictions and with 21-cm power spectrum measurements at higher redshifts. Our results also disfavour the presence of extremely cold HI regions at $z < 5.84$ and open the way to future 21-cm absorption from high-redshift sources.}

\keywords{Galaxies: quasars: absorption lines; Cosmology: dark ages, reionization, first stars; Cosmology: diffuse radiation; Galaxies: intergalactic medium; Techniques: interferometric}
\maketitle
    
\section{Introduction}

The epoch of reionization (EoR) remains one of the least known periods of cosmic history. During this epoch, the light emitted by the first stars and galaxies reionized the neutral Hydrogen in the InterGalactic Medium (IGM). Different observations have brought complementary information on the duration and timing of cosmic reionization over the last decade. Ly$\alpha$ observations of high-redshift quasars have historically played a significant role in constraining the evolution of the IGM neutral fraction \citep{GunnPeterson1965ApJ...142.1633G}, first showing that reionization ended at $z \sim 5 - 6$ \citep{Becker2015PASA...32...45B}. Recent observations show that an inhomogeneous ultraviolet background persists until $z \sim 5.3$ \citep{Becker2015PASA...32...45B,Bosman2022MNRAS.514...55B,Eilers2018ApJ...867...30E}, indicating that reionization is largely complete by then \citep{Kulkarni2019MNRAS.485L..24K}. 

Constraints on the onset of reionization are more uncertain, as observations of cosmic probes that rely on Ly$\alpha$ have larger uncertainties and the best constraints likely remain from the measurements of the CMB optical depth \citep{Planck2020A&A...641A...6P}. Once the various probes are combined, however, reionization can be timed quite accurately, and the IGM neutral fraction drops from 50\% at $z = 7.65$ to 1\% at $z = 5.44$ \citep{Qin2025PASA...42...49Q}.

The redshifted 21-cm line is an alternative probe of cosmic reionization, tracing directly the evolution of the neutral IGM gas \citep[see][for a classic review]{Furlanetto2006PhR...433..181F}. Such observations require, however, exquisite sensitivity and control over systematic effects, and have only recently started to place first significant upper limits, although more meaningful on IGM temperature rather than ionization fraction \citep{HERA2023ApJ...945..124H,Mertens2025A&A...698A.186M,Nunhokee2025ApJ...989...57N}.

Most of the theory and observations of the redshifted 21-cm line have been carried out considering its contrast against the CMB background, where the 21-cm emission can appear in either absorption or emission against the CMB, depending upon their relative temperatures. It was early recognised, however, that a bright radio source at high redshift could be equally used to observe the 21-cm line in absorption, due to the intervening cold, neutral IGM \citep{Carilli2002ApJ...577...22C}. Detecting 21-cm absorption against bright radio background sources has traditionally been regarded as difficult, primarily due to the limited number of suitable high-redshift background sources. This picture is changing quickly, however, as the census of known high-redshift quasars grows thanks to modern surveys \citep[e.g.,][]{Matsuoka2026arXiv260610160M, Euclid2026arXiv260703432Y} and their associated radio follow-up campaigns \citep[e.g.,][]{Keller2024MNRAS.528.5692K}. \cite{Soltinsky2025MNRAS.537..364S} compiled a sample of approximately 30 such quasars and evaluated the feasibility of detecting 21-cm absorption, considering both individual sight lines and a stacked statistical approach.

In this work, we present early 21-cm absorption observations against the radio-loud quasar PSO~J352.4034–15.3373 (hereafter J2329-1520) at $z = 5.84 \pm 0.02$ \citep{Banados2018ApJ...861L..14B} and derive upper limits on the 21-cm optical depth at the end of reionization. 

The paper is organised as follows. We define the modelling framework for the redshifted 21-cm absorption and present the forecast through simulations in Section~\ref{sec: Modeling framework}; we describe the observations and our data reduction in Section~\ref{sec: Observation and data reduction}; we present the radio spectrum of J2329-1520 and derive constraints on the 21-cm optical depth in section~\ref{sec: The spectrum of J2329-1520 and upper limits on the 21-cm absorption}; we summarise and conclude in Section~\ref{sec: Summary and Conclusions}.

\section{Modelling framework}\label{sec: Modeling framework}

Most theoretical models of 21-cm absorption focused on high-redshift sources -- typically $z > 8$ -- where the IGM is expected to be significantly neutral and cold. In addition to a broad absorption profile from the intervening neutral IGM, deep, narrow-band absorption troughs are expected owing to small-scale fluctuations in the gas temperature and density, generating the so-called 21-cm forest \citep[e.g.,][]{Carilli2002ApJ...577...22C, Ciardi2013MNRAS.428.1755C}. 
\citet{Thyagarajan:2020nch} and \citet{Soltinsky2025MNRAS.537..364S} investigated the detectability of the line-of-sight (one-dimensional) power spectrum, which can be related to physical parameters such as the X-ray galaxy luminosity responsible for the IGM heating. In this work, we adopted a different approach: we assumed a simplified parameterization of the 21-cm absorption while simultaneously fitting the continuum source spectrum, as this model resembles actual observations more closely.

The absorption of the continuum spectrum due to an intervening neutral medium can be written as:
\begin{equation}\label{eqn: power-law flux model}
    S_\text{o}(\nu) = S_\text{src}(\nu) \, \text{e}^{-\tau_{21}(\nu)} =  S_{200}\left(\frac{\nu}{\nu_\star}\right)^{\alpha}  \, \text{e}^{-\tau_{21} (\nu)}\, ,
\end{equation}
where $S_\text{o}$ is the observed (hence absorbed) flux density as a function of the observed frequency $\nu$, $S_\text{src}$ is the intrinsic source flux density, and $\tau_{21}$ is the 21-cm line optical depth. We assume that the intrinsic spectrum of the background radio source follows a power law spectrum and define $S_{200}$ as the source flux density at the reference frequency $\nu_\star=200$~MHz and $\alpha$ as the spectral index. 

The 21-cm optical depth of the IGM is described by \citep[e.g.,][]{Furlanetto2006PhR...433..181F}:
\begin{equation}\label{eqn: 21cm optical depth}
    \begin{split}
        \tau_{21}(\nu) = \tau_{21}\left( \frac{\nu_{21,0}}{1 + z} \right) & = \frac{3}{32\pi} \, \frac{hc^3A_{10}}{k_\text{B}T_\text{s}\nu_{21,0}} \, \frac{x_\text{HI}n_\text{H}}{(1+z)(\text{d}v_{\|}/\text{d}r_{\|})} \\
        & \approx 0.0092 \left( \frac{1+\delta(z)}{T_\text{s}(z)} \right) \, x_{\text{HI}}(z) \, \left( 1+z \right)^{3/2} \, ,
    \end{split}
\end{equation}
where $\nu_{21,0}=1420$~MHz is the rest frequency of the hyperfine transition line, $A_{10}=2.85 \times 10^{-15}\,\text{s}^{-1}$ is the spontaneous emission coefficient of the 21-cm hyperfine transition, $T_\text{s}$ is the spin temperature\footnote{This temperature reflects the population ratio of excited atoms to de-excited atoms through the Boltzmann distribution.}, $x_\text{HI}$ is the hydrogen neutral fraction, $n_\text{H}$ is the total hydrogen number density, $\delta$ is the matter density contrast, and $\text{d}v_\|/\text{d}r_\|$ is the gradient of the proper velocity along the line of sight, which describes the redshift-space distortion. 
The redshift-space distortion is a second-order effect, and we ignored it in the second line of Eq.~\eqref{eqn: 21cm optical depth} by assuming $\text{d}v_\|/\text{d}r_\|=H(z)/(1+z)$, with $H(z)$ as the Hubble parameter.
We discuss two different models for the 21-cm optical depth, which we labelled as the \textit{global model} and the \textit{island model}.

\subsection{Global model}\label{sec: Global Model}

We assumed that the IGM along the line of sight follows the global redshift evolution of the EoR. That is, we assume that $x_{\text{HI}}$ and $T_s$ vary continuously and monotonically with redshift $z$. We used a hyperbolic tangent ($\tanh$) model for the redshift evolution of the neutral fraction $x_\text{HI}$:
\begin{equation}\label{eqn:HI_neutral_fraction}
x_\text{HI}(z; \Delta z, z_r)=\frac{1}{2}\left[\tanh{\left(\frac{y(z)-y(z_r)}{\Delta _y}\right)}+1\right]\, ,
\end{equation}
where $\Delta z$ is the duration of reionization, $z_r$ the redshift at which $x_\text{HI} = 0.5$, $y(z)=(1+z)^{3/2}$ and $\Delta_y=(1+z)^{1/2}\Delta z$. This empirical model is used in the analysis of CMB polarisation data \citep[e.g.,][]{Lewis2006MNRAS.373..561L,PlanckEoR2016A&A...596A.108P} and 21-cm global signal observations \citep[e.g.,][]{Monsalve2018ApJ...863...11M,Spinelli2019MNRAS.489.4007S}. 

The evolution of the spin temperature throughout reionization remains uncertain, and only recent observations have begun to place meaningful constraints on it \citep[e.g.,][]{HERA2022ApJ...924...51A,HERA2023ApJ...945..124H,Ghara2024MNRAS.530..191G,Ceccotti2025A&A...696A..56C}. Here we assumed a fiducial behaviour where the spin temperature is heated above the CMB temperature at $z \approx 12$ and reaches $T_S = T_{K} \approx 500$~K at $ z\sim 6$ \citep[e.g.,][]{Furlanetto:2006jb, Ciardi2013MNRAS.428.1755C, Thyagarajan:2020nch}. Under the common assumption that the spin temperature is coupled to the gas kinetic temperature, we derived the temperature evolution by log-interpolating the temperature values between these two redshift points and obtaining:
\begin{equation}\label{eqn: spin temperature evolution}
        T_s(z)=T_\star\cdot10^{-\frac{2}{7}(z-z_\star)}\, ,
\end{equation}
where $T_\star \equiv T_\text{s}(z=z_\star)$ is the spin temperature at $z=z_\star$, with $z_\star = 6$. In the remainder of the paper, we will assume $T_\star$ is a parameter to be constrained by observations. We acknowledge that this is a simplified, toy model for the temperature evolution; however, it is sufficient as a proof of concept of our current work. In future work, we will explore more realistic temperature evolutions from simulations
Finally, we assumed $\delta=0$, which represents a realistic or at least conservative choice \citep[e.g.,][]{Soltinsky2025MNRAS.537..364S}.

\subsection{Island Model}\label{sec: Island Model}

The second model describes a simplified scenario for the late stages of reionization where small, cold HI islands may still exist, surrounded by a largely ionized and hot IGM \citep[e.g.,][]{CODA-I2016MNRAS.463.1462O, CODA-II2020MNRAS.496.4087O}.
In this case, a forest-like absorption is expected, with discontinuous absorption features appearing across the frequency band, rather than the continuous absorption along the line of sight expected at higher redshift.

The 21-cm optical depth $\tau_{21,d}$ corresponding to such an HI island can be approximated using a rectangular function:
\begin{equation}
    \tau_{21,d}(z)=
    \begin{cases}
        \tau_{21}(z)\,\Big |_{x_\text{HI}=1} & ;\;|\nu-\nu_p|<\frac{\Delta\nu}{2}\\
        0 & ;\;\text{otherwise},
    \end{cases}
\end{equation}
where $\nu_p$ is the central frequency of the absorption feature and $\Delta\nu$ is its frequency width. We assumed $x_\text{HI} = 1$, i.e., a fully neutral hydrogen region, and the magnitude of the absorption optical depth is therefore set by the ratio $\frac{1+\delta}{T_s}$ (see Eq.~\ref{eqn: 21cm optical depth}).

\subsection{Simulations, parameter estimation and forecasts}\label{sec:Simulation_forecast}

We carried out simulations to assess how well the \textit{global model} parameters can be constrained using 21-cm absorption observations. We first considered a hypothetical quasar at redshift $z = 8$, assumed to be as bright as GLEAM J085614+022359 (hereafter J0856+0224), the brightest radio quasar known at $z > 5.5$ \citep{Drouart2020PASA...37...26D}. The source has a flux density at 200~MHz $S_{200} = 613.1 \pm 5.2$~mJy and spectral index $\alpha = 0.99 \pm 0.02$ in the $100 - 1400$~MHz range, and adopted its nominal parameter values for our simulation. We used a global 21-cm model in which the midpoint of reionization is $z_r = 7$ and its duration is $\Delta z = 1$, broadly consistent with CMB measurements \citep{Planck2020A&A...641A...6P} and constraints from galaxy luminosity functions and Ly$\alpha$ spectra \citep{Qin2025PASA...42...49Q} 
We assumed the IGM temperature to be $T_* = 500$~K at $z = 6$ and no contribution from overdensities, i.e., $\delta  = 0$. Tab.~\ref{tab: Fiducial value and Prior} summarises the nominal parameters of the model and the prior range used in the simulation.

\begin{table}[t]
\caption{Fiducial value and prior range for the 21-cm absorption model parameters.}
\centering
\begin{tabular}{lll}
\toprule\toprule
Parameters & Simulation's input value & Uniform prior range \\
\midrule
$z_r$ & $7$ & $[0,9]$ \\
$\Delta z$ & $1$ & $[0,9]$ \\
$(1+\delta)\,T_\star^{-1}$ & $0.002~\text{K}^{-1}$ & $[0,2]~\text{K}^{-1}$ \\
$\nu_p$ & - & $[207.6,222.5]$~MHz \\
$\Delta\nu$ & - & $[0.39,5]$~MHz \\
\bottomrule
\end{tabular}
\label{tab: Fiducial value and Prior}
\end{table}

\begin{figure}[t]
    \centering
    \includegraphics[width=\linewidth]{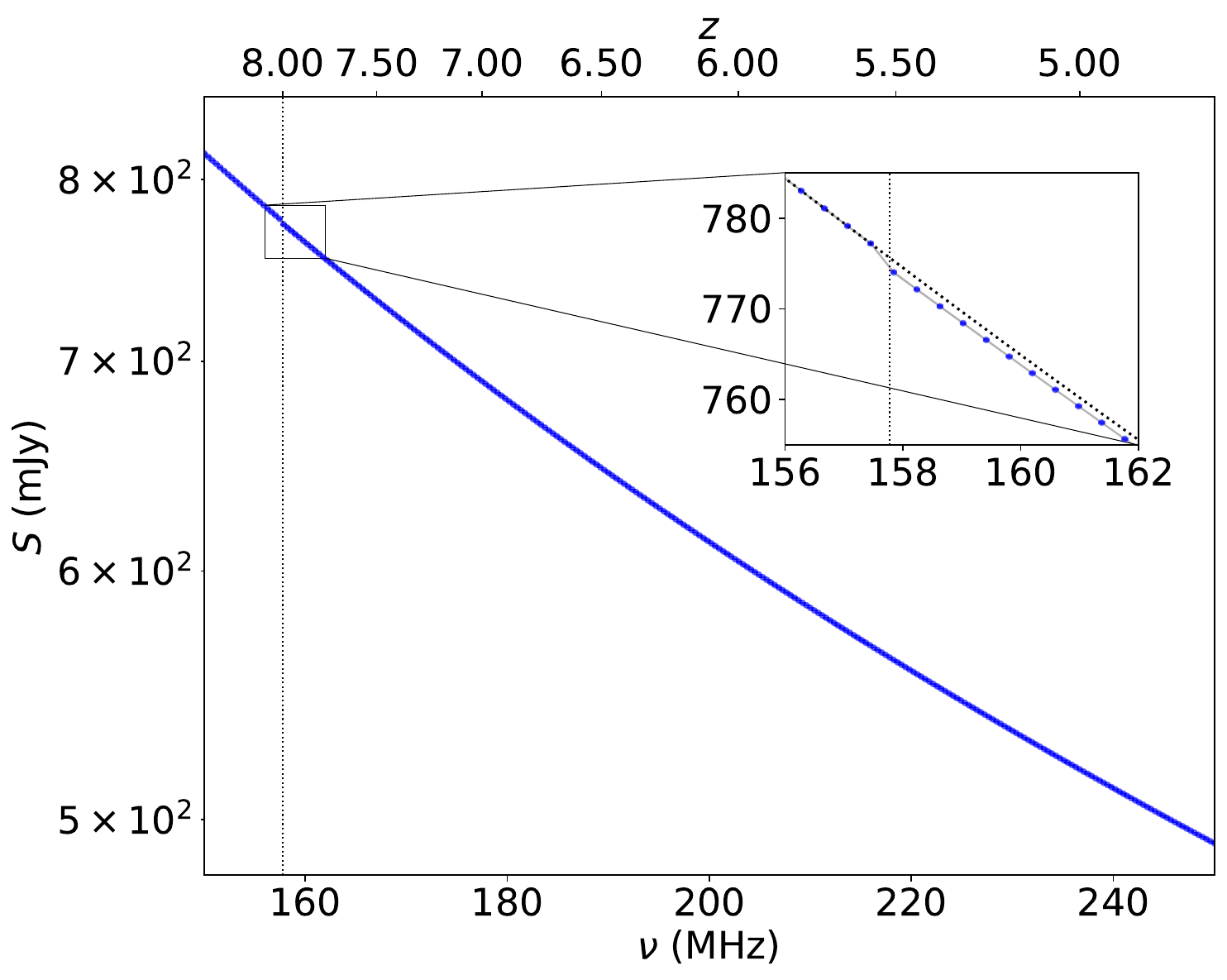}
    \caption{Simulated spectrum of the hypothetical radio loud source at $z = 8$ -- see text for details. The inset zooms into the absorption at frequencies redward of the 21-cm line at the source's redshift (vertical dotted line). The blue data points are the simulated spectrum with 100 \textmu Jy uncertainties ($2\sigma$) per output channel (achievable by a $\sim 40$~h observation with SKAO AA*). In the inset, the dotted black line indicates the radio continuum, while the thin grey line between data points is the fitted absorbed spectrum.}
    \label{fig:sim1_spectrum}
\end{figure}

We added Gaussian noise to the simulated spectrum, assuming a noise standard deviation of 50~$\mu$Jy per 390~kHz channel, and considered 256 output channels covering the $150-250$~MHz frequency range. This simulation setup was designed to resemble a future observation with the Square Kilometre Array Observatory (SKAO) of 40~h in the AA* configuration\footnote{The noise standard deviation was estimated using the SKAO sensitivity calculator, \url{https://sensitivity-calculator.skao.int/}}.
Fig.~\ref{fig:sim1_spectrum} shows the simulated spectrum.

We assumed a standard Gaussian likelihood function $\mathcal L$ to fit the model parameters to the simulated data:
\begin{equation}
    \begin{split}
        \mathcal{L}(\textbf{S}(\nu);\theta,\textbf{C})=&\frac{1}{(2\pi)^{N/2} |\textbf{C}|^{1/2}} \, e^{ -\frac{1}{2} [\textbf{S}(\nu) - \textbf{S}_{\text{o}}(\nu,\theta)]^\top \textbf{C}\,[\bm{S}(\nu) - \bm{S}_{\text{o}}(\nu,\theta)]} ,
    \end{split}
    \label{likelihood_function}
\end{equation}
where $\textbf{S}$ is a vector of the measured source flux densities as a function of frequency, $\theta=\{(1+\delta) \, T_\star^{-1},z_r,\Delta z\}$ is the set of parameters describing the 21-cm absorption model, and $\textbf{C}$  is the frequency-dependent covariance matrix. We assumed the covariance matrix to be diagonal and equal to the noise variance per frequency channel. 

\begin{figure}[t]
    \centering
    \includegraphics[width=\linewidth]{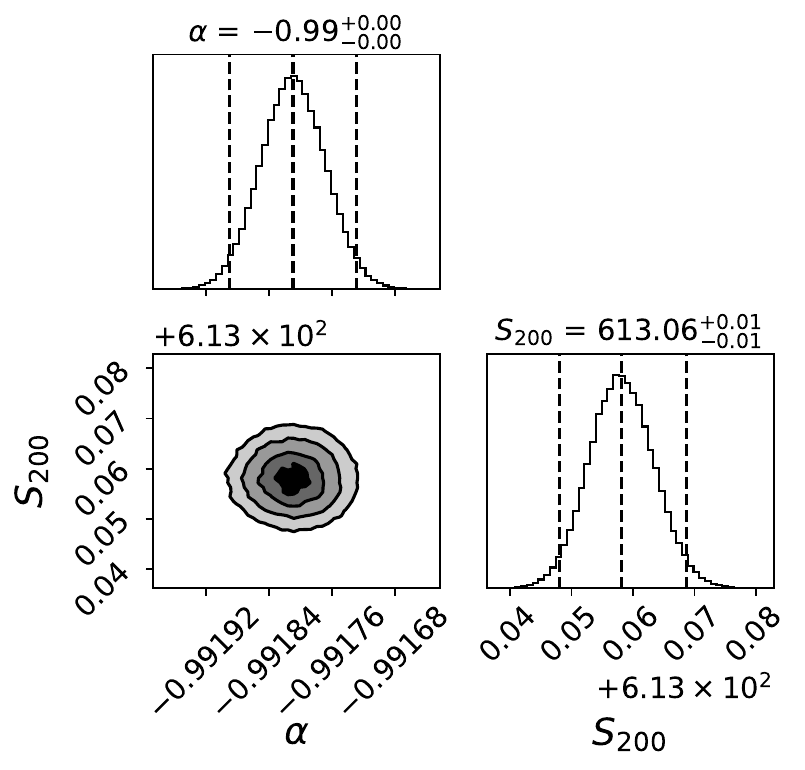}
    \caption{Constraints on the flux density at 200~MHz (in mJy) and spectral index of the hypothetical source at $z = 8$. Contours are drawn at the 2, 1.5, 1, and 0.5$\sigma$ levels. Dashed vertical lines mark the 95\%~C.L., with the central dashed line drawn at the 50$^{\rm th}$ percentile. The corresponding values are reported above the histograms.}
    \label{fig:sim1_continuum_parameter_estimation}
\end{figure}

\begin{figure}[t]
    \centering
    \includegraphics[width=\linewidth]{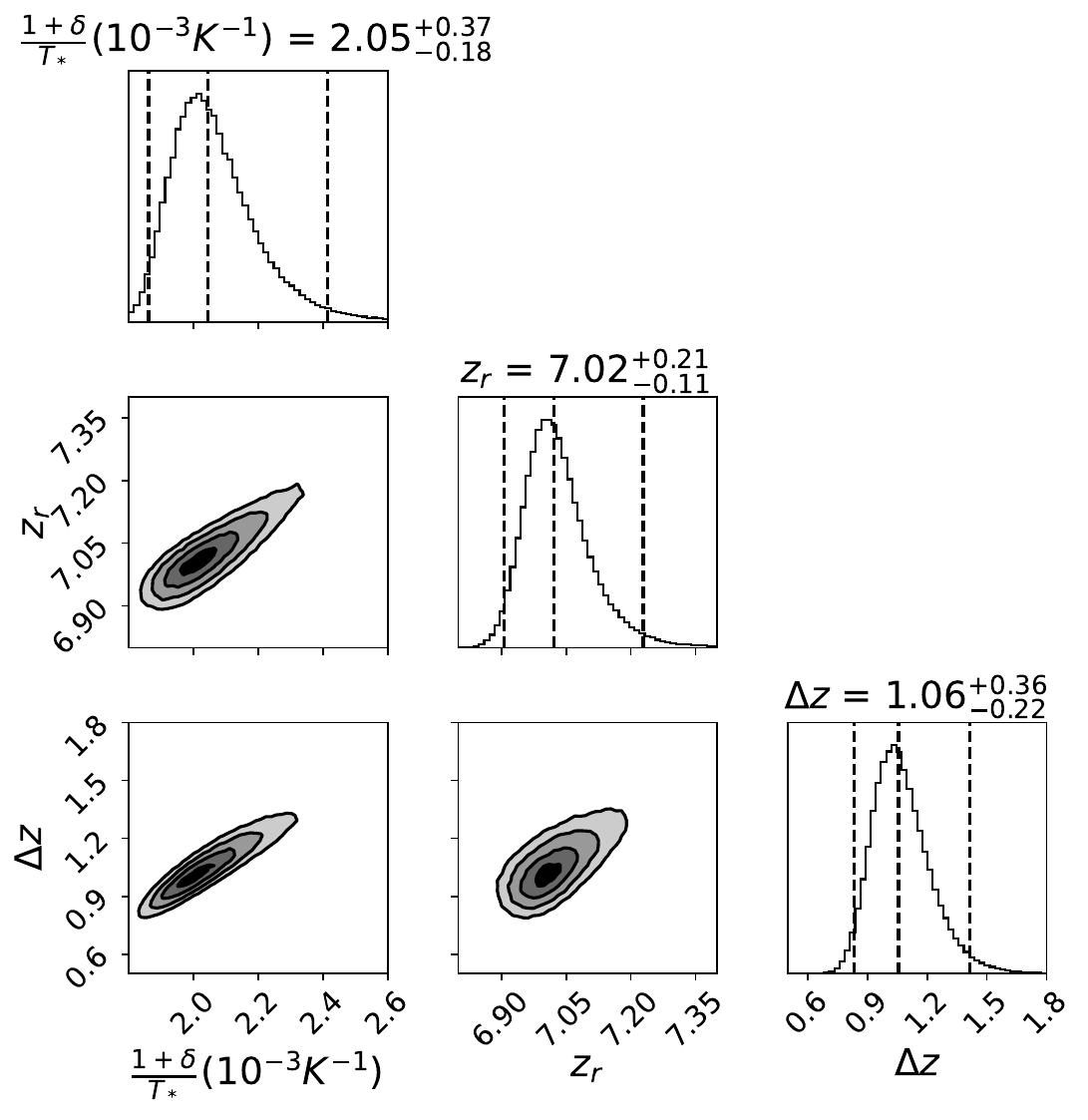}
    \caption{Constraints on the 21-cm optical depth parameters in the \textit{global model} case from the simulated spectrum of the hypothetical source at $z=8$. Contours are drawn at the 2, 1.5, 1, and 0.5$\sigma$ levels. Dashed vertical lines mark the 95\%~C.L., with the central dashed line drawn at the 50$^{\rm th}$ percentile. The corresponding values are reported above the histograms.}
    \label{fig:sim1_parameter_estimation}
\end{figure}

\begin{figure}[t]
    \centering
    \includegraphics[width=\linewidth]{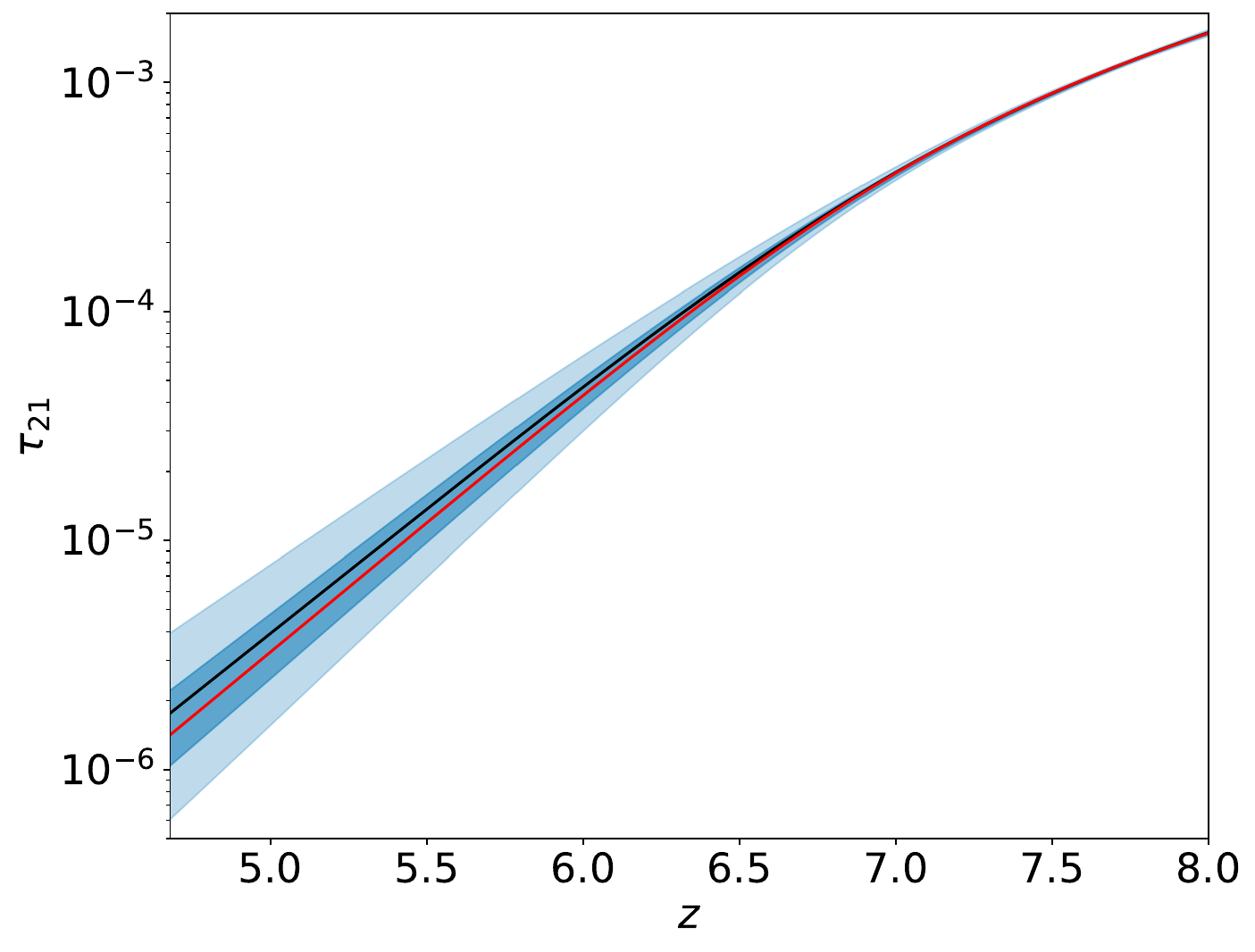}
    \caption{Reconstructed 21-cm optical depth ($\tau_{21}$) in the \textit{global model} case as a function of redshift for the hypothetical source $z = 8$. The red solid line is the input $\tau_{21}$ profile, the black solid line is the best-fit optical depth profile and the light and dark blue shaded regions represent the 68\% and 95\%~C.I., respectively.}
    \label{fig:sim1_tau}
\end{figure}

\begin{figure}[t]
    \centering
    \includegraphics[width=\linewidth]{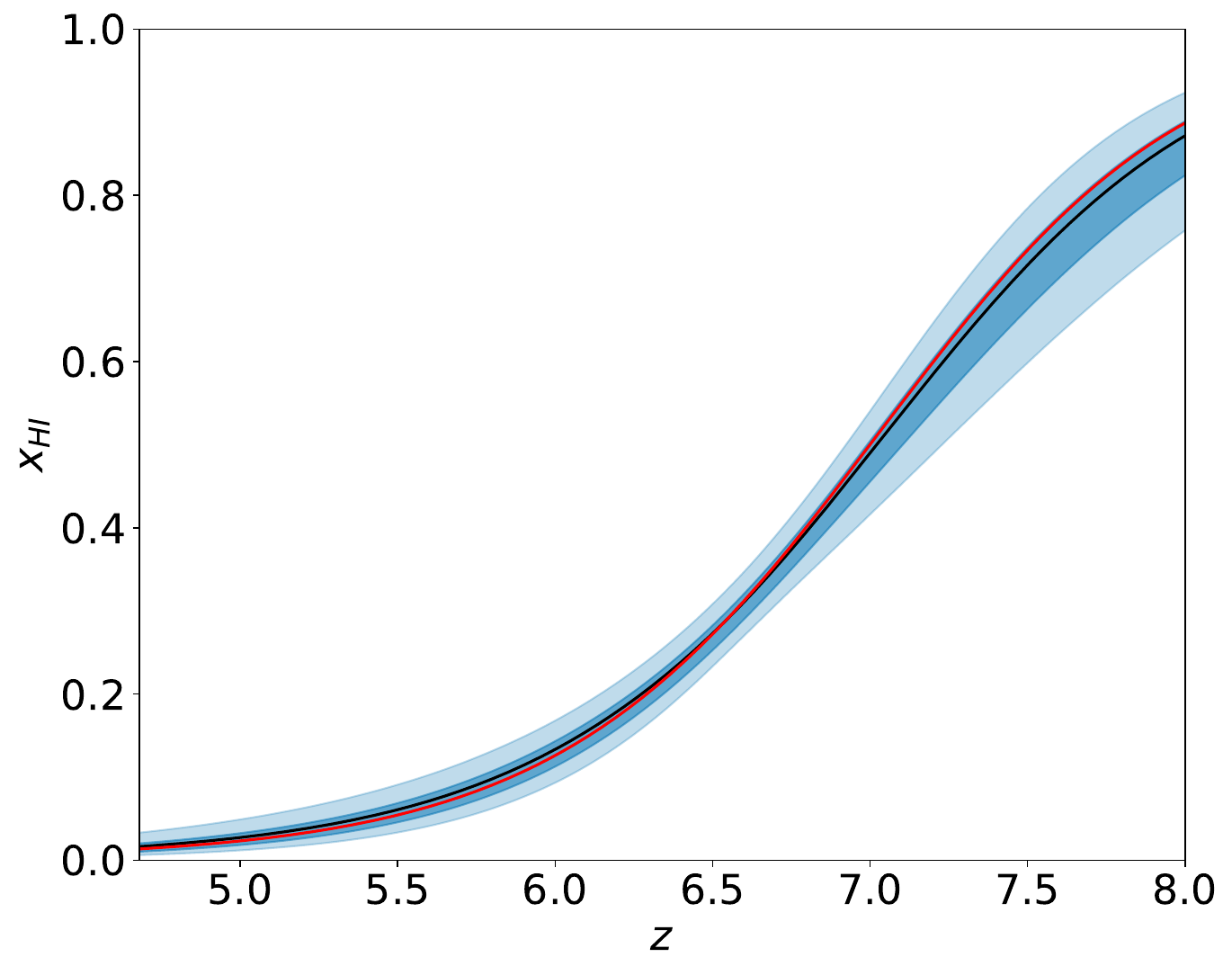}
    \caption{Reconstructed HI fraction ($x_\text{HI}$) as a function of redshift for the simulated source at $z = 8$. The red solid line is the input profile used to simulate the absorbed spectrum. The black solid line is the best-fit profile estimated from the simulation. The light and dark blue shaded regions represent the 68\% and 95\%~C.I., respectively.}
    \label{fig:sim1_x_HI}
\end{figure}

We sampled the posterior distribution via the Markov Chain Monte Carlo (MCMC) method implemented in the \texttt{EMCEE} Python package \footnote{The GitHub page of the package can be found at this link: \url{https://emcee.readthedocs.io/en/stable/}.} \citep{Foreman-Mackey2013PASP..125..306F}. We note that we fitted the continuum and 21-cm absorption parameters simultaneously.
Fig.~\ref{fig:sim1_continuum_parameter_estimation}\footnote{The posterior distributions are separated into continuum and 21-cm optical depth parameters just for easier visualization, but the posterior is always, even in the rest of the paper, sampled jointly.} and~\ref{fig:sim1_parameter_estimation} show the constraints on the input parameters. We found that the continuum spectrum parameters -- namely, the spectral index and the source flux density -- were extremely well constrained, and all 21-cm parameters were also well constrained, with uncertainties of the order of 10\% or better. 
The redshift evolution of the 21-cm optical depth and the neutral hydrogen fraction were also accurately reconstructed (Fig.~\ref{fig:sim1_tau} and \ref{fig:sim1_x_HI}, respectively).

\begin{figure}[t]
    \centering
    \includegraphics[width=\linewidth]{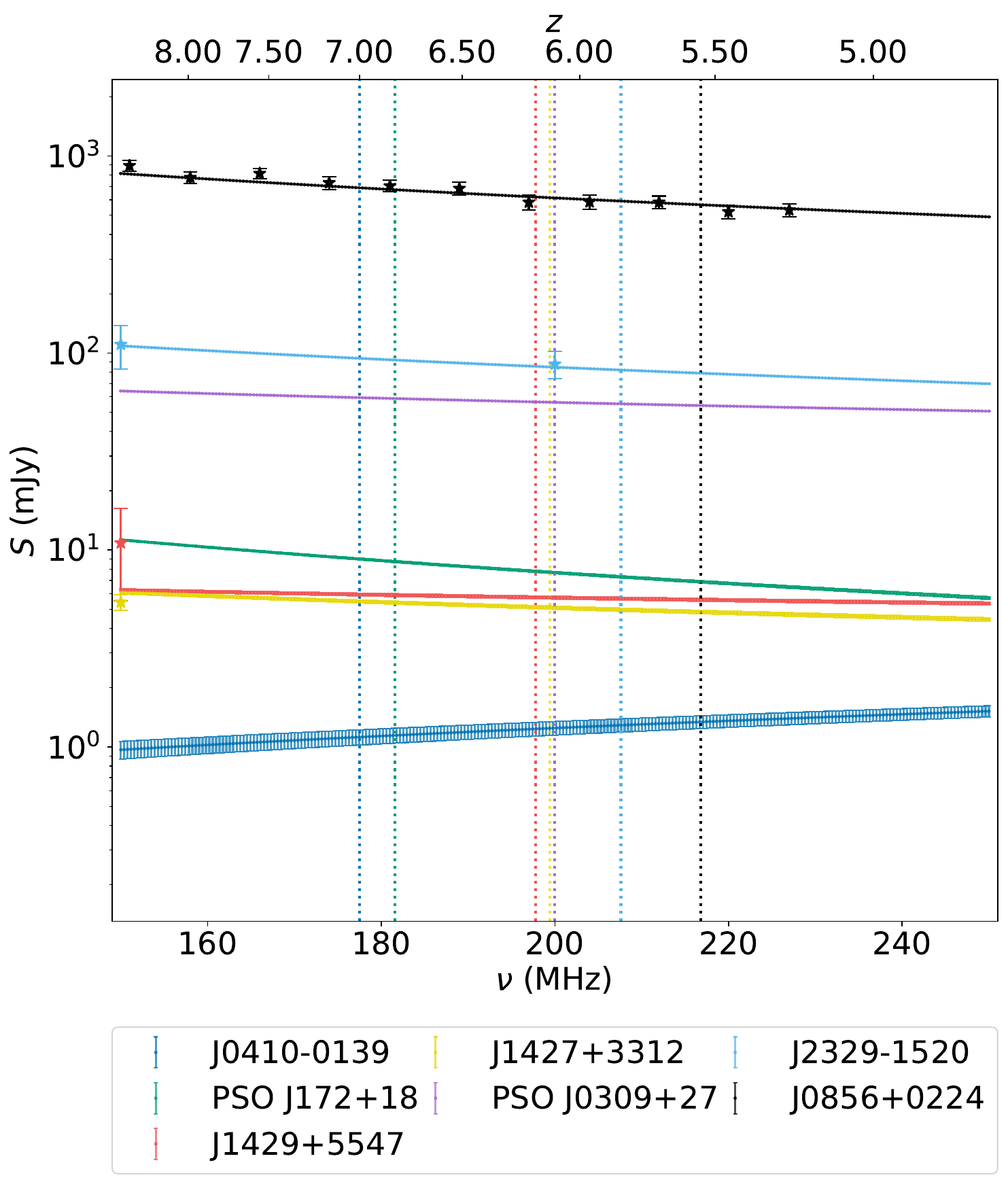}
    \caption{Simulated spectra of radio-loud sources in Tab.~\ref{tab:radio-loud quasars sample}. The spectra are simulated in the $150-250$~MHz frequency range with 256 output channels (channel width $\sim0.39$ MHz). Different colours indicate different source spectra.  Small data points are the simulated spectra with a $2\sigma$ uncertainty of 100 \textmu Jy per output channel (achievable by SKAO AA* $\sim 40$~h of observation). The larger star data points are the continuum flux measurements in the literature with their 2σ uncertainties (see Appendix~\ref{apx:High_redshift_Quasars_Sample_and_their_Continuum_Spectra}). The vertical dotted lines indicate the source's redshifts. The redshifted 21-cm absorption lines are expected blueward, i.e., to the right, of those lines.}
    \label{fig:sim2_simulated_spectrum}
\end{figure}

We then extended our simulations to a more realistic scenario where we simultaneously fitted the 21-cm optical depth in the \textit{global model} case using a sample of high-redshift quasars \citep[Tab.~\ref{tab:radio-loud quasars sample};][]{Soltinsky2025MNRAS.537..364S}. 
\cite{Thyagarajan:2020nch} showed that the one-dimensional power spectrum of the 21-cm absorption can be better constrained by stacking the signal from multiple lines of sight; here, we assessed how well the 21-cm optical depth in the \textit{global model} case can be constrained using the sample of currently known sources (Tab.~\ref{tab:radio-loud quasars sample}).
The highest-redshift source in the sample is at $z = 7.00$ and the lowest at $z = 5.55$, allowing us to sample the reionization tail. We assumed the same noise level used for the single-source simulation for each source in the sample. Simulated spectra are shown in Fig.~\ref{fig:sim2_simulated_spectrum}.

\begin{table*}[t]
    \centering
    \caption{Sample of radio-loud sources.}
    \begin{tabular}{llccccc}
    \toprule \toprule
    \textbf{Source Name} & \textbf{Shorten Name} & Ref. & $z$ & $\nu_{21cm}$ (MHz) & $\alpha_{\text{fit}}$ & $S_{200\,\text{MHz}}$ (mJy) \\
    \toprule
    VLASS J041009.05-013919.88 & J0410-0139 & (1) & $7.00$ & $177.5$ & $0.88\pm0.01$ & $1.25\pm0.02$ \\
    PSO J172.3556+18.7734 & PSO J172+18 & (2) & 6.82 & 181.5 & $-1.33\pm0.08$ & $7.68\pm1.36$ \\
    CFHQS J142952+544717 & J1429+5547 & (3) & 6.18 & 197.7 & $-0.54\pm0.11$ & $9.29\pm2.01$ \\
    FIRST J1427385+331241 & J1427+3312 & (4) & 6.12 & 199.4 & $-0.60\pm0.03$ & $4.63\pm0.19$ \\
    PSO J030947.49+271757.31 & PSO J0309+27 & (5) & 6.10 & 200.0 & $-0.50\pm0.05$ & $60.19\pm4.18$ \\
    PSO J352.4034-15.3373 & J2329-1520 & (6) & 5.84 & 207.6 & $-0.87\pm0.03$ & $84.73\pm5.26$ \\
    GLEAM J085614+022359 & J0856+0224 & (7) & 5.55 & 216.8 & $-0.99\pm0.02$ & $613.06\pm5.19$ \\
    \bottomrule
    \end{tabular}
    \tablefoot{ The radio continuum parameters -- the spectral index $\alpha_\text{fit}$ and the flux density at $200\,$MHz $S_{200}$ -- were fitted to measurements from previous radio surveys (see Appendix \ref{apx:High_redshift_Quasars_Sample_and_their_Continuum_Spectra}). The reported uncertainty is the standard deviation ($1\sigma$).
    }
    \tablebib{
    (1)~\citet{Banados2025NatAs...9..293B}; (2)~\citet{Banados2021ApJ...909...80B}; (3)~\citet{Willott2010AJ....139..906W}; (4)~\citet{McGreer2006ApJ...652..157M}; (5)~\citet{Belladitta2020A&A...635L...7B}; (6)~\citet{Banados2018ApJ...861L..14B}; (7)~\citet{Drouart2020PASA...37...26D}.
    }
    \label{tab:radio-loud quasars sample}
\end{table*}

\begin{figure*}[t]
    \centering
    \begin{tabular}{ccc}
        \begin{subfigure}{0.3\textwidth}
            \includegraphics[width=\linewidth]{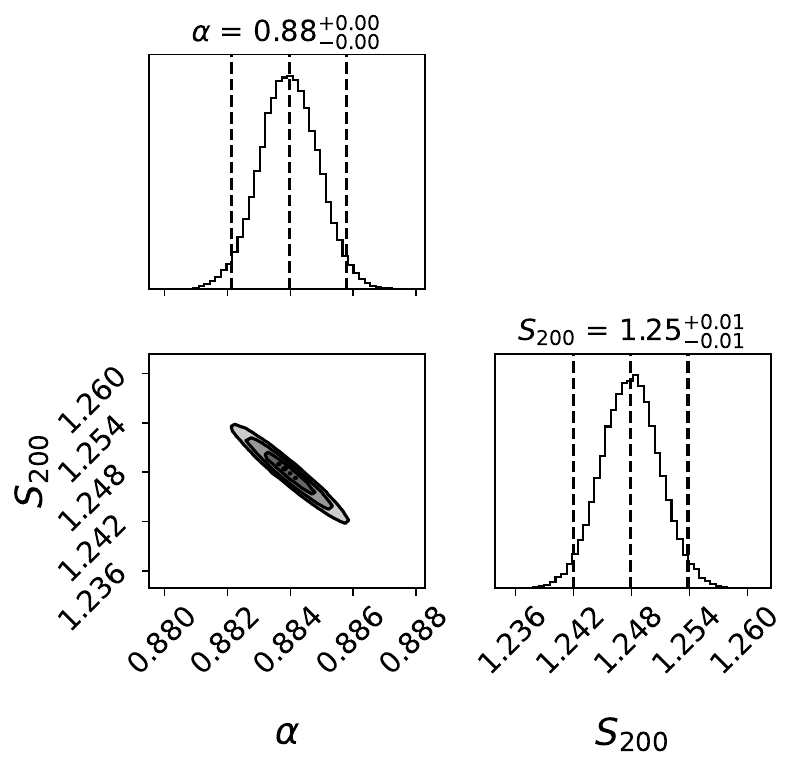}
            \subcaption{J0410-0139}
        \end{subfigure} &
        \begin{subfigure}{0.3\textwidth}
            \includegraphics[width=\linewidth]{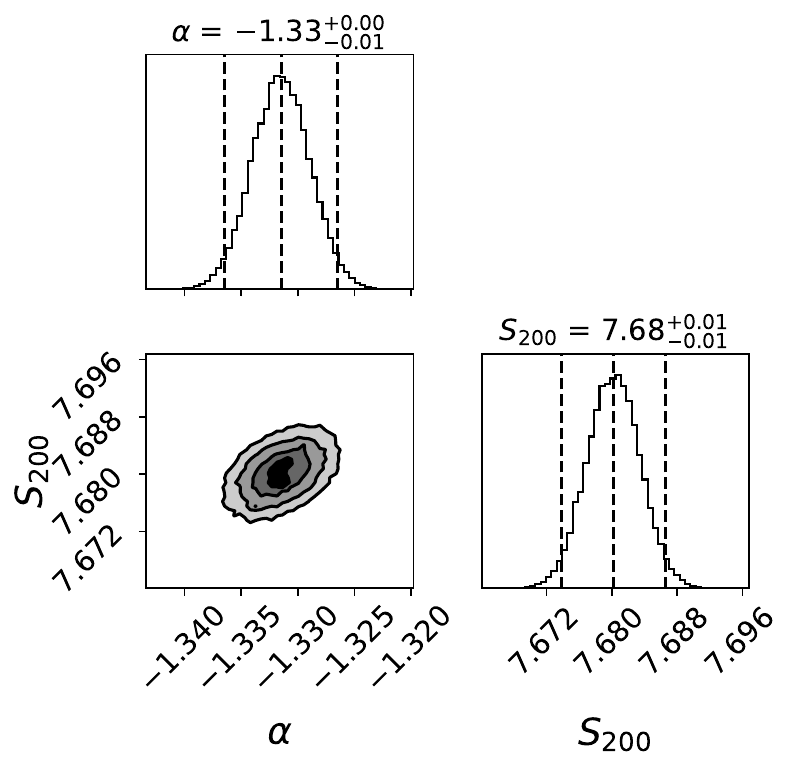}
            \subcaption{PSO J172+18}
        \end{subfigure} &
        \begin{subfigure}{0.3\textwidth}
            \includegraphics[width=\linewidth]{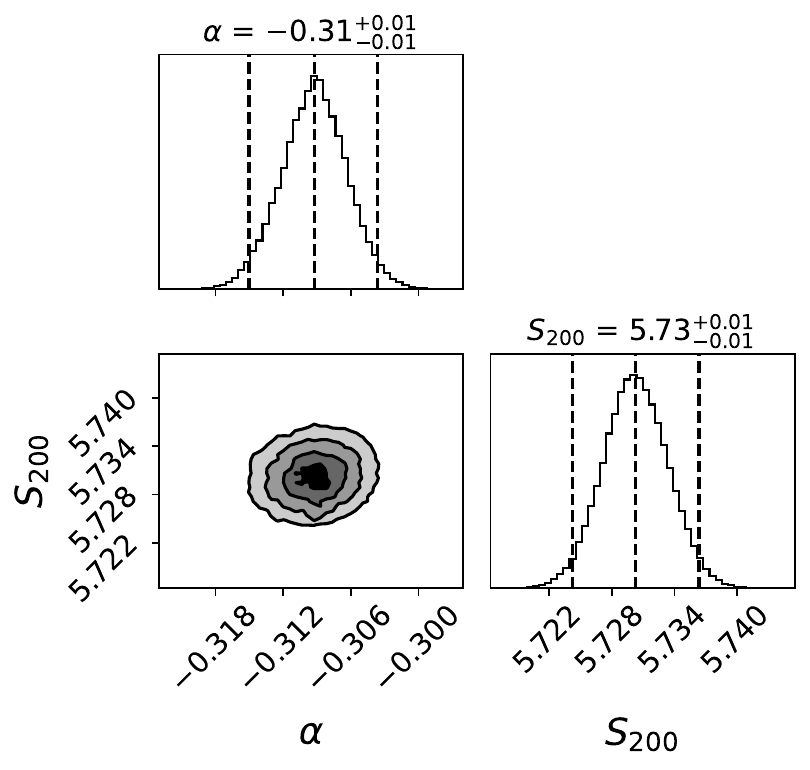}
            \subcaption{J1429+5547}
        \end{subfigure} \\

        \begin{subfigure}{0.3\textwidth}
            \includegraphics[width=\linewidth]{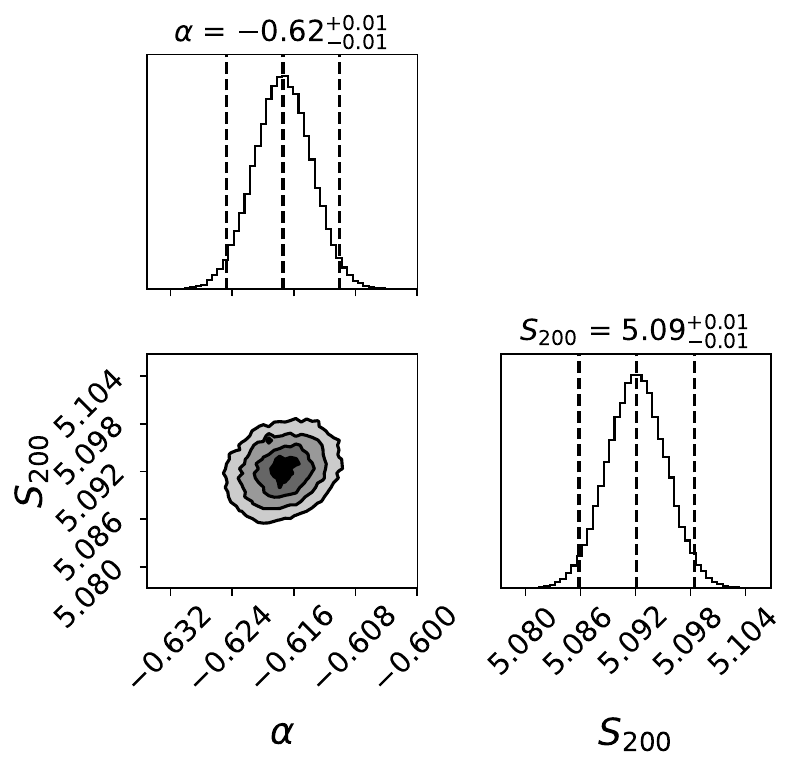}
            \subcaption{J1427+3312}
        \end{subfigure} &
        \begin{subfigure}{0.3\textwidth}
            \includegraphics[width=\linewidth]{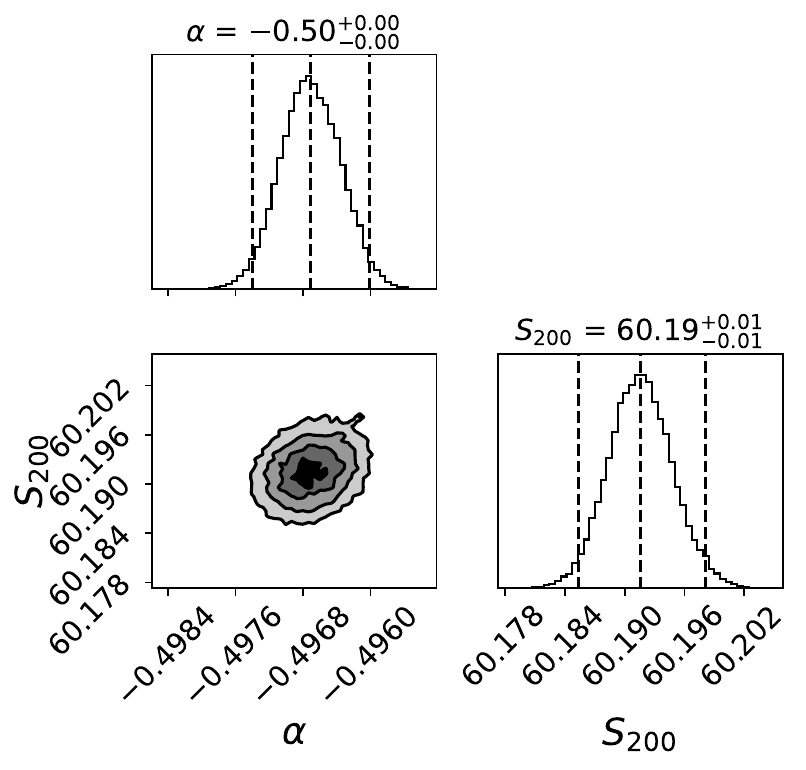}
            \subcaption{PSO J0309+27}
        \end{subfigure} &
        \begin{subfigure}{0.3\textwidth}
            \includegraphics[width=\linewidth]{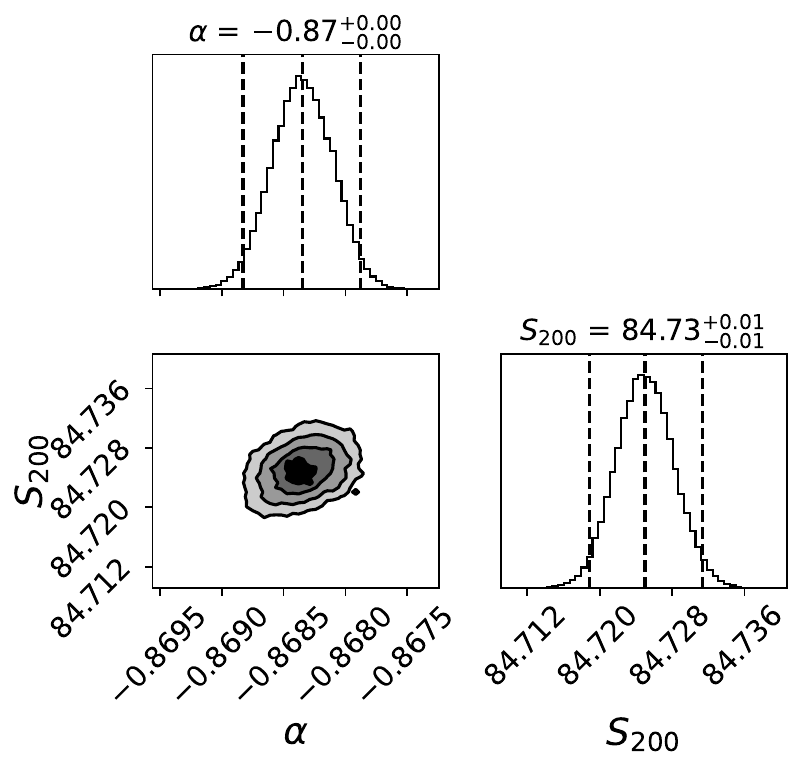}
            \subcaption{J2329-1520}
        \end{subfigure} \\

        \begin{subfigure}{0.3\textwidth}
            \includegraphics[width=\linewidth]{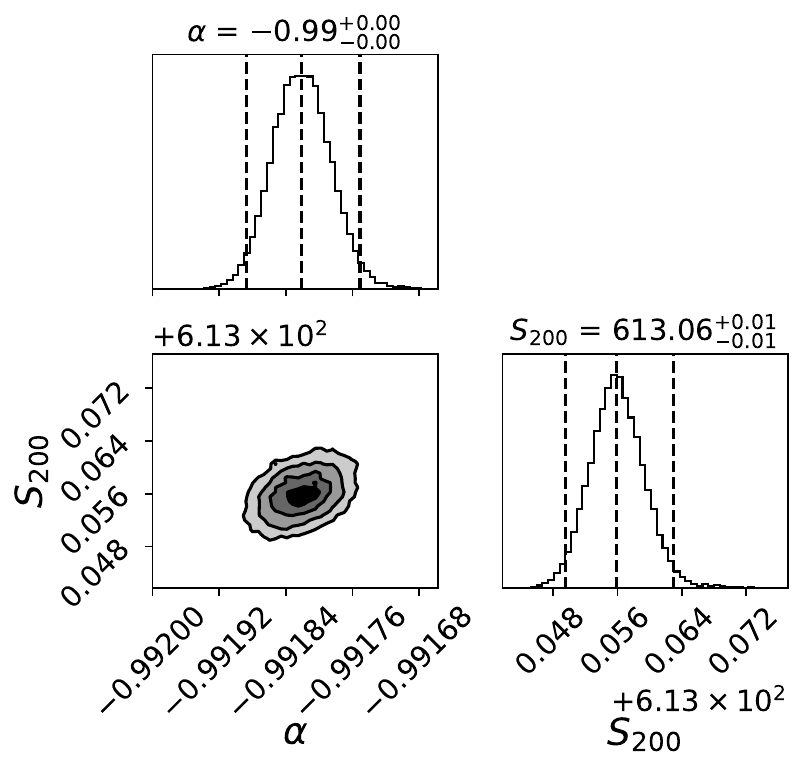}
            \subcaption{J0856+0224}
        \end{subfigure} 
    \end{tabular}
    \caption{Same as Fig.~\ref{fig:sim1_continuum_parameter_estimation}, but for each source in the sample of Tab.~\ref{tab:radio-loud quasars sample}.}  
\label{fig:sim2_continuum_parameter_estimation}
\end{figure*}

\begin{figure}[t]
\centering
\includegraphics[width=0.45\textwidth]{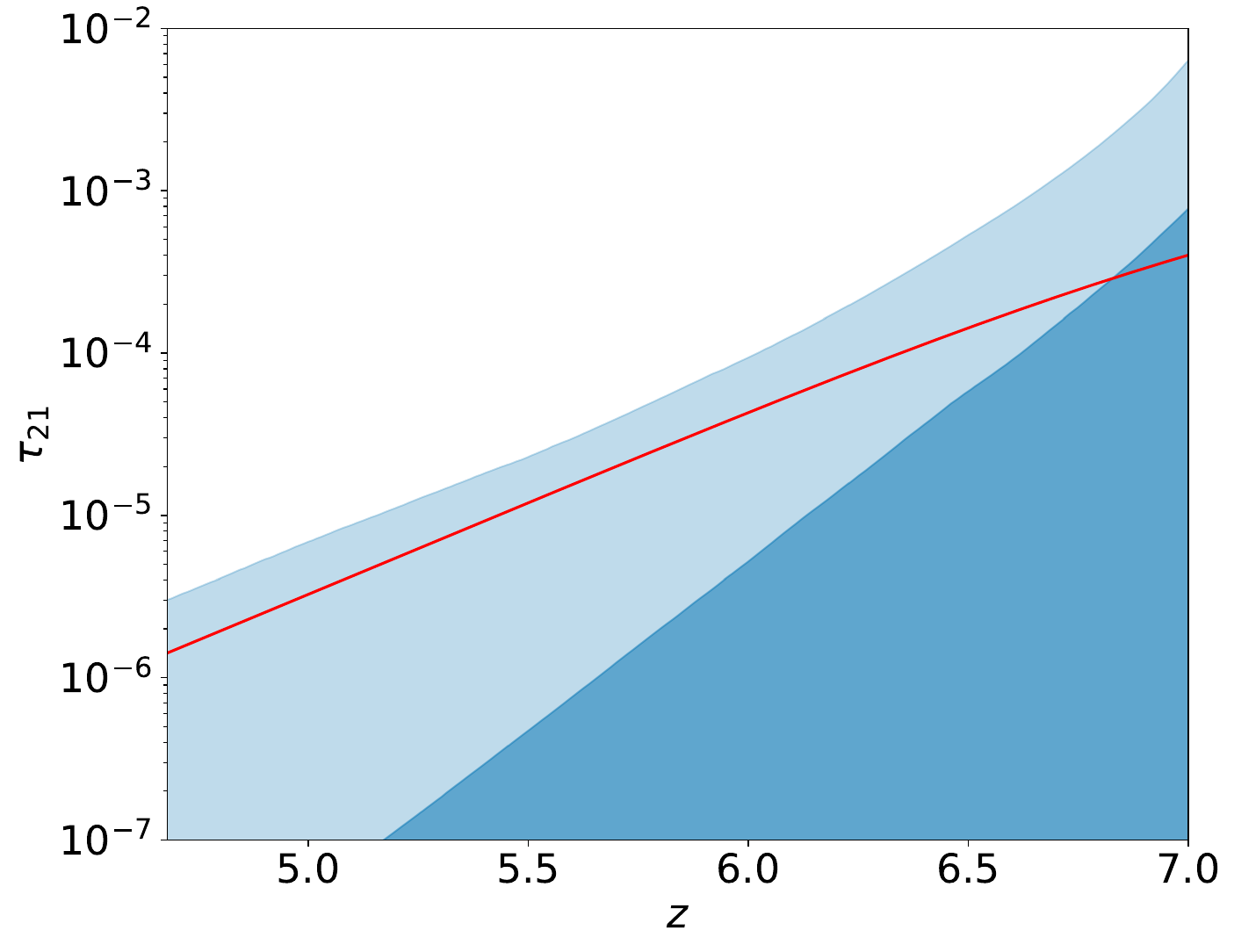}
\caption{Constraints on the 21-cm optical depth from simulated spectra of the high-redshift quasar sample in Tab.~\ref{tab:radio-loud quasars sample}. The red line is the simulated optical depth. The light and dark blue shaded regions represent the 68\% and 95\%~C.I., respectively. The redshift range is from the highest redshift in the sample ($z=7.00$; J0410-0139) to the lowest-frequency edge of the simulated observation ($z=4.68$; 250~MHz).}
\label{fig:sim2_tau}
\end{figure}

\begin{figure}[t]
\centering
\includegraphics[width=0.45\textwidth]{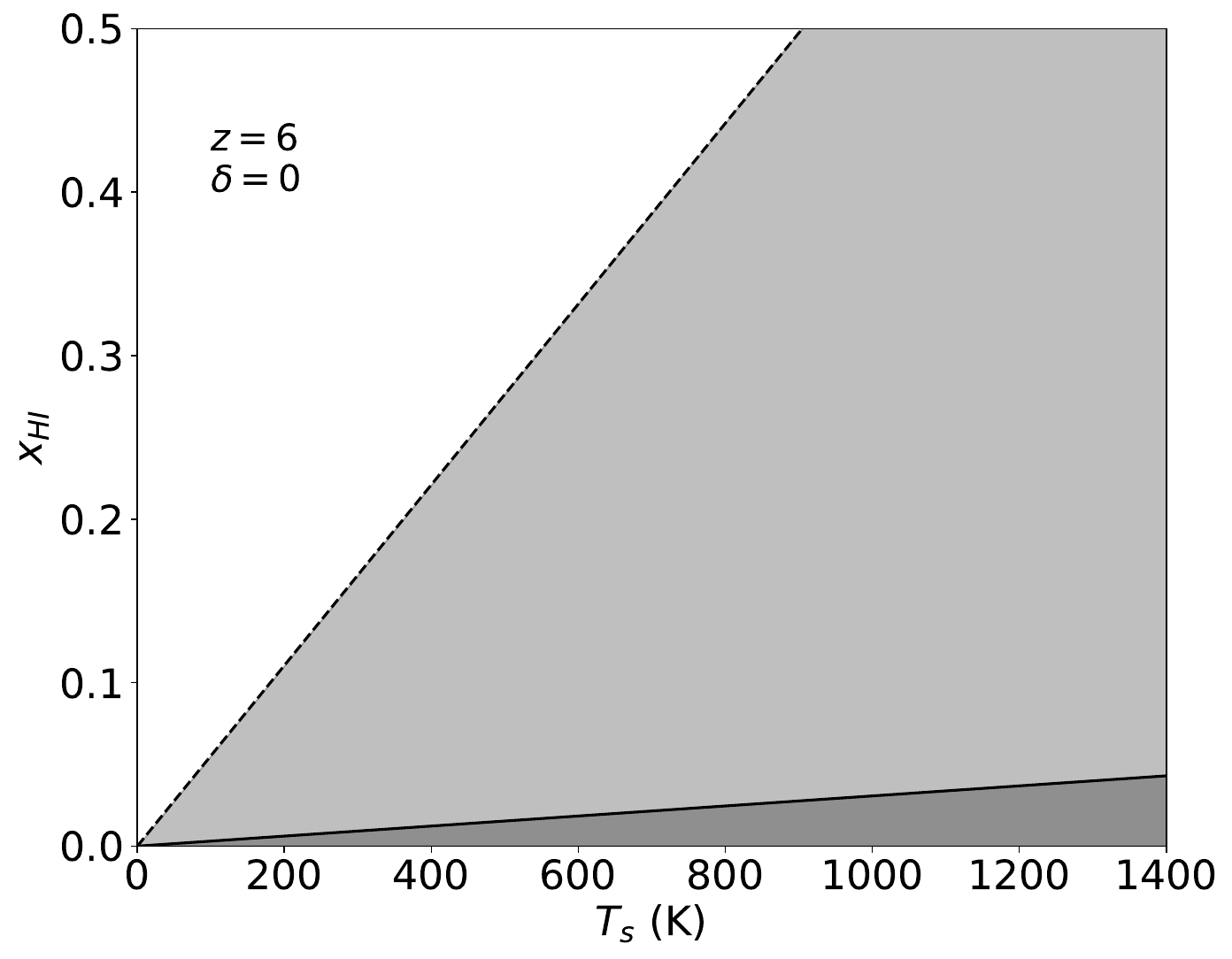}
\caption{Constraints in the $T_s-x_\text{HI}$ plane from simulated spectra of the high-redshift quasar sample in Tab.~\ref{tab:radio-loud quasars sample}. The upper limit on the optical depth in the null detection case can be translated to the upper limit of the ratio $T_s$ and $x_{\text{HI}}$, or the slope in this parameter space. The light and dark grey shaded regions represent the 95\% and 68\%~C.L., respectively. The upper limit is derived at redshift $z=6$ and assuming that the universe is at the average matter density along the line of sight, $\delta=0$.}
\label{fig:sim2_Ts-xHI}
\end{figure}

We jointly fitted the continuum spectrum parameters for each source and the 21-cm optical depth parameters. Although the continuum parameters were well fitted (Fig.~\ref{fig:sim2_continuum_parameter_estimation}), we could only place an upper limit on the 21-cm optical depth, which is nevertheless consistent with the input value (Fig.~\ref{fig:sim2_tau}). We note that, with the noise level and the source sample considered in the simulation, we can place constraints on the 21-cm optical depth at the $10^{-4}$ level at $z = 6$, consistent with the values expected from simulations \citep[e.g.,][]{Thyagarajan:2020nch}. 
The optical depth constraint can be, in turn, translated into limits in the $T_s-x_\text{HI}$ plane (Fig.~\ref{fig:sim2_Ts-xHI}): assuming a neutral fraction of $\sim 5 - 10\%$, the gas would need to have been heated above $100 - 200$~K.

\begin{figure}[t]
    \centering
    \includegraphics[width=\linewidth]{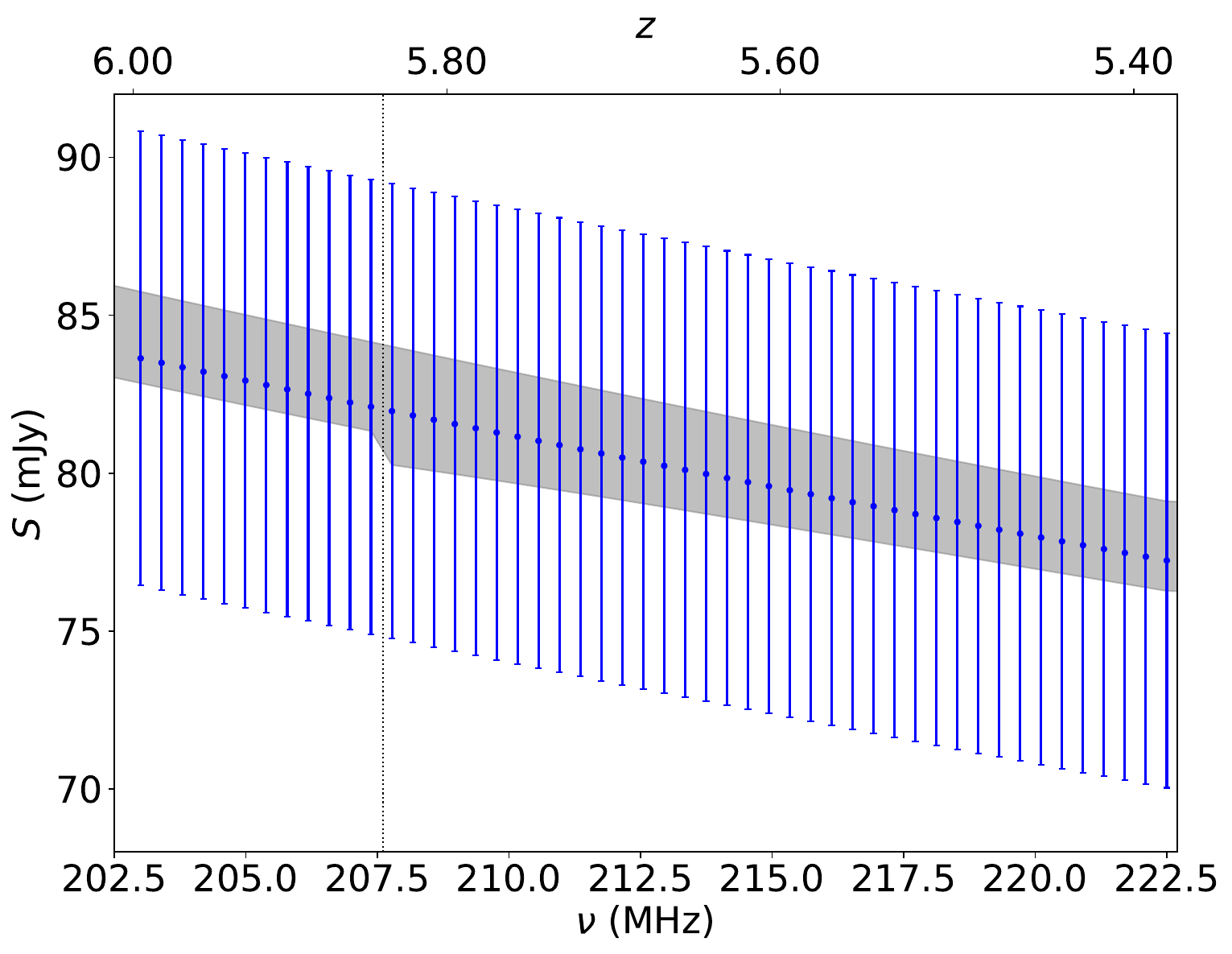}
    \caption{The simulated spectrum of J2329-1520 (blue data points) with $2\sigma$ uncertainty of 7.2~mJy per 390~kHz output channel, mimicing the real observed spectrum. The grey area is the 95\%~C.I. of the best fit spectrum, including the 21-cm optical depth upper limit. The vertical dotted line indicates the source's redshift.}
    \label{fig:sim3_spectrum}
\end{figure}

\begin{figure}[t]
    \centering
    \includegraphics[width=\linewidth]{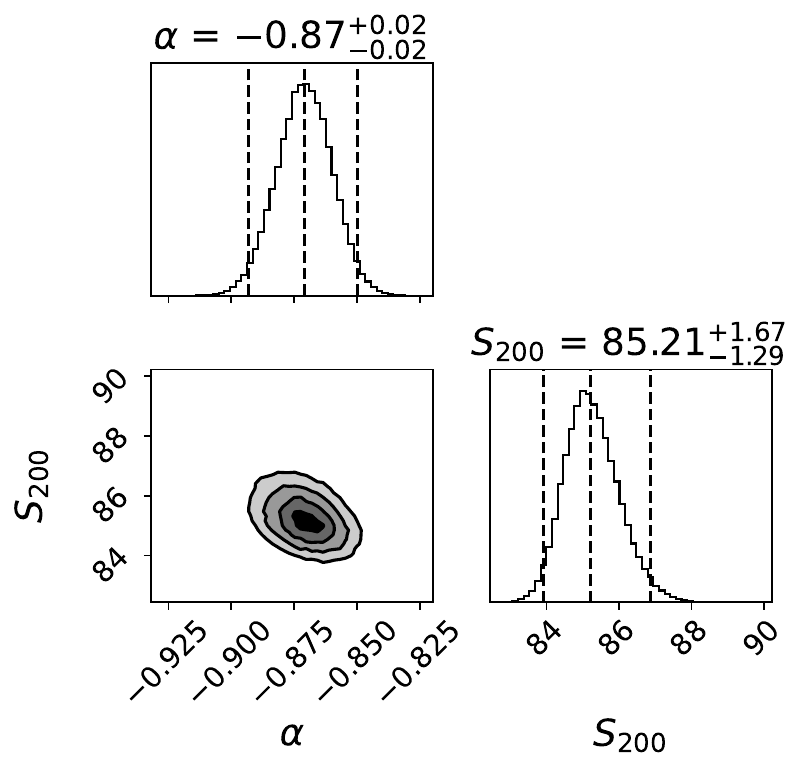}
    \caption{Same as Fig.~\ref{fig:sim1_continuum_parameter_estimation}, but for the J2329-1520 simulated spectrum.}
     \label{fig:sim3_continuum_parameter_estimation}
\end{figure}

\begin{figure}[t]
\centering
\includegraphics[width=0.45\textwidth]{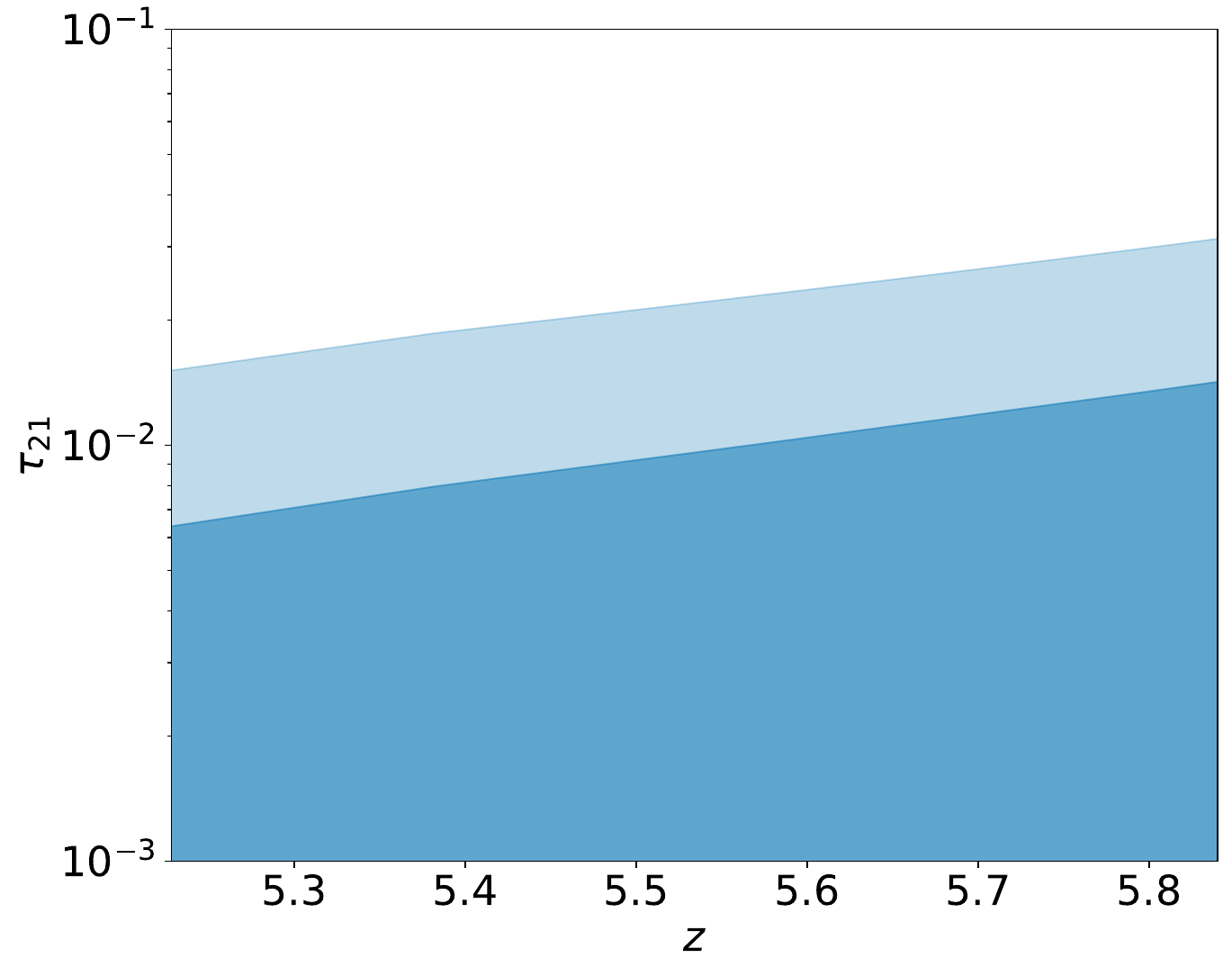}
\caption{Same as Fig.~\ref{fig:sim2_tau}, but for  simulated observations of J2329-1520. The redshift range, from 5.38 to 5.84, corresponds to the lower edge of the frequency window and the source's redshift.}

\label{fig:sim3_tau}
\end{figure}

\begin{figure}[t]
    \centering
    \includegraphics[width=\linewidth]{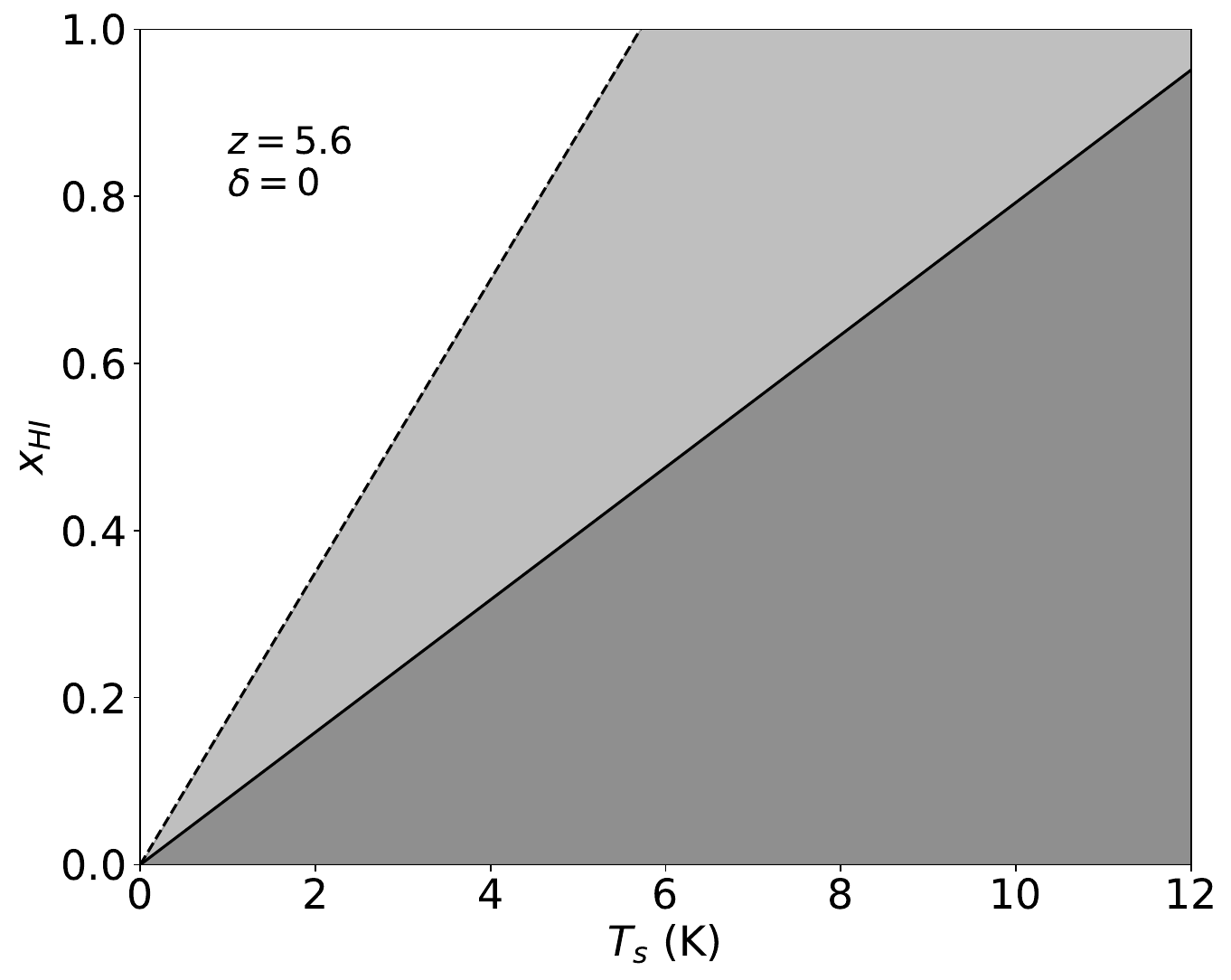}
    \caption{Same as Fig.~\ref{fig:sim2_Ts-xHI} but for the simulated spectrum of J2329-1520. The upper limit is derived at redshift $z=5.6$ and assuming that the universe is at the average matter density along the line of sight, $\delta=0$.}

    \label{fig:sim3_Ts-xHI}
\end{figure}

As a final simulation case, we considered J2329-1520, a radio-loud quasar at $z = 5.84$ \citep{Banados2018ApJ...861L..14B} - whose observations will be presented in the next section. Tab.~\ref{tab:J2329-1520_External_measurement} lists the main literature properties of the source. We simulated its spectrum in the $203{-}225.5$~MHz range, divided into 50 channels of 390~kHz width, with a noise standard deviation of 3.6~mJy~beam$^{-1}$. This setup was designed to mimic actual observations. The simulated spectrum and best-fit models are shown in Figures~\ref{fig:sim3_spectrum}, \ref{fig:sim3_continuum_parameter_estimation} and~\ref{fig:sim3_tau}, where, again, the continuum and 21-cm optical depth parameters were fitted jointly. As in the previous case, the continuum spectrum parameters were very well constrained, whereas we could only place an upper limit on the 21-cm optical depth, which is two to three orders of magnitude above the input values (see Fig.~\ref{fig:sim2_tau}). As before, this upper limit is translated into limits in the $T_s-x_\text{HI}$ plane in Fig.~\ref{fig:sim3_Ts-xHI}.

\begin{table}[t]
\centering
\caption{Literature measurements for the J2329-1520 quasar.}
\begin{tabular}{lcc}
\toprule\toprule
& \textbf{Frequency} & \textbf{Flux Density (mJy)} \\
\midrule
VLA-S\tablefootmark{a} & $3.0$~GHz & $8.20 \pm 0.25$ \\
NVSS\tablefootmark{b} & $1.4$~GHz & $14.9 \pm 0.7$ \\
GLEAM WIDE\tablefootmark{c} & $200$~MHz & $87.8 \pm 6.9$ \\
TGSS\tablefootmark{d} & $147.5$~MHz & $110.6 \pm 13.8$ \\
\midrule\midrule
R.A. (J2000) & \multicolumn{2}{c}{$23^\text{h} \, 29^\text{m} \, 36\rlap{.}^\text{s}8363$} \\
Dec. (J2000) & \multicolumn{2}{c}{$-15^\circ \, 20' \, 14\rlap{.}''460$} \\
Redshift ($z$) & \multicolumn{2}{c}{$5.84 \pm 0.02$} \\
$S_{200}$ & \multicolumn{2}{c}{$84.7 \pm 5.3$~mJy} \\
$\alpha$ & \multicolumn{2}{c}{$-0.87 \pm 0.03$} \\
\bottomrule
\end{tabular}
\tablefoot{Coordinates and flux density measurements of J2329-1520 from the past radio surveys, including \tablefoottext{a}{\citet{Banados2018ApJ...861L..14B}}, \tablefoottext{b}{\citet{NVSS1998AJ....115.1693C}}, \tablefoottext{c}{\citet{GLEAM2017MNRAS.464.1146H}}, and \tablefoottext{d}{\citet{TGSS2017A&A...598A..78I}}. The peak flux density at 200~MHz ($S_{200}$) and the spectral index ($\alpha$) defined in Eq.~\ref{eqn: power-law flux model} are fitted from the literature measurements mentioned above.}
\label{tab:J2329-1520_External_measurement}
\end{table}

\section{Observations and data reduction}\label{sec: Observation and data reduction}

\begin{table*}[t]
    \centering
    \caption{Details of the uGMRT observations on J2329-1520.}
    \begin{tabular}{lcccccc}
    \toprule \toprule
    Proposal Code/ &  \multicolumn{2}{c}{Date/Time (TAI)} & On-source Time & Integration Time & Frequency \\
    \cline{2-3}
     Run & Start & End & (min) & (s) & (MHz) \\
    \toprule
    ddtC007/1 & 23-06-2018/20:11:44.8 & 24-06-2018/00:34:32.7 & 166 & 8.05 & 203.0-228.0 \\
    ddtC007/2 & 29-06-2018/20:30:57.9 & 30-06-2018/01:32:51.9 & 186 & 10.70 & 203.0-228.0 \\
    ddtC219/1 & 16-07-2022/18:55:58.0 & 17-07-2022/03:56:57.2 & 360 & 5.37 & 197.5-222.5 \\
    ddtC219/2 & 27-08-2022/16:37:14.5 & 27-08-2022/20:49:23.5 & 165 & 5.37 & 197.5-222.5 \\
    ddtC219/3 & 28-11-2022/14:03:53.1 & 28-11-2022/18:26:57.1 & 173 & 5.37 & 197.5-222.5 \\ 
    \bottomrule
    \end{tabular}
    \tablefoot{The date/time format is dd-mm-yyyy/hh:mm:ss. On-source time refers to the total observing duration on J2329-1520 in each observing run. The International Atomic Time (TAI; UTC+00:00:37.0) is around five and a half hours behind the uGMRT local time, which is the Indian Standard Time (IST; UTC+05:30:00.0).}
    \label{tab:observations}
\end{table*}

We analysed observations of the radio-loud quasar PSO J352.4034–15.3373, located at redshift $z=5.84 \pm 0.02$ \citep[see][]{Banados2018ApJ...861L..14B}, taken with the upgraded Giant Metrewave Radio Telescope (uGMRT, proposal code: ddtC007). In particular, we used archival data from observations carried out on 23 and 29 June 2018, each with 3~h of on-source time. Source 3C48 was observed at the end of each run as a bandpass calibrator. Source J2321-163 was observed for 6~min every 30~min as a secondary (phase) calibrator. Visibilities were recorded with the integration time of 9.05~s and 10.7~s in the two observing runs, respectively, covering the $203-228$~MHz range with 16\,384 channels, each 1.5~kHz wide.

    We also analysed a second archive observation (proposal code: ddtC219), which included three separate runs on 17 July 2022, 27 August 2022, and 28 November 2022, respectively. The on-source times were 6, 3, and 3~h, respectively. The same set of calibrators was used, with the target and the phase calibrator (J2321-163) observed alternately. The integration time was $5.37$~s in this second set of observations. The $197.5{-}222.5$~MHz bandwidth was covered with 16\,384 channels, with the same frequency resolution as the first observing run.

    We used the Common Astronomy Software Application \citep[\texttt{CASA};][]{CASA2022PASP..134k4501C} for the data reduction. The five observing runs were calibrated separately and then imaged jointly. Frequency band edges ($\sim10\%$ in total) were flagged, together with the first and last 10~s of each run. Radio Frequency Interference (RFI) was identified and flagged using \texttt{aoflagger} \citep{Offringa2012A&A...539A..95O}.
    
    We used the primary calibrator to solve for frequency-dependent, complex gains at the highest frequency resolution, using the CASA \texttt{bandpass} task. We assumed a third-order polylogarithmic model for the flux density of 3C48 where the coefficients are taken from Table 6 of \cite{Perley&Butler2017ApJS..230....7P}.After calibration, we further flagged outliers in the distribution of the residual visibilities of the primary calibrator using the \texttt{rflag} and \texttt{tfcrop} algorithms. We concluded flagging with another round of \texttt{aoflagger}. We then repeated the calibration and flagging until the solutions converged.
    \begin{figure}[t]
    \centering
    \begin{tikzpicture}[
    box/.style={draw, rectangle, minimum width=4cm, minimum height=1cm, align=center},
    arrow/.style={->, thick},
    doublearrow/.style={<->, thick},
    node distance=0.5cm
    ]
    
    \node[box] (flux) {Flux Calibrator: 3C48};
    \node[box, below=of flux] (bandpass) {Bandpass Solution};
    \node[box, below=of bandpass] (phasecal) {Phase Calibrator: 2321-163};
    \node[box, below=of phasecal] (gainphase) {Gain (phase-only) Solution};
    \node[box, below=of gainphase, yshift=-0.5cm] (target) {Target: J2329-1520};
    
    \node[box, below left=1cm and -1.7cm of target] (selfphase) {Gain (phase-only)\\ self-calibration solution};
    \node[box, below right=1cm and -1.7cm of target] (selfamp) {Gain (phase \& amplitude)\\ self-calibration solution};
    
    \draw[arrow] (flux) -- node[right]{calibrate} (bandpass);
    \draw[arrow] (bandpass) -- node[right]{apply} (phasecal);
    \draw[arrow] (phasecal) -- node[right]{calibrate} (gainphase);
    \draw[arrow] (gainphase) -- node[right]{apply} (target);
    
    \draw[arrow] (bandpass.west) -- ++(-0.5cm,0) |- node[left]{apply} (target.west);

    \draw[doublearrow] (target.south west) .. controls +(0,-0.5cm) .. 
    node[left]{$1^\text{st}$\,round} (selfphase.north);
    
    \draw[doublearrow] (target.south east) .. controls +(0,-0.5cm) .. 
    node[right]{$2^\text{nd}$\,round} (selfamp.north);
    
    \draw[decorate, decoration={brace, amplitude=8pt}]
    ([xshift=0.6cm]flux.north east) --
    ([xshift=0.6cm]target.south east)
    node[midway, xshift=0.6cm, rotate=270] {For individual measurement set};
    
    \draw [decorate, decoration={brace, amplitude=8pt, mirror}]
    ([yshift=-0.6cm]selfphase.south west) --
    ([yshift=-0.6cm]selfamp.south east)
    node[midway, yshift=-0.6cm] {With a sky model from joint imaging};
    \end{tikzpicture}
    \caption{Calibration and self-calibration workflow.}
    \label{fig:Calibration and self-calibration workflow.}
\end{figure}
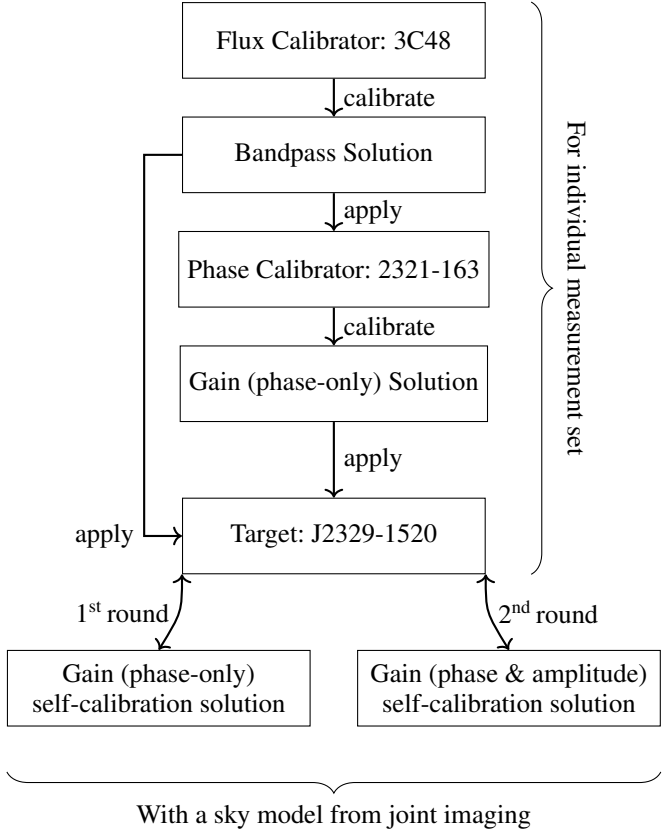

Bandpass solutions were applied to the secondary calibrator whose visibilities were then flagged. We derived phase solutions from the secondary calibrator every 60~s using the \texttt{gaincal} task, iterating between flagging and solving for phases until solutions converged.

We then applied both calibration solutions to the target field, followed by flagging visibilities using \texttt{rflag}, \texttt{tfcrop}, and \texttt{aoflagger}. At this point, we averaged visibilities over 8 channels to reduce data volume and imaged all observations (18~h in total) jointly into a single, multi-frequency synthesised image using the \texttt{wsclean} software \citep{offringa-wsclean-2014}. We used Briggs weighting with a robustness parameter of 0, yielding a $15.60'' \times 11.59''$ synthesised beam. We generated a $2.5^\circ \times 2.5^\circ$ image, covering an area slightly larger than the uGMRT field of view at these frequencies and deconvolved it using the multiscale cleaning algorithm \citep{Cornwell2008arXiv0806.2228C,Ofringa2017MNRAS.471..301O} with both the auto-mask and auto-threshold options. The auto-mask option limits the area where CLEAN components are searched for, and the auto-threshold option sets the component threshold based on the local noise rms estimate. We set the auto-mask threshold to $4\sigma$ and the auto-threshold to $1\sigma$. 

We constructed a sky model that we then used for self-calibration. We solved only for phases, averaging across frequencies, with a solution interval equal to the integration time (see Tab.~\ref{tab:observations}). After applying the phase solutions, we flagged outliers of the residual visibility distribution and repeated the calibration until no data were flagged.

\begin{figure}[t]
    \centering
    \includegraphics[width=0.45\textwidth]{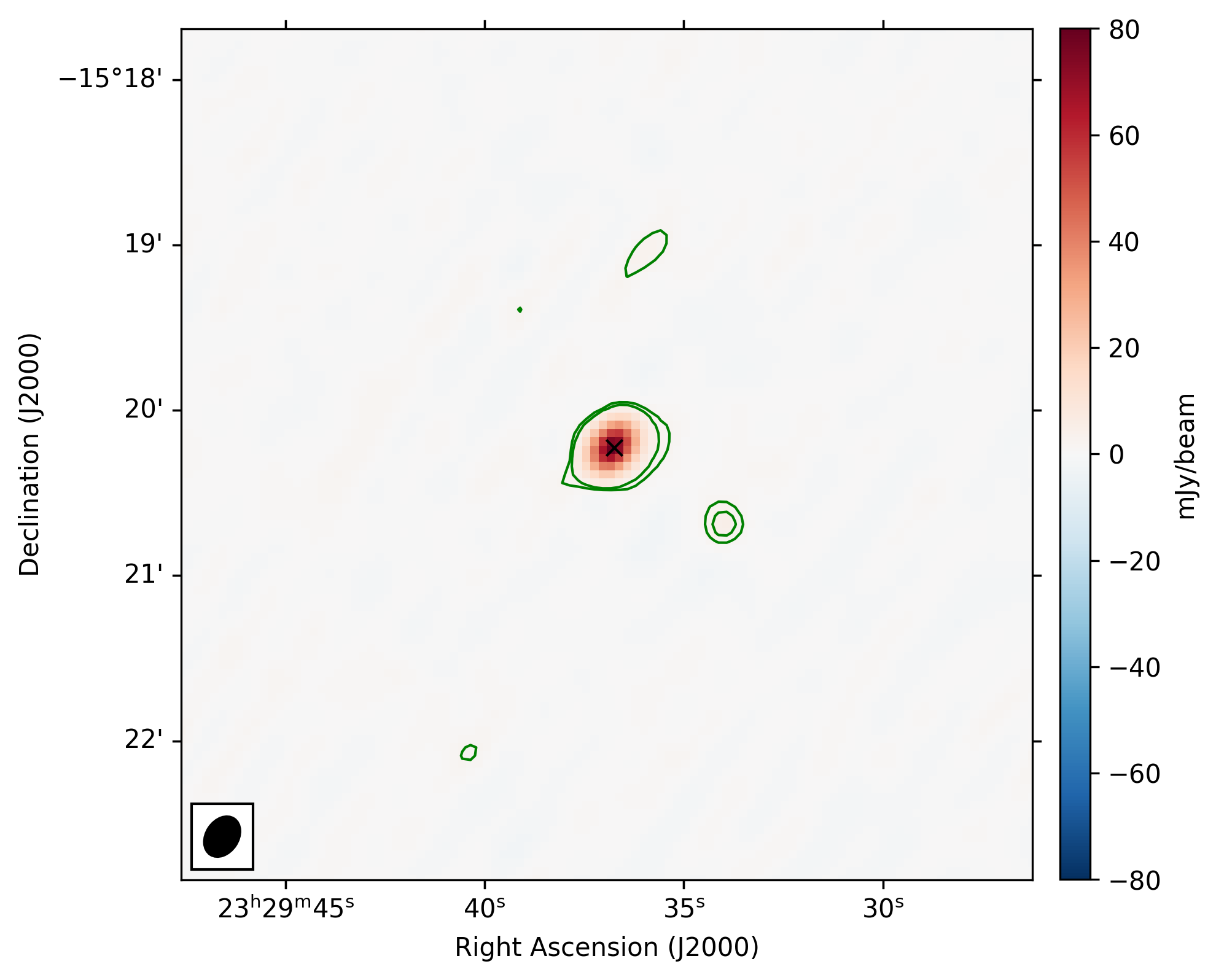}
    \caption{Continuum image of the J2329-1520  (PSO J352.4034–15.3373) quasar at 212.75~MHz. The black cross marks the pixel with peak flux density. Green contours are drawn at 3$\sigma$ and 5$\sigma$ levels, respectively, with the noise rms of $\sigma = 0.66$~mJy~beam$^{-1}$. The image angular resolution is $15.6'' \times 11.6''$.}
    \label{fig:J2329-1520_continuum_image}
    \end{figure}

We imaged the various nights separately and found flux density variations of J2329-1520 that we attributed to temporal amplitude gain variations. To correct for them, we used the CASA task \texttt{gaincal} with an integration time solution interval (see Tab.~\ref{tab:cal_solutions_ddtc007}) and where we imposed a fix on the overall normalisation - therefore preserving the source absolute flux density and the possible magnitude of the 21-cm absorption. Fig.~\ref{fig:Calibration and self-calibration workflow.} provides a flow chart of the calibration procedure.
\begin{table}[t]
    \centering
    \caption{Calibration solution intervals.}
    \begin{tabular}{lcc}
    \hline\hline
    Calibration Solution & \multicolumn{2}{c}{Solution interval} \\
    \cline{2-3}
     & Time & Frequency \\
    \hline
    bandpass & $10$~min\tablefootmark{a} & $1.5$~KHz \\
    gain (secondary) & $60$~s & $25$~MHz\tablefootmark{c} \\
    gain (self-cal) & integration time \tablefootmark{b} & $25$~MHz\tablefootmark{c} \\
    \hline
    \end{tabular}
    \tablefoot{The integration times are $5.37$, $8.05$, and $10.7$~s for three observing runs of ddtC219, first, and second observing runs of ddtC007, respectively.
    \tablefoottext{a}{Total observing time per scan}
    \tablefoottext{b}{Integration time}
    \tablefoottext{c}{Bandwidth.}
    }
    \label{tab:cal_solutions_ddtc007}
    \end{table}
The final continuum image of the target quasar is shown in Fig.~\ref{fig:J2329-1520_continuum_image}, where we achieved a noise standard deviation of 0.66~mJy~beam$^{-1}$.  
    
    \section{The spectrum of J2329-1520 and upper limits on the 21-cm absorption}\label{sec: The spectrum of J2329-1520 and upper limits on the 21-cm absorption}

\begin{figure}[t]
    \centering
    \includegraphics[width=\linewidth]{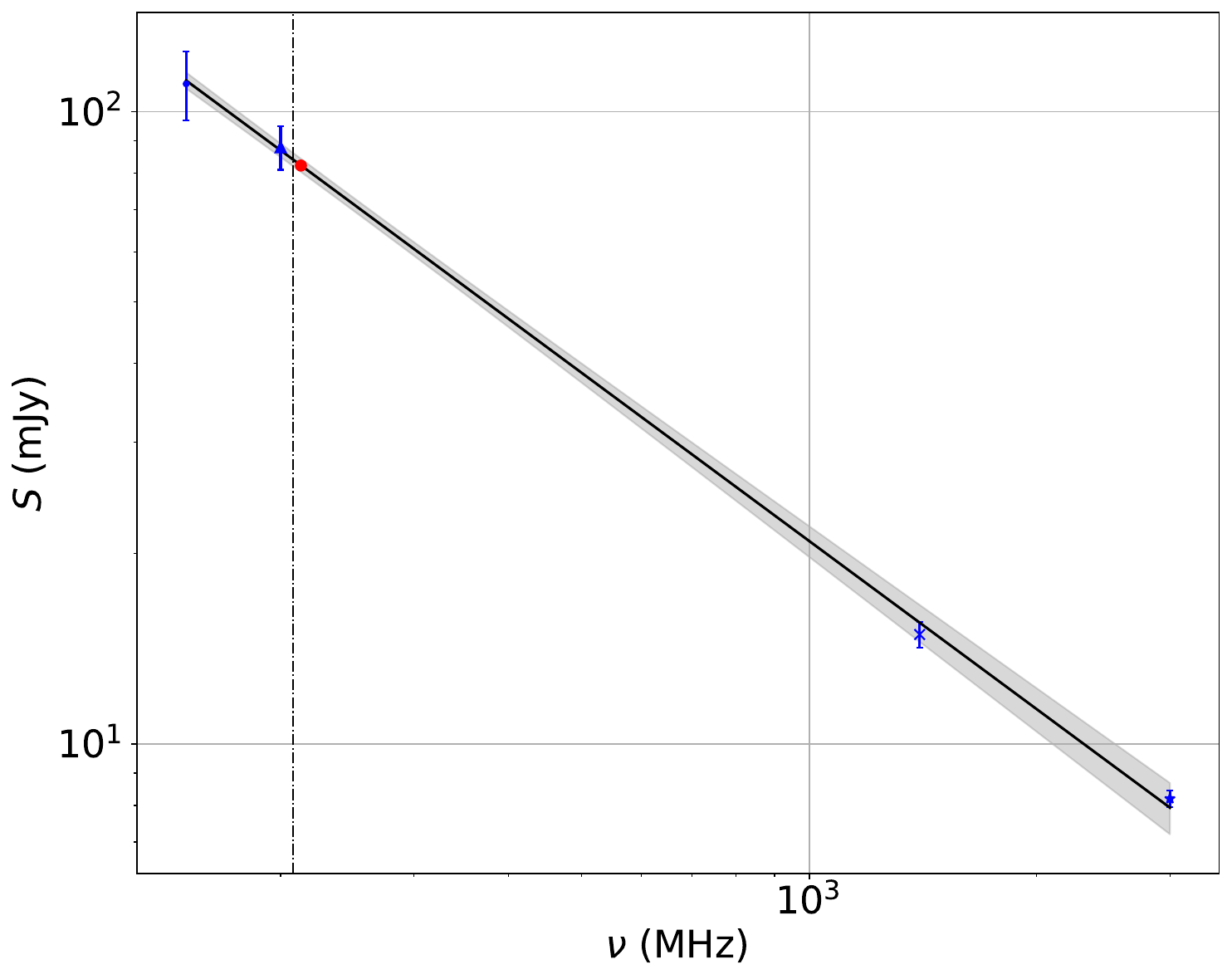} 
    \caption{The radio continuum spectrum of J2329-1520 from 150~MHz to 3.0~GHz. Blue points are the continuum measurements in the literature (Tab.~\ref{tab:J2329-1520_External_measurement}), while the red one is from this work. The power-law best fit of the radio continuum is the black line, where the grey contour shows the 95\% C.L. of the fit.}
    \label{fig:J2329-1520_continuum_spectrum_plot}
    \end{figure}

\begin{figure}[t]
    \centering
    \includegraphics[width=\linewidth]{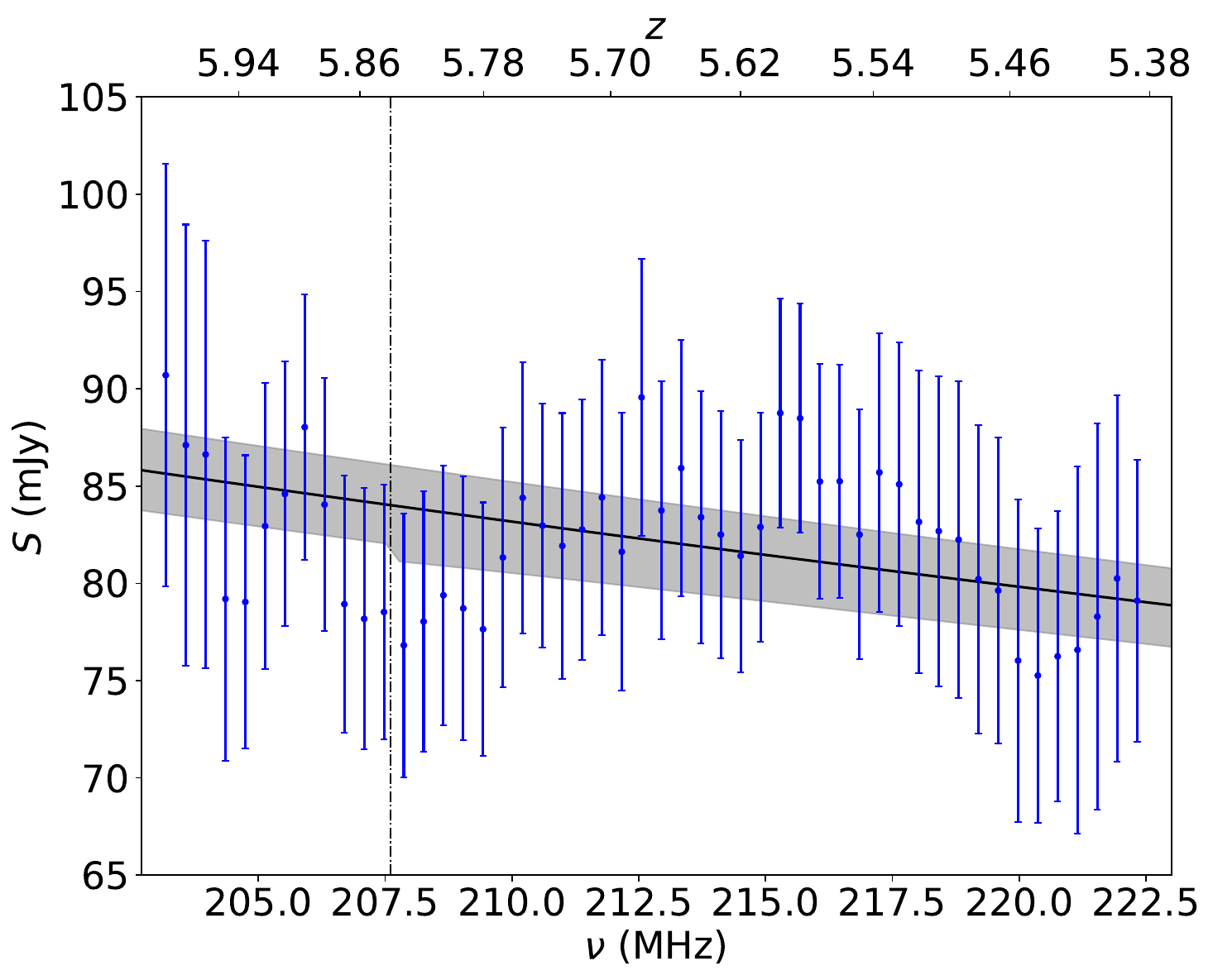} 
    \caption{Spectrum of the J2329-1520 quasar. The frequency band $203 - 222.5$~MHz is divided into 64 output channels (each 0.39~MHz wide). The blue data points are the fitted flux densities with 2$\sigma$ uncertainties in each output channel image estimated from \texttt{PyBDSF} (see text for details). The power-law fit of the radio continuum is the black line. The grey contour indicates the 95\%~C.I. of the best-fit 21-cm absorption \textit{global model} 
    (see text for details). The vertical dashed line indicates the source's redshift.}
    \label{fig:J2329-1520_absorption_contour}
    \end{figure}

    The \texttt{wsclean} package jointly outputs a multifrequency-synthesised, deconvolved image and the single-channel, deconvolved images that, in our case, we set to be 390~kHz wide. To generate a spectrum of the J2329-1530 source, we ran the \texttt{pyBDSF} source finder \citep{Mohan2015ascl.soft02007M} on each channel image to perform a Gaussian fit on the target source. The frequency-averaged source flux density is shown in Fig.~\ref{fig:J2329-1520_continuum_spectrum_plot} along with its broad-band spectrum that includes literature measurements. The source spectrum from the uGMRT data alone is shown in Fig.~\ref{fig:J2329-1520_absorption_contour}. Note that the first and last 6 output channels are excluded since they fall outside the overlapping frequency range of the two observations. Therefore, the effective number of output channels is 50 rather than 64. The noise rms per output channel in Fig.~\ref{fig:J2329-1520_absorption_contour} is taken from \texttt{pyBDSF}  uncertainty estimates of the source flux density; this estimate gives an uncertainty of $\sim 3.55$~mJy~beam$^{-1}$ per channel, consistent with the  $3.53$~mJy~beam$^{-1}$ median rms noise per channel. The median rms noise per channel was estimated as the median of the rms noise distributions, where the rms noise was obtained from the central $100 \times 100$ pixels ($300''\times300''$) of the channel residual image - after subtracting the sky model. We note that if we scale down the per-channel rms noise using the radiometer equation to estimate the noise in the multi-frequency synthesised image, we obtain a 0.45~mJy~beam$^{-1}$ rms noise, a factor of 50\% difference compared to what we actually measured (Fig.~\ref{fig:J2329-1520_continuum_image}). Future work will be devoted to understanding this discrepancy, although it is possible that residual RFI and sidelobe noise are not completely averaged down in the multifrequency averaged image.

    Finally, we provided an independent noise estimate from the Stokes V image, which is expected to contain no sky signal. The multi-frequency synthesised Stokes V image has a 0.3~mJy~beam$^{-1}$ rms noise (see Fig.~\ref{fig:StokesV_histogram}) that can be scaled to a channel noise of 2.4~mJy~beam$^{-1}$, again 50\% smaller than what we estimated from the Stokes I image cube. Our final noise assumption of 3.6~mJy~beam$^{-1}$ for a 390~kHz channel width may therefore appear somewhat conservative, although in line with estimates obtained with different methods.
    
\begin{figure}[t]
        \centering
        \includegraphics[width=\linewidth]{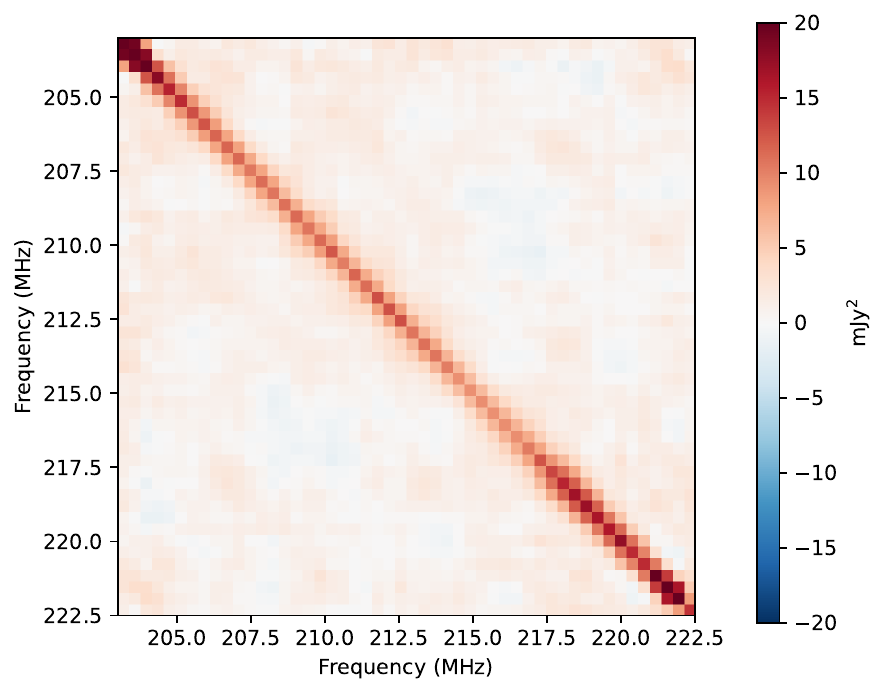}
        \caption{Covariance matrix estimated from the J2329-1520 spectrum.}
        \label{fig:J2329-1520_covariance}
    \end{figure}

    We fitted the observed spectrum using Eq.~\ref{eqn: power-law flux model}, i.e. a joint fit of the continuum and 21-cm optical depth parameters. In sampling the likelihood (Eq.~\ref{likelihood_function}), we relaxed the assumption of a diagonal covariance matrix, and estimated the full covariance matrix from the data themselves. We followed \cite{deOliveira-Costa2008MNRAS.388..247D} and \cite{bernardi13} and estimated the covariance matrix from the image cube, using a $100\times100$ pixels ($300''\times300''$) region per channel centred on the quasar's position in the residual image:
    \begin{equation}
        \bm{C}=\frac{1}{N_\text{pix}}\sum_{n=1}^{N_\text{pix}} \bm{x}_n\bm{x}_n^T,
    \end{equation}
    where $N_\text{pix}=10^4$ is the number of image pixels, and 
    $\bm{x}_n$ is a vector of $n^{\text{th}}$-pixel flux density in the cropped residual image in each frequency output channel. $\bm{C}$ is therefore a $50\times50$ matrix (Fig.~\ref{fig:J2329-1520_covariance}). The median of square-rooted diagonal terms on the covariance matrix is the rms noise from the residual image cube $3.53$~mJy~beam$^{-1}$ per $390$~kHz output channel, which is consistent with the PyBDSF estimates as mentioned before. The off-diagonal terms are non-zero but significantly smaller, indicating that the flux density measurements in each channel are not significantly correlated with one another.
    
    \subsection{Constraints on the 21-cm absorption optical depth: \textit{global model}}\label{sec: EoR Constraint: Continuous Absorption}

\begin{figure}[t]
        \centering
        \includegraphics[width=\linewidth]{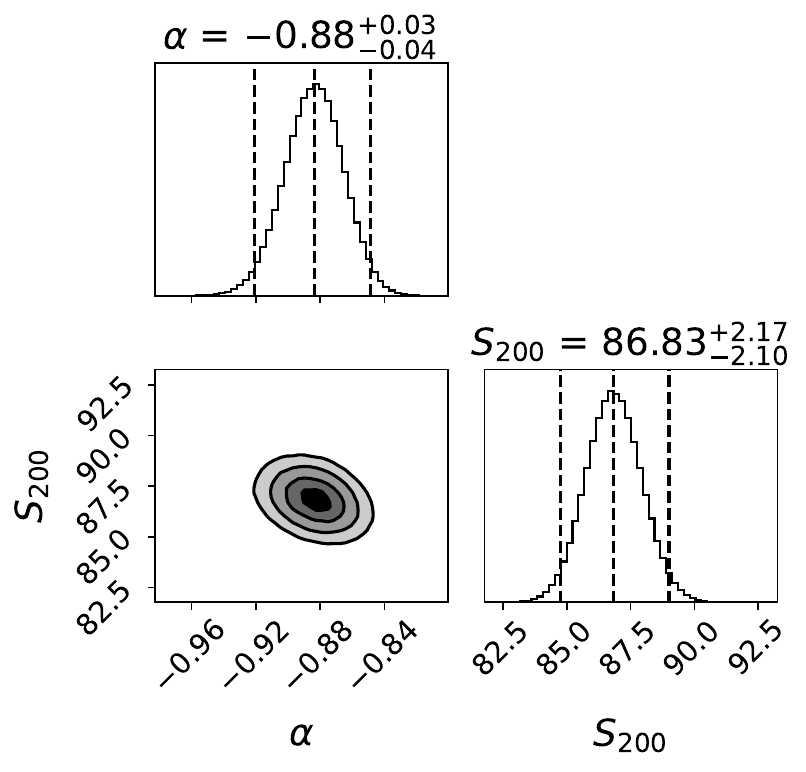}
        \caption{Same as Fig.~\ref{fig:sim1_continuum_parameter_estimation}, but for the J2329-1520 observed spectrum (also including external literature measurements in Tab.~\ref{tab:J2329-1520_External_measurement}).}
         \label{fig:J2329-1520_continuum_parameter_estimation}
    \end{figure}

    We fitted the \textit{global model} of the 21-cm optical depth (see Section \ref{sec: Global Model}) to the observed spectrum, following the same procedure as for the simulated cases. We obtained better measurements of the continuum spectrum (Fig.~\ref{fig:J2329-1520_continuum_parameter_estimation}): compared to the fit using the archival measurements alone in Tab.~\ref{tab:radio-loud quasars sample}), we tightened the 95\% C.I. on the flux density normalisation, $S_{200}$, from $[74.21,95.25]$ to $[84.73,89.00]$~mJy, and the spectral index, $\alpha$, from $[-0.93,-0.81]$ to $[-0.92,-0.85]$.

\begin{figure}[t]
    \includegraphics[width=\linewidth]{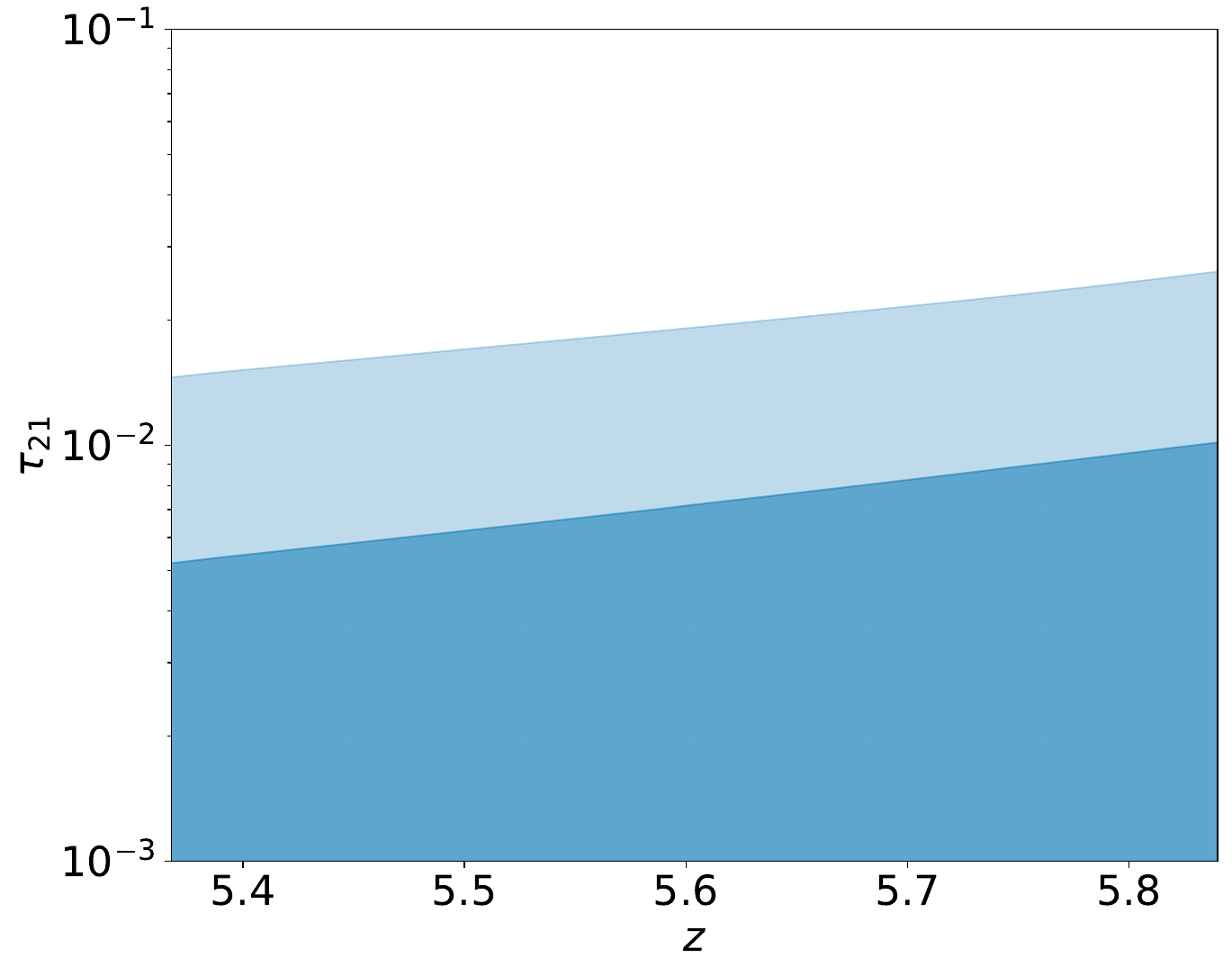}
    \caption{Same as Fig.~\ref{fig:sim2_tau}, but for the J2329-1520 observed spectrum. The redshift range, from 5.38 to 5.84, corresponds to the lower edge of the frequency window (222.5~MHz) and the redshift of J2329-1520.}
    \label{fig:J2329-1520_tau}
    \end{figure}

    We placed an upper limit on the 21-cm optical depth in the $5.38 < z < 5.84$ range, ranging from $2.6\times10^{-2}$ at $z = 5.84$ down to $1.5\times10^{-2}$ at $z = 5.38$ at $95\%$~C.L. (Fig.~\ref{fig:J2329-1520_tau}). The $95\%$~C.I. of the best fit model is shown as a grey contour in Fig.~\ref{fig:J2329-1520_absorption_contour}. We note that constraints obtained on both the continuum and 21-cm optical depth parameters are consistent with what is presented in the simulated case (Section~\ref{sec:Simulation_forecast}), indicating a negligible contribution from systematic errors.
    
    \subsection{Constraints on the 21 cm absorption optical depth: \textit{island}}

    We fitted the \textit{island model} (Section~\ref{sec: Island Model}) to the J2329-1520 spectrum, aiming to constrain the presence of a residual, neutral island gas at the end of reionization. We assumed a uniform prior on the island frequency centre $\nu_p$ within the observed frequency range (Tab.~\ref{tab: Fiducial value and Prior}) and a uniform prior on $\Delta\nu$ between $0.39$~MHz and 5~MHz, corresponding to a maximum comoving size of $17$~Mpc for the neutral IGM island.
    
\begin{figure}[t]
        \centering
        \includegraphics[width=\linewidth]{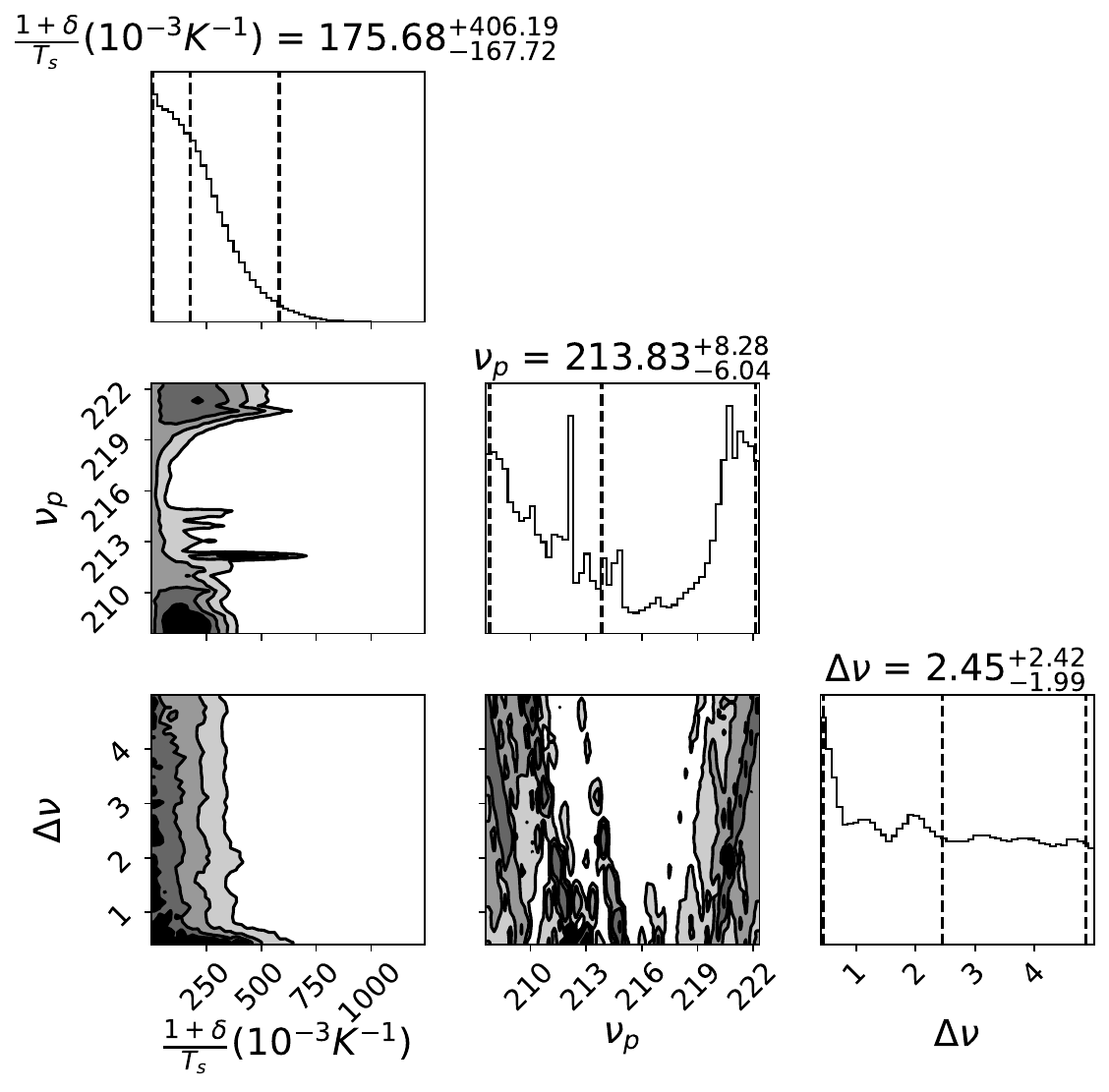}
        \caption{Posterior probability distributions for the 21-cm optical depth, \textit{island model} parameters obtained from the J2329-1520 spectrum (see text for details). The constraints on the ratio between the density contrast and the spin temperature ($\frac{1+\delta}{T_s}$), the midpoint ($\nu_p$), and the width of the absorption ($\Delta\nu$), which are the parameters of the absorption model, are obtained by fitting the observed spectrum of J2329-1520 and the external measurements in Tab.~\ref{tab:J2329-1520_External_measurement} jointly with the continuum parameters ($S_\text{200}$ and $\alpha$) as nuisance parameters. Dashed vertical lines mark the 95\%~C.L., with the central dashed line drawn at the 50$^{\rm th}$ percentile. The corresponding values are reported above the histograms.}
        \label{fig:J2329-1520_discrete_parameter_estimation}
    \end{figure}

\begin{figure}[t]
    \centering
    \includegraphics[width=\linewidth]{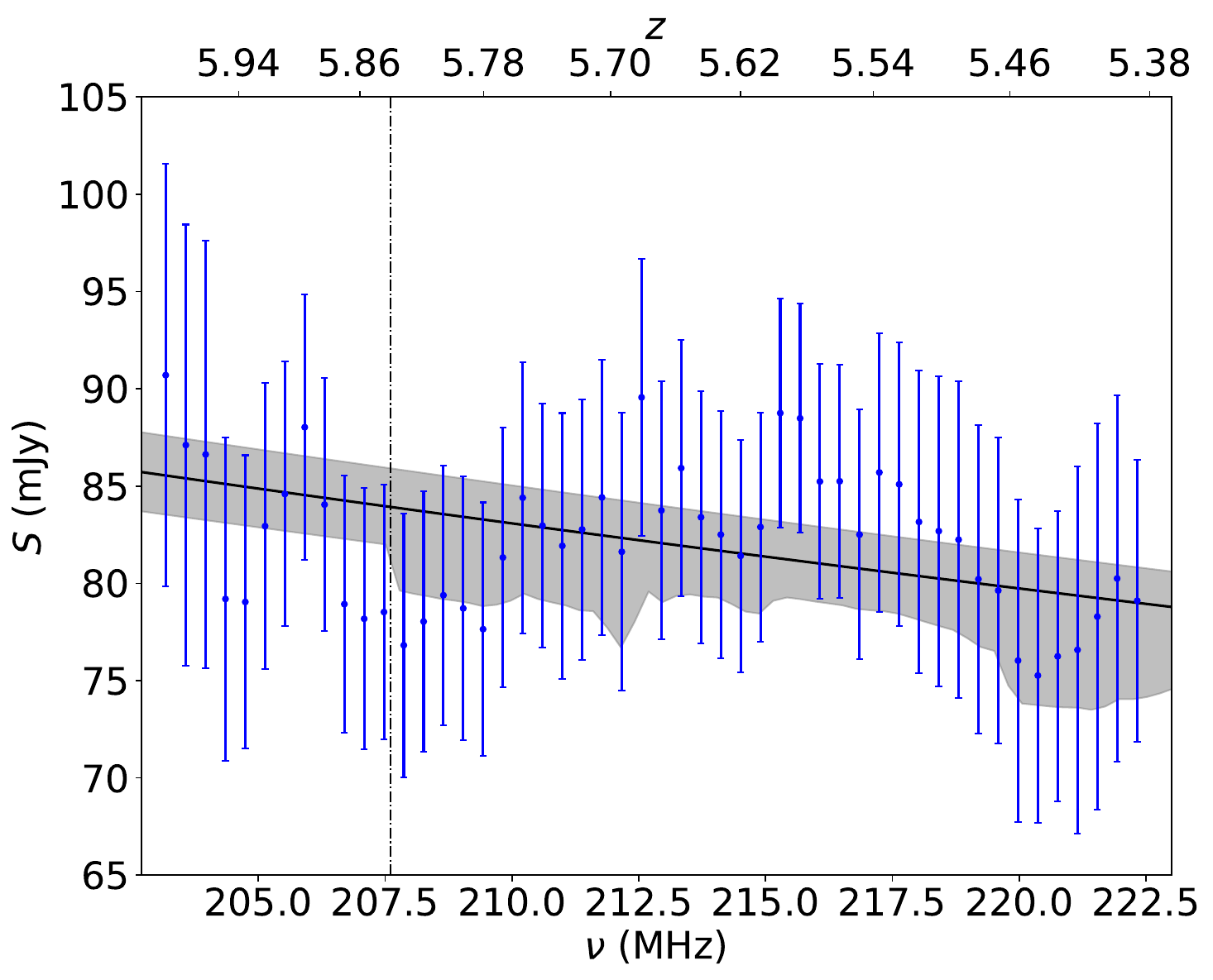} 
    \caption{Same as Fig.~\ref{fig:J2329-1520_absorption_contour}, but for the \textit{island model}. Like the case of the \textit{global model}, the best fit model - i.e. the best fit continuum spectrum with the limits on the 21-cm absorption - are consistent with each frequency bin of the observed spectrum.}
    \label{fig:J2329-1520_discrete_absorption_contour}
    \end{figure}

    The resulting posterior distributions are shown in Fig.~\ref{fig:J2329-1520_discrete_parameter_estimation}, and the $95\%$~C.L. best fit model is plotted on top of the observed spectrum in Fig.~\ref{fig:J2329-1520_discrete_absorption_contour}. We found no evidence for HI regions: the distribution of the width and position of the HI island are rather uniform across the prior range. The posterior distribution of the optical depth magnitude, which depends essentially upon the spin temperature, shows a monotonically decreasing behaviour with decreasing spin temperature (top panel of Fig.~\ref{fig:J2329-1520_discrete_parameter_estimation}), disfavouring very cold HI regions. We placed a 95\% C.L. lower limit on the temperature of the HI region to be $T_s > 1.73$~K (assuming $\delta = 0$). We note that the best fit model - e.g. a power law continuum with no absorption - is consistent with all the spectrum bins within the 95\% uncertainties.

    We finally computed the adiabatic cooling temperature limit at $5.38<z<5.84$ as: 
    
    \begin{equation}
        T_c = 2.73 \, \frac{(1 + z)^2}{1 + z_d} = 0.74{-}0.85~\text{K},
    \end{equation}
    
    where we assumed the redshift of thermal decoupling between the IGM and the CMB to be $z_d = 150$. We can see that our limits provide independent evidence for IGM heating at $z > 5.84$, as largely expected by simulations and consistent with EoR observations at higher redshift \citep{HERA2023ApJ...945..124H,Ghara2024MNRAS.530..191G,Nunhokee2025ApJ...989...57N}.
   
\section{Summary and Conclusions}\label{sec: Summary and Conclusions}

In this paper, we investigated the detectability of the 21-cm absorption against high-redshift radio sources in order to probe the IGM at $z > 5.5$ and presented the 21-cm absorption spectrum against the bright radio quasar J2329-1520 in the $5.38 < z < 5.84$ range.

We first use simulations of 21-cm absorption spectra against high-redshift quasars: for the first time, we jointly fit the continuum, synchrotron spectrum, and the 21-cm absorption depth, for which we constructed a model that resembles the redshift evolution of the volume-average IGM HI and temperature - what we labelled as a \textit{global model}. We validated our method in the case of a hypothetical $\sim 600$~mJy source at $z = 8$, where we found that it is possible to reconstruct the HI and temperature evolution of the IGM over the $5 < z < 8$ range, assuming a 50~$\mu$Jy rms noise per 390~kHz. 
We then showed that, with the same noise level, it is only possible to place an upper limit on the 21-cm optical depth when a sample of seven known quasars at $z > 5.5$ is used. As the noise estimate is tailored to what can be achieved in 40~h with the AA* SKAO configuration, our simulations were intended to guide early 21-cm absorption projects with the SKAO.

We then derived the J2329-1520 spectrum from the analysis of uGMRT archive observations in the $203 - 222.5$~MHz range. We obtained a continuum image with a 0.66~mJy~beam$^{-1}$ rms noise and presented a spectrum binned in 390~kHz-wide channels, with a 3.6~mJy~beam$^{-1}$ rms noise per channel. We use a 0 robust weight in generating both the continuum and the channel images, yielding a $15.6'' \times 11.6''$ resolution - a choice of weighting scheme that optimally trades angular resolution for sensitivity. We fitted both the \textit{global model} and the \textit{island model} of the 21-cm optical depth to the observed spectrum. The \textit{island model} describes the 21-cm optical depth that can arise from a patch of HI that survived the reionization process.

In both cases, we constrained the continuum spectrum accurately, without finding evidence for 21-cm absorption. In the case of the \textit{global model}, we found that the 21-cm optical depth is lower than $1.5 \times 10^{-2}$ at $5.38<z<5.84$ at 95\% C.L. In the \textit{Island model} case, the upper limit on the optical depth can be turned into a lower limit on the HI region temperature, and we found that such HI regions are warmer than 1.73~K at 95\% C.L. 

\begin{acknowledgements}
      Part of this work was supported by the German \emph{Deut\-sche For\-schungs\-ge\-mein\-schaft, DFG\/} project
      number Ts~17/2--1. EC and GB acknowledge support from the Ministry of Universities and Research (MUR) through the PRIN project ‘Optimal inference from radio images of the epoch of reionization’. CK acknowledges support from the INAF SKAlow programme. The authors thank Tom\'a\v s \v Soltinsk\' y for useful comments and discussion.
\end{acknowledgements}

\bibliographystyle{aa}
\bibliography{references}

@ARTICLE{Perley&Butler2017ApJS..230....7P,
       author = {{Perley}, R.~A. and {Butler}, B.~J.},
        title = "{An Accurate Flux Density Scale from 50 MHz to 50 GHz}",
      journal = {\apjs},
         year = 2017,
        month = may,
       volume = {230},
       number = {1},
          eid = {7},
        pages = {7},
          doi = {10.3847/1538-4365/aa6df9},
archivePrefix = {arXiv},
       eprint = {1609.05940},
 primaryClass = {astro-ph.IM},
       adsurl = {https://ui.adsabs.harvard.edu/abs/2017ApJS..230....7P}
}

@ARTICLE{Banados2018ApJ...861L..14B,
       author = {{Ba{\~n}ados}, Eduardo and {Carilli}, Chris and {Walter}, Fabian and {Momjian}, Emmanuel and {Decarli}, Roberto and {Farina}, Emanuele P. and {Mazzucchelli}, Chiara and {Venemans}, Bram P.},
        title = "{A Powerful Radio-loud Quasar at the End of Cosmic Reionization}",
      journal = {\apjl},
         year = 2018,
        month = jul,
       volume = {861},
       number = {2},
          eid = {L14},
        pages = {L14},
          doi = {10.3847/2041-8213/aac511},
archivePrefix = {arXiv},
       eprint = {1807.02531},
 primaryClass = {astro-ph.GA},
       adsurl = {https://ui.adsabs.harvard.edu/abs/2018ApJ...861L..14B}
}

@ARTICLE{Furlanetto2006PhR...433..181F,
       author = {{Furlanetto}, Steven R. and {Oh}, S. Peng and {Briggs}, Frank H.},
        title = "{Cosmology at low frequencies: The 21 cm transition and the high-redshift Universe}",
      journal = {\physrep},
         year = 2006,
        month = oct,
       volume = {433},
       number = {4-6},
        pages = {181-301},
          doi = {10.1016/j.physrep.2006.08.002},
archivePrefix = {arXiv},
       eprint = {astro-ph/0608032},
 primaryClass = {astro-ph},
       adsurl = {https://ui.adsabs.harvard.edu/abs/2006PhR...433..181F}
}

@ARTICLE{Planck2020A&A...641A...6P,
       author = {{Planck Collaboration} and {Aghanim}, N. and {Akrami}, Y. and {Ashdown}, M. and {Aumont}, J. and {Baccigalupi}, C. and {Ballardini}, M. and {Banday}, A.~J. and {Barreiro}, R.~B. and {Bartolo}, N. and {Basak}, S. and {Battye}, R. and {Benabed}, K. and {Bernard}, J. -P. and {Bersanelli}, M. and {Bielewicz}, P. and {Bock}, J.~J. and {Bond}, J.~R. and {Borrill}, J. and {Bouchet}, F.~R. and {Boulanger}, F. and {Bucher}, M. and {Burigana}, C. and {Butler}, R.~C. and {Calabrese}, E. and {Cardoso}, J. -F. and {Carron}, J. and {Challinor}, A. and {Chiang}, H.~C. and {Chluba}, J. and {Colombo}, L.~P.~L. and {Combet}, C. and {Contreras}, D. and {Crill}, B.~P. and {Cuttaia}, F. and {de Bernardis}, P. and {de Zotti}, G. and {Delabrouille}, J. and {Delouis}, J. -M. and {Di Valentino}, E. and {Diego}, J.~M. and {Dor{\'e}}, O. and {Douspis}, M. and {Ducout}, A. and {Dupac}, X. and {Dusini}, S. and {Efstathiou}, G. and {Elsner}, F. and {En{\ss}lin}, T.~A. and {Eriksen}, H.~K. and {Fantaye}, Y. and {Farhang}, M. and {Fergusson}, J. and {Fernandez-Cobos}, R. and {Finelli}, F. and {Forastieri}, F. and {Frailis}, M. and {Fraisse}, A.~A. and {Franceschi}, E. and {Frolov}, A. and {Galeotta}, S. and {Galli}, S. and {Ganga}, K. and {G{\'e}nova-Santos}, R.~T. and {Gerbino}, M. and {Ghosh}, T. and {Gonz{\'a}lez-Nuevo}, J. and {G{\'o}rski}, K.~M. and {Gratton}, S. and {Gruppuso}, A. and {Gudmundsson}, J.~E. and {Hamann}, J. and {Handley}, W. and {Hansen}, F.~K. and {Herranz}, D. and {Hildebrandt}, S.~R. and {Hivon}, E. and {Huang}, Z. and {Jaffe}, A.~H. and {Jones}, W.~C. and {Karakci}, A. and {Keih{\"a}nen}, E. and {Keskitalo}, R. and {Kiiveri}, K. and {Kim}, J. and {Kisner}, T.~S. and {Knox}, L. and {Krachmalnicoff}, N. and {Kunz}, M. and {Kurki-Suonio}, H. and {Lagache}, G. and {Lamarre}, J. -M. and {Lasenby}, A. and {Lattanzi}, M. and {Lawrence}, C.~R. and {Le Jeune}, M. and {Lemos}, P. and {Lesgourgues}, J. and {Levrier}, F. and {Lewis}, A. and {Liguori}, M. and {Lilje}, P.~B. and {Lilley}, M. and {Lindholm}, V. and {L{\'o}pez-Caniego}, M. and {Lubin}, P.~M. and {Ma}, Y. -Z. and {Mac{\'\i}as-P{\'e}rez}, J.~F. and {Maggio}, G. and {Maino}, D. and {Mandolesi}, N. and {Mangilli}, A. and {Marcos-Caballero}, A. and {Maris}, M. and {Martin}, P.~G. and {Martinelli}, M. and {Mart{\'\i}nez-Gonz{\'a}lez}, E. and {Matarrese}, S. and {Mauri}, N. and {McEwen}, J.~D. and {Meinhold}, P.~R. and {Melchiorri}, A. and {Mennella}, A. and {Migliaccio}, M. and {Millea}, M. and {Mitra}, S. and {Miville-Desch{\^e}nes}, M. -A. and {Molinari}, D. and {Montier}, L. and {Morgante}, G. and {Moss}, A. and {Natoli}, P. and {N{\o}rgaard-Nielsen}, H.~U. and {Pagano}, L. and {Paoletti}, D. and {Partridge}, B. and {Patanchon}, G. and {Peiris}, H.~V. and {Perrotta}, F. and {Pettorino}, V. and {Piacentini}, F. and {Polastri}, L. and {Polenta}, G. and {Puget}, J. -L. and {Rachen}, J.~P. and {Reinecke}, M. and {Remazeilles}, M. and {Renzi}, A. and {Rocha}, G. and {Rosset}, C. and {Roudier}, G. and {Rubi{\~n}o-Mart{\'\i}n}, J.~A. and {Ruiz-Granados}, B. and {Salvati}, L. and {Sandri}, M. and {Savelainen}, M. and {Scott}, D. and {Shellard}, E.~P.~S. and {Sirignano}, C. and {Sirri}, G. and {Spencer}, L.~D. and {Sunyaev}, R. and {Suur-Uski}, A. -S. and {Tauber}, J.~A. and {Tavagnacco}, D. and {Tenti}, M. and {Toffolatti}, L. and {Tomasi}, M. and {Trombetti}, T. and {Valenziano}, L. and {Valiviita}, J. and {Van Tent}, B. and {Vibert}, L. and {Vielva}, P. and {Villa}, F. and {Vittorio}, N. and {Wandelt}, B.~D. and {Wehus}, I.~K. and {White}, M. and {White}, S.~D.~M. and {Zacchei}, A. and {Zonca}, A.},
        title = "{Planck 2018 results. VI. Cosmological parameters}",
      journal = {\aap},
         year = 2020,
        month = sep,
       volume = {641},
          eid = {A6},
        pages = {A6},
          doi = {10.1051/0004-6361/201833910},
archivePrefix = {arXiv},
       eprint = {1807.06209},
 primaryClass = {astro-ph.CO},
       adsurl = {https://ui.adsabs.harvard.edu/abs/2020A&A...641A...6P}
}

@ARTICLE{Foreman-Mackey2013PASP..125..306F,
       author = {{Foreman-Mackey}, Daniel and {Hogg}, David W. and {Lang}, Dustin and {Goodman}, Jonathan},
        title = "{emcee: The MCMC Hammer}",
      journal = {\pasp},
         year = 2013,
        month = mar,
       volume = {125},
       number = {925},
        pages = {306},
          doi = {10.1086/670067},
archivePrefix = {arXiv},
       eprint = {1202.3665},
 primaryClass = {astro-ph.IM},
       adsurl = {https://ui.adsabs.harvard.edu/abs/2013PASP..125..306F}
}

@software{Mohan2015ascl.soft02007M,
       author = {{Mohan}, Niruj and {Rafferty}, David},
        title = "{PyBDSF: Python Blob Detection and Source Finder}",
 howpublished = {Astrophysics Source Code Library, record ascl:1502.007},
         year = 2015,
        month = feb,
          eid = {ascl:1502.007},
       adsurl = {https://ui.adsabs.harvard.edu/abs/2015ascl.soft02007M}
}

@ARTICLE{Soltinsky2025MNRAS.537..364S,
       author = {{{\v{S}}oltinsk{\'y}}, Tom{\'a}{\v{s}} and {Kulkarni}, Girish and {Tendulkar}, Shriharsh P. and {Bolton}, James S.},
        title = "{Prospects of a statistical detection of the 21-cm forest and its potential to constrain the thermal state of the neutral IGM during reionization}",
      journal = {\mnras},
         year = 2025,
        month = feb,
       volume = {537},
       number = {1},
        pages = {364-378},
          doi = {10.1093/mnras/staf026},
archivePrefix = {arXiv},
       eprint = {2412.06879},
 primaryClass = {astro-ph.CO},
       adsurl = {https://ui.adsabs.harvard.edu/abs/2025MNRAS.537..364S}
}

@article{Thyagarajan:2020nch,
    author = "Thyagarajan, Nithyanandan",
    title = "{Statistical Detection of IGM Structures during Cosmic Reionization using Absorption of the Redshifted 21 cm line by HI against Compact Background Radio Sources}",
    eprint = "2006.10070",
    archivePrefix = "arXiv",
    primaryClass = "astro-ph.CO",
    doi = "10.3847/1538-4357/ab9e6d",
    journal = "Astrophys. J.",
    volume = "899",
    number = "1",
    pages = "16",
    year = "2020"
}

@article{offringa-wsclean-2014,
  author = {Offringa, A. R. and McKinley, B. and Hurley-Walker and others},
  title = {{WSClean: an implementation of a fast, generic wide-field imager for radio astronomy}},
  volume = {444},
  number = {1},
  pages = {606-619},
  year = {2014},
  doi = {10.1093/mnras/stu1368},
  journal = {MNRAS}
}

@ARTICLE{NVSS1998AJ....115.1693C,
       author = {{Condon}, J.~J. and {Cotton}, W.~D. and {Greisen}, E.~W. and {Yin}, Q.~F. and {Perley}, R.~A. and {Taylor}, G.~B. and {Broderick}, J.~J.},
        title = "{The NRAO VLA Sky Survey}",
      journal = {\aj},
         year = 1998,
        month = may,
       volume = {115},
       number = {5},
        pages = {1693-1716},
          doi = {10.1086/300337},
       adsurl = {https://ui.adsabs.harvard.edu/abs/1998AJ....115.1693C}
}

@ARTICLE{GLEAM2017MNRAS.464.1146H,
       author = {{Hurley-Walker}, N. and {Callingham}, J.~R. and {Hancock}, P.~J. and {Franzen}, T.~M.~O. and {Hindson}, L. and {Kapi{\'n}ska}, A.~D. and {Morgan}, J. and {Offringa}, A.~R. and {Wayth}, R.~B. and {Wu}, C. and {Zheng}, Q. and {Murphy}, T. and {Bell}, M.~E. and {Dwarakanath}, K.~S. and {For}, B. and {Gaensler}, B.~M. and {Johnston-Hollitt}, M. and {Lenc}, E. and {Procopio}, P. and {Staveley-Smith}, L. and {Ekers}, R. and {Bowman}, J.~D. and {Briggs}, F. and {Cappallo}, R.~J. and {Deshpande}, A.~A. and {Greenhill}, L. and {Hazelton}, B.~J. and {Kaplan}, D.~L. and {Lonsdale}, C.~J. and {McWhirter}, S.~R. and {Mitchell}, D.~A. and {Morales}, M.~F. and {Morgan}, E. and {Oberoi}, D. and {Ord}, S.~M. and {Prabu}, T. and {Shankar}, N. Udaya and {Srivani}, K.~S. and {Subrahmanyan}, R. and {Tingay}, S.~J. and {Webster}, R.~L. and {Williams}, A. and {Williams}, C.~L.},
        title = "{GaLactic and Extragalactic All-sky Murchison Widefield Array (GLEAM) survey - I. A low-frequency extragalactic catalogue}",
      journal = {\mnras},
         year = 2017,
        month = jan,
       volume = {464},
       number = {1},
        pages = {1146-1167},
          doi = {10.1093/mnras/stw2337},
archivePrefix = {arXiv},
       eprint = {1610.08318},
 primaryClass = {astro-ph.GA},
       adsurl = {https://ui.adsabs.harvard.edu/abs/2017MNRAS.464.1146H}
}

@ARTICLE{TGSS2017A&A...598A..78I,
       author = {{Intema}, H.~T. and {Jagannathan}, P. and {Mooley}, K.~P. and {Frail}, D.~A.},
        title = "{The GMRT 150 MHz all-sky radio survey. First alternative data release TGSS ADR1}",
      journal = {\aap},
         year = 2017,
        month = feb,
       volume = {598},
          eid = {A78},
        pages = {A78},
          doi = {10.1051/0004-6361/201628536},
archivePrefix = {arXiv},
       eprint = {1603.04368},
 primaryClass = {astro-ph.CO},
       adsurl = {https://ui.adsabs.harvard.edu/abs/2017A&A...598A..78I}
}

@ARTICLE{Carilli2002ApJ...577...22C,
       author = {{Carilli}, C.~L. and {Gnedin}, N.~Y. and {Owen}, F.},
        title = "{H I 21 Centimeter Absorption beyond the Epoch of Reionization}",
      journal = {\apj},
         year = 2002,
        month = sep,
       volume = {577},
       number = {1},
        pages = {22-30},
          doi = {10.1086/342179},
archivePrefix = {arXiv},
       eprint = {astro-ph/0205169},
 primaryClass = {astro-ph},
       adsurl = {https://ui.adsabs.harvard.edu/abs/2002ApJ...577...22C}
}

@ARTICLE{Ciardi2013MNRAS.428.1755C,
       author = {{Ciardi}, B. and {Labropoulos}, P. and {Maselli}, A. and {Thomas}, R. and {Zaroubi}, S. and {Graziani}, L. and {Bolton}, J.~S. and {Bernardi}, G. and {Brentjens}, M. and {de Bruyn}, A.~G. and {Daiboo}, S. and {Harker}, G.~J.~A. and {Jelic}, V. and {Kazemi}, S. and {Koopmans}, L.~V.~E. and {Martinez}, O. and {Mellema}, G. and {Offringa}, A.~R. and {Pandey}, V.~N. and {Schaye}, J. and {Veligatla}, V. and {Vedantham}, H. and {Yatawatta}, S.},
        title = "{Prospects for detecting the 21 cm forest from the diffuse intergalactic medium with LOFAR}",
      journal = {\mnras},
         year = 2013,
        month = jan,
       volume = {428},
       number = {2},
        pages = {1755-1765},
          doi = {10.1093/mnras/sts156},
archivePrefix = {arXiv},
       eprint = {1209.2615},
 primaryClass = {astro-ph.CO},
       adsurl = {https://ui.adsabs.harvard.edu/abs/2013MNRAS.428.1755C}
}

@ARTICLE{Lewis2006MNRAS.373..561L,
       author = {{Lewis}, Antony and {Weller}, Jochen and {Battye}, Richard},
        title = "{The cosmic microwave background and the ionization history of the Universe}",
      journal = {\mnras},
         year = 2006,
        month = dec,
       volume = {373},
       number = {2},
        pages = {561-570},
          doi = {10.1111/j.1365-2966.2006.10983.x},
archivePrefix = {arXiv},
       eprint = {astro-ph/0606552},
 primaryClass = {astro-ph},
       adsurl = {https://ui.adsabs.harvard.edu/abs/2006MNRAS.373..561L}
}

@ARTICLE{Banados2025NatAs...9..293B,
       author = {{Ba{\~n}ados}, Eduardo and {Momjian}, Emmanuel and {Connor}, Thomas and {Belladitta}, Silvia and {Decarli}, Roberto and {Mazzucchelli}, Chiara and {Venemans}, Bram P. and {Walter}, Fabian and {Wang}, Feige and {Xie}, Zhang-Liang and {Barth}, Aaron J. and {Eilers}, Anna-Christina and {Fan}, Xiaohui and {Khusanova}, Yana and {Schindler}, Jan-Torge and {Stern}, Daniel and {Yang}, Jinyi and {Andika}, Irham Taufik and {Carilli}, Christopher L. and {Farina}, Emanuele P. and {Fabian}, Andrew and {Hennawi}, Joseph F. and {Pensabene}, Antonio and {Rojas-Ruiz}, Sof{\'\i}a},
        title = "{A blazar in the epoch of reionization}",
      journal = {Nature Astronomy},
         year = 2025,
        month = feb,
       volume = {9},
        pages = {293-301},
          doi = {10.1038/s41550-024-02431-4},
archivePrefix = {arXiv},
       eprint = {2407.07236},
 primaryClass = {astro-ph.GA},
       adsurl = {https://ui.adsabs.harvard.edu/abs/2025NatAs...9..293B}
}

@ARTICLE{Banados2021ApJ...909...80B,
       author = {{Ba{\~n}ados}, Eduardo and {Mazzucchelli}, Chiara and {Momjian}, Emmanuel and {Eilers}, Anna-Christina and {Wang}, Feige and {Schindler}, Jan-Torge and {Connor}, Thomas and {Andika}, Irham Taufik and {Barth}, Aaron J. and {Carilli}, Chris and {Davies}, Frederick B. and {Decarli}, Roberto and {Fan}, Xiaohui and {Farina}, Emanuele Paolo and {Hennawi}, Joseph F. and {Pensabene}, Antonio and {Stern}, Daniel and {Venemans}, Bram P. and {Wenzl}, Lukas and {Yang}, Jinyi},
        title = "{The Discovery of a Highly Accreting, Radio-loud Quasar at z = 6.82}",
      journal = {\apj},
         year = 2021,
        month = mar,
       volume = {909},
       number = {1},
          eid = {80},
        pages = {80},
          doi = {10.3847/1538-4357/abe239},
archivePrefix = {arXiv},
       eprint = {2103.03295},
 primaryClass = {astro-ph.CO},
       adsurl = {https://ui.adsabs.harvard.edu/abs/2021ApJ...909...80B}
}

@ARTICLE{Drouart2020PASA...37...26D,
       author = {{Drouart}, Guillaume and {Seymour}, Nick and {Galvin}, Tim J. and {Afonso}, Jose and {Callingham}, Joseph R. and {De Breuck}, Carlos and {Johnston-Hollitt}, Melanie and {Kapi{\'n}ska}, Anna D. and {Lehnert}, Matthew D. and {Vernet}, Jo{\"e}l},
        title = "{The GLEAMing of the first supermassive black holes}",
      journal = {\pasa},
         year = 2020,
        month = jul,
       volume = {37},
          eid = {e026},
        pages = {e026},
          doi = {10.1017/pasa.2020.6},
archivePrefix = {arXiv},
       eprint = {2111.08104},
 primaryClass = {astro-ph.GA},
       adsurl = {https://ui.adsabs.harvard.edu/abs/2020PASA...37...26D}
}

@ARTICLE{Gloudemans2023,
       author = {{Gloudemans}, A.~J. and {Saxena}, A. and {Intema}, H. and {Callingham}, J.~R. and {Duncan}, K.~J. and {R{\"o}ttgering}, H.~J.~A. and {Belladitta}, S. and {Hardcastle}, M.~J. and {Harikane}, Y. and {Spingola}, C.},
        title = "{No strong radio absorption detected in the low-frequency spectra of radio-loud quasars at z > 5.6}",
      journal = {\aap},
         year = 2023,
        month = oct,
       volume = {678},
          eid = {A128},
        pages = {A128},
          doi = {10.1051/0004-6361/202347582},
archivePrefix = {arXiv},
       eprint = {2309.03936},
 primaryClass = {astro-ph.GA},
       adsurl = {https://ui.adsabs.harvard.edu/abs/2023A&A...678A.128G}
}

@ARTICLE{Offringa2012A&A...539A..95O,
       author = {{Offringa}, A.~R. and {van de Gronde}, J.~J. and {Roerdink}, J.~B.~T.~M.},
        title = "{A morphological algorithm for improving radio-frequency interference detection}",
      journal = {\aap},
         year = 2012,
        month = mar,
       volume = {539},
          eid = {A95},
        pages = {A95},
          doi = {10.1051/0004-6361/201118497},
archivePrefix = {arXiv},
       eprint = {1201.3364},
 primaryClass = {astro-ph.IM},
       adsurl = {https://ui.adsabs.harvard.edu/abs/2012A&A...539A..95O}
}

@ARTICLE{Keller2024MNRAS.528.5692K,
       author = {{Keller}, Pascal M. and {Thyagarajan}, Nithyanandan and {Kumar}, Ajay and {Kanekar}, Nissim and {Bernardi}, Gianni},
        title = "{The radio-loud fraction of quasars at z > 6}",
      journal = {\mnras},
         year = 2024,
        month = mar,
       volume = {528},
       number = {4},
        pages = {5692-5702},
          doi = {10.1093/mnras/stae418},
archivePrefix = {arXiv},
       eprint = {2402.08732},
 primaryClass = {astro-ph.GA},
       adsurl = {https://ui.adsabs.harvard.edu/abs/2024MNRAS.528.5692K}
}

@ARTICLE{Frey2011AA...531L...5F,
       author = {{Frey}, S. and {Paragi}, Z. and {Gurvits}, L.~I. and {Gab{\'a}nyi}, K. {\'E}. and {Cseh}, D.},
        title = "{Into the central 10 pc of the most distant known radio quasar. VLBI imaging observations of J1429+5447 at z = 6.21}",
      journal = {\aap},
         year = 2011,
        month = jul,
       volume = {531},
          eid = {L5},
        pages = {L5},
          doi = {10.1051/0004-6361/201117341},
archivePrefix = {arXiv},
       eprint = {1106.0717},
 primaryClass = {astro-ph.CO},
       adsurl = {https://ui.adsabs.harvard.edu/abs/2011A&A...531L...5F}
}

@ARTICLE{LoTSS2017AA...598A.104S,
       author = {{Shimwell}, T.~W. and {R{\"o}ttgering}, H.~J.~A. and {Best}, P.~N. and {Williams}, W.~L. and {Dijkema}, T.~J. and {de Gasperin}, F. and {Hardcastle}, M.~J. and {Heald}, G.~H. and {Hoang}, D.~N. and {Horneffer}, A. and {Intema}, H. and {Mahony}, E.~K. and {Mandal}, S. and {Mechev}, A.~P. and {Morabito}, L. and {Oonk}, J.~B.~R. and {Rafferty}, D. and {Retana-Montenegro}, E. and {Sabater}, J. and {Tasse}, C. and {van Weeren}, R.~J. and {Br{\"u}ggen}, M. and {Brunetti}, G. and {Chy{\.z}y}, K.~T. and {Conway}, J.~E. and {Haverkorn}, M. and {Jackson}, N. and {Jarvis}, M.~J. and {McKean}, J.~P. and {Miley}, G.~K. and {Morganti}, R. and {White}, G.~J. and {Wise}, M.~W. and {van Bemmel}, I.~M. and {Beck}, R. and {Brienza}, M. and {Bonafede}, A. and {Calistro Rivera}, G. and {Cassano}, R. and {Clarke}, A.~O. and {Cseh}, D. and {Deller}, A. and {Drabent}, A. and {van Driel}, W. and {Engels}, D. and {Falcke}, H. and {Ferrari}, C. and {Fr{\"o}hlich}, S. and {Garrett}, M.~A. and {Harwood}, J.~J. and {Heesen}, V. and {Hoeft}, M. and {Horellou}, C. and {Israel}, F.~P. and {Kapi{\'n}ska}, A.~D. and {Kunert-Bajraszewska}, M. and {McKay}, D.~J. and {Mohan}, N.~R. and {Orr{\'u}}, E. and {Pizzo}, R.~F. and {Prandoni}, I. and {Schwarz}, D.~J. and {Shulevski}, A. and {Sipior}, M. and {Smith}, D.~J.~B. and {Sridhar}, S.~S. and {Steinmetz}, M. and {Stroe}, A. and {Varenius}, E. and {van der Werf}, P.~P. and {Zensus}, J.~A. and {Zwart}, J.~T.~L.},
        title = "{The LOFAR Two-metre Sky Survey. I. Survey description and preliminary data release}",
      journal = {\aap},
         year = 2017,
        month = feb,
       volume = {598},
          eid = {A104},
        pages = {A104},
          doi = {10.1051/0004-6361/201629313},
archivePrefix = {arXiv},
       eprint = {1611.02700},
 primaryClass = {astro-ph.IM},
       adsurl = {https://ui.adsabs.harvard.edu/abs/2017A&A...598A.104S}
}

@ARTICLE{LOFARBootes2016MNRAS.460.2385W,
       author = {{Williams}, W.~L. and {van Weeren}, R.~J. and {R{\"o}ttgering}, H.~J.~A. and {Best}, P. and {Dijkema}, T.~J. and {de Gasperin}, F. and {Hardcastle}, M.~J. and {Heald}, G. and {Prandoni}, I. and {Sabater}, J. and {Shimwell}, T.~W. and {Tasse}, C. and {van Bemmel}, I.~M. and {Br{\"u}ggen}, M. and {Brunetti}, G. and {Conway}, J.~E. and {En{\ss}lin}, T. and {Engels}, D. and {Falcke}, H. and {Ferrari}, C. and {Haverkorn}, M. and {Jackson}, N. and {Jarvis}, M.~J. and {Kapi{\'n}ska}, A.~D. and {Mahony}, E.~K. and {Miley}, G.~K. and {Morabito}, L.~K. and {Morganti}, R. and {Orr{\'u}}, E. and {Retana-Montenegro}, E. and {Sridhar}, S.~S. and {Toribio}, M.~C. and {White}, G.~J. and {Wise}, M.~W. and {Zwart}, J.~T.~L.},
        title = "{LOFAR 150-MHz observations of the Bo{\"o}tes field: catalogue and source counts}",
      journal = {\mnras},
         year = 2016,
        month = aug,
       volume = {460},
       number = {3},
        pages = {2385-2412},
          doi = {10.1093/mnras/stw1056},
archivePrefix = {arXiv},
       eprint = {1605.01531},
 primaryClass = {astro-ph.CO},
       adsurl = {https://ui.adsabs.harvard.edu/abs/2016MNRAS.460.2385W}
}

@ARTICLE{Coppejans2015MNRAS.450.1477C,
       author = {{Coppejans}, Rocco and {Cseh}, David and {Williams}, Wendy L. and {van Velzen}, Sjoert and {Falcke}, Heino},
        title = "{Megahertz peaked-spectrum sources in the Bo{\"o}tes field I - a route towards finding high-redshift AGN}",
      journal = {\mnras},
         year = 2015,
        month = jun,
       volume = {450},
       number = {2},
        pages = {1477-1485},
          doi = {10.1093/mnras/stv681},
       adsurl = {https://ui.adsabs.harvard.edu/abs/2015MNRAS.450.1477C}
}

@ARTICLE{deVries2002AJ....123.1784D,
       author = {{de Vries}, W.~H. and {Morganti}, R. and {R{\"o}ttgering}, H.~J.~A. and {Vermeulen}, R. and {van Breugel}, W. and {Rengelink}, R. and {Jarvis}, M.~J.},
        title = "{Deep Westerbork 1.4 GHz Imaging of the Bootes Field}",
      journal = {\aj},
         year = 2002,
        month = mar,
       volume = {123},
       number = {3},
        pages = {1784-1800},
          doi = {10.1086/338906},
archivePrefix = {arXiv},
       eprint = {astro-ph/0111543},
 primaryClass = {astro-ph},
       adsurl = {https://ui.adsabs.harvard.edu/abs/2002AJ....123.1784D}
}

@ARTICLE{Momjian2008AJ....136..344M,
       author = {{Momjian}, Emmanuel and {Carilli}, Christopher L. and {McGreer}, Ian D.},
        title = "{Very Large Array and Very Long Baseline Array Observations of the Highest Redshift Radio-Loud QSO J1427+3312 at Z = 6.12}",
      journal = {\aj},
         year = 2008,
        month = jul,
       volume = {136},
       number = {1},
        pages = {344-349},
          doi = {10.1088/0004-6256/136/1/344},
archivePrefix = {arXiv},
       eprint = {0805.2897},
 primaryClass = {astro-ph},
       adsurl = {https://ui.adsabs.harvard.edu/abs/2008AJ....136..344M}
}

@ARTICLE{PlanckEoR2016A&A...596A.108P,
       author = {{Planck Collaboration} and {Adam}, R. and {Aghanim}, N. and {Ashdown}, M. and {Aumont}, J. and {Baccigalupi}, C. and {Ballardini}, M. and {Banday}, A.~J. and {Barreiro}, R.~B. and {Bartolo}, N. and {Basak}, S. and {Battye}, R. and {Benabed}, K. and {Bernard}, J.-P. and {Bersanelli}, M. and {Bielewicz}, P. and {Bock}, J.~J. and {Bonaldi}, A. and {Bonavera}, L. and {Bond}, J.~R. and {Borrill}, J. and {Bouchet}, F.~R. and {Boulanger}, F. and {Bucher}, M. and {Burigana}, C. and {Calabrese}, E. and {Cardoso}, J.-F. and {Carron}, J. and {Chiang}, H.~C. and {Colombo}, L.~P.~L. and {Combet}, C. and {Comis}, B. and {Couchot}, F. and {Coulais}, A. and {Crill}, B.~P. and {Curto}, A. and {Cuttaia}, F. and {Davis}, R.~J. and {de Bernardis}, P. and {de Rosa}, A. and {de Zotti}, G. and {Delabrouille}, J. and {Di Valentino}, E. and {Dickinson}, C. and {Diego}, J.~M. and {Dor{\'e}}, O. and {Douspis}, M. and {Ducout}, A. and {Dupac}, X. and {Elsner}, F. and {En{\ss}lin}, T.~A. and {Eriksen}, H.~K. and {Falgarone}, E. and {Fantaye}, Y. and {Finelli}, F. and {Forastieri}, F. and {Frailis}, M. and {Fraisse}, A.~A. and {Franceschi}, E. and {Frolov}, A. and {Galeotta}, S. and {Galli}, S. and {Ganga}, K. and {G{\'e}nova-Santos}, R.~T. and {Gerbino}, M. and {Ghosh}, T. and {Gonz{\'a}lez-Nuevo}, J. and {G{\'o}rski}, K.~M. and {Gruppuso}, A. and {Gudmundsson}, J.~E. and {Hansen}, F.~K. and {Helou}, G. and {Henrot-Versill{\'e}}, S. and {Herranz}, D. and {Hivon}, E. and {Huang}, Z. and {Ili{\'c}}, S. and {Jaffe}, A.~H. and {Jones}, W.~C. and {Keih{\"a}nen}, E. and {Keskitalo}, R. and {Kisner}, T.~S. and {Knox}, L. and {Krachmalnicoff}, N. and {Kunz}, M. and {Kurki-Suonio}, H. and {Lagache}, G. and {L{\"a}hteenm{\"a}ki}, A. and {Lamarre}, J.-M. and {Langer}, M. and {Lasenby}, A. and {Lattanzi}, M. and {Lawrence}, C.~R. and {Le Jeune}, M. and {Levrier}, F. and {Lewis}, A. and {Liguori}, M. and {Lilje}, P.~B. and {L{\'o}pez-Caniego}, M. and {Ma}, Y.-Z. and {Mac{\'\i}as-P{\'e}rez}, J.~F. and {Maggio}, G. and {Mangilli}, A. and {Maris}, M. and {Martin}, P.~G. and {Mart{\'\i}nez-Gonz{\'a}lez}, E. and {Matarrese}, S. and {Mauri}, N. and {McEwen}, J.~D. and {Meinhold}, P.~R. and {Melchiorri}, A. and {Mennella}, A. and {Migliaccio}, M. and {Miville-Desch{\^e}nes}, M.-A. and {Molinari}, D. and {Moneti}, A. and {Montier}, L. and {Morgante}, G. and {Moss}, A. and {Naselsky}, P. and {Natoli}, P. and {Oxborrow}, C.~A. and {Pagano}, L. and {Paoletti}, D. and {Partridge}, B. and {Patanchon}, G. and {Patrizii}, L. and {Perdereau}, O. and {Perotto}, L. and {Pettorino}, V. and {Piacentini}, F. and {Plaszczynski}, S. and {Polastri}, L. and {Polenta}, G. and {Puget}, J.-L. and {Rachen}, J.~P. and {Racine}, B. and {Reinecke}, M. and {Remazeilles}, M. and {Renzi}, A. and {Rocha}, G. and {Rossetti}, M. and {Roudier}, G. and {Rubi{\~n}o-Mart{\'\i}n}, J.~A. and {Ruiz-Granados}, B. and {Salvati}, L. and {Sandri}, M. and {Savelainen}, M. and {Scott}, D. and {Sirri}, G. and {Sunyaev}, R. and {Suur-Uski}, A.-S. and {Tauber}, J.~A. and {Tenti}, M. and {Toffolatti}, L. and {Tomasi}, M. and {Tristram}, M. and {Trombetti}, T. and {Valiviita}, J. and {Van Tent}, F. and {Vielva}, P. and {Villa}, F. and {Vittorio}, N. and {Wandelt}, B.~D. and {Wehus}, I.~K. and {White}, M. and {Zacchei}, A. and {Zonca}, A.},
        title = "{Planck intermediate results. XLVII. Planck constraints on reionization history}",
      journal = {\aap},
         year = 2016,
        month = dec,
       volume = {596},
          eid = {A108},
        pages = {A108},
          doi = {10.1051/0004-6361/201628897},
archivePrefix = {arXiv},
       eprint = {1605.03507},
 primaryClass = {astro-ph.CO},
       adsurl = {https://ui.adsabs.harvard.edu/abs/2016A&A...596A.108P}
}

@ARTICLE{Willott2010AJ....139..906W,
       author = {{Willott}, Chris J. and {Delorme}, Philippe and {Reyl{\'e}}, C{\'e}line and {Albert}, Loic and {Bergeron}, Jacqueline and {Crampton}, David and {Delfosse}, Xavier and {Forveille}, Thierry and {Hutchings}, John B. and {McLure}, Ross J. and {Omont}, Alain and {Schade}, David},
        title = "{The Canada-France High-z Quasar Survey: Nine New Quasars and the Luminosity Function at Redshift 6}",
      journal = {\aj},
         year = 2010,
        month = mar,
       volume = {139},
       number = {3},
        pages = {906-918},
          doi = {10.1088/0004-6256/139/3/906},
archivePrefix = {arXiv},
       eprint = {0912.0281},
 primaryClass = {astro-ph.CO},
       adsurl = {https://ui.adsabs.harvard.edu/abs/2010AJ....139..906W}
}

@ARTICLE{McGreer2006ApJ...652..157M,
       author = {{McGreer}, Ian D. and {Becker}, Robert H. and {Helfand}, David J. and {White}, Richard L.},
        title = "{Discovery of a z = 6.1 Radio-Loud Quasar in the NOAO Deep Wide Field Survey}",
      journal = {\apj},
         year = 2006,
        month = nov,
       volume = {652},
       number = {1},
        pages = {157-162},
          doi = {10.1086/507767},
archivePrefix = {arXiv},
       eprint = {astro-ph/0607278},
 primaryClass = {astro-ph},
       adsurl = {https://ui.adsabs.harvard.edu/abs/2006ApJ...652..157M}
}

@ARTICLE{Belladitta2020A&A...635L...7B,
       author = {{Belladitta}, S. and {Moretti}, A. and {Caccianiga}, A. and {Spingola}, C. and {Severgnini}, P. and {Della Ceca}, R. and {Ghisellini}, G. and {Dallacasa}, D. and {Sbarrato}, T. and {Cicone}, C. and {Cassar{\`a}}, L.~P. and {Pedani}, M.},
        title = "{The first blazar observed at z > 6}",
      journal = {\aap},
         year = 2020,
        month = mar,
       volume = {635},
          eid = {L7},
        pages = {L7},
          doi = {10.1051/0004-6361/201937395},
archivePrefix = {arXiv},
       eprint = {2002.05178},
 primaryClass = {astro-ph.CO},
       adsurl = {https://ui.adsabs.harvard.edu/abs/2020A&A...635L...7B}
}

@ARTICLE{Shao2022A&A...659A.159S,
       author = {{Shao}, Yali and {Wagg}, Jeff and {Wang}, Ran and {Momjian}, Emmanuel and {Carilli}, Chris L. and {Walter}, Fabian and {Riechers}, Dominik A. and {Intema}, Huib T. and {Weiss}, Axel and {Brunthaler}, Andreas and {Menten}, Karl M.},
        title = "{The radio spectral turnover of radio-loud quasars at z > 5}",
      journal = {\aap},
         year = 2022,
        month = mar,
       volume = {659},
          eid = {A159},
        pages = {A159},
          doi = {10.1051/0004-6361/202142489},
archivePrefix = {arXiv},
       eprint = {2112.03133},
 primaryClass = {astro-ph.GA},
       adsurl = {https://ui.adsabs.harvard.edu/abs/2022A&A...659A.159S}
}

@ARTICLE{Shao2020A&A...641A..85S,
       author = {{Shao}, Yali and {Wagg}, Jeff and {Wang}, Ran and {Carilli}, Chris L. and {Riechers}, Dominik A. and {Intema}, Huib T. and {Weiss}, Axel and {Menten}, Karl M.},
        title = "{Observations by GMRT at 323 MHz of radio-loud quasars at z > 5}",
      journal = {\aap},
         year = 2020,
        month = sep,
       volume = {641},
          eid = {A85},
        pages = {A85},
          doi = {10.1051/0004-6361/202038469},
archivePrefix = {arXiv},
       eprint = {2006.13762},
 primaryClass = {astro-ph.GA},
       adsurl = {https://ui.adsabs.harvard.edu/abs/2020A&A...641A..85S}
}

@ARTICLE{Qin2025PASA...42...49Q,
       author = {{Qin}, Yuxiang and {Mesinger}, Andrei and {Prelogovi{\'c}}, David and {Becker}, George and {Bischetti}, Manuela and {Bosman}, Sarah and {Davies}, Frederick and {D'Odorico}, Valentina and {Gaikwad}, Prakash and {Haehnelt}, Martin and {Keating}, Laura and {Lai}, Samuel and {Ryan-Weber}, Emma and {Satyavolu}, Sindhu and {Walter}, Fabian and {Zhu}, Yongda},
        title = "{Percent-level timing of reionisation: Self-consistent, implicit-likelihood inference from XQR-30+ Ly{\ensuremath{\alpha}} forest data}",
      journal = {\pasa},
         year = 2025,
        month = apr,
       volume = {42},
          eid = {e049},
        pages = {e049},
          doi = {10.1017/pasa.2025.35},
archivePrefix = {arXiv},
       eprint = {2412.00799},
 primaryClass = {astro-ph.CO},
       adsurl = {https://ui.adsabs.harvard.edu/abs/2025PASA...42...49Q}
}

@ARTICLE{CODA-I2016MNRAS.463.1462O,
       author = {{Ocvirk}, Pierre and {Gillet}, Nicolas and {Shapiro}, Paul R. and {Aubert}, Dominique and {Iliev}, Ilian T. and {Teyssier}, Romain and {Yepes}, Gustavo and {Choi}, Jun-Hwan and {Sullivan}, David and {Knebe}, Alexander and {Gottl{\"o}ber}, Stefan and {D'Aloisio}, Anson and {Park}, Hyunbae and {Hoffman}, Yehuda and {Stranex}, Timothy},
        title = "{Cosmic Dawn (CoDa): the First Radiation-Hydrodynamics Simulation of Reionization and Galaxy Formation in the Local Universe}",
      journal = {\mnras},
         year = 2016,
        month = dec,
       volume = {463},
       number = {2},
        pages = {1462-1485},
          doi = {10.1093/mnras/stw2036},
archivePrefix = {arXiv},
       eprint = {1511.00011},
 primaryClass = {astro-ph.GA},
       adsurl = {https://ui.adsabs.harvard.edu/abs/2016MNRAS.463.1462O}
}

@ARTICLE{CODA-II2020MNRAS.496.4087O,
       author = {{Ocvirk}, Pierre and {Aubert}, Dominique and {Sorce}, Jenny G. and {Shapiro}, Paul R. and {Deparis}, Nicolas and {Dawoodbhoy}, Taha and {Lewis}, Joseph and {Teyssier}, Romain and {Yepes}, Gustavo and {Gottl{\"o}ber}, Stefan and {Ahn}, Kyungjin and {Iliev}, Ilian T. and {Hoffman}, Yehuda},
        title = "{Cosmic Dawn II (CoDa II): a new radiation-hydrodynamics simulation of the self-consistent coupling of galaxy formation and reionization}",
      journal = {\mnras},
         year = 2020,
        month = aug,
       volume = {496},
       number = {4},
        pages = {4087-4107},
          doi = {10.1093/mnras/staa1266},
archivePrefix = {arXiv},
       eprint = {1811.11192},
 primaryClass = {astro-ph.GA},
       adsurl = {https://ui.adsabs.harvard.edu/abs/2020MNRAS.496.4087O}
}

@ARTICLE{CASA2022PASP..134k4501C,
       author = {{CASA Team} and {Bean}, Ben and {Bhatnagar}, Sanjay and {Castro}, Sandra and {Donovan Meyer}, Jennifer and {Emonts}, Bjorn and {Garcia}, Enrique and {Garwood}, Robert and {Golap}, Kumar and {Gonzalez Villalba}, Justo and {Harris}, Pamela and {Hayashi}, Yohei and {Hoskins}, Josh and {Hsieh}, Mingyu and {Jagannathan}, Preshanth and {Kawasaki}, Wataru and {Keimpema}, Aard and {Kettenis}, Mark and {Lopez}, Jorge and {Marvil}, Joshua and {Masters}, Joseph and {McNichols}, Andrew and {Mehringer}, David and {Miel}, Renaud and {Moellenbrock}, George and {Montesino}, Federico and {Nakazato}, Takeshi and {Ott}, Juergen and {Petry}, Dirk and {Pokorny}, Martin and {Raba}, Ryan and {Rau}, Urvashi and {Schiebel}, Darrell and {Schweighart}, Neal and {Sekhar}, Srikrishna and {Shimada}, Kazuhiko and {Small}, Des and {Steeb}, Jan-Willem and {Sugimoto}, Kanako and {Suoranta}, Ville and {Tsutsumi}, Takahiro and {van Bemmel}, Ilse M. and {Verkouter}, Marjolein and {Wells}, Akeem and {Xiong}, Wei and {Szomoru}, Arpad and {Griffith}, Morgan and {Glendenning}, Brian and {Kern}, Jeff},
        title = "{CASA, the Common Astronomy Software Applications for Radio Astronomy}",
      journal = {\pasp},
         year = 2022,
        month = nov,
       volume = {134},
       number = {1041},
          eid = {114501},
        pages = {114501},
          doi = {10.1088/1538-3873/ac9642},
archivePrefix = {arXiv},
       eprint = {2210.02276},
 primaryClass = {astro-ph.IM},
       adsurl = {https://ui.adsabs.harvard.edu/abs/2022PASP..134k4501C}
}

@ARTICLE{Cornwell2008arXiv0806.2228C,
       author = {{Cornwell}, T.~J.},
        title = "{Multi-Scale CLEAN deconvolution of radio synthesis images}",
      journal = {arXiv e-prints},
         year = 2008,
        month = jun,
          eid = {arXiv:0806.2228},
        pages = {arXiv:0806.2228},
          doi = {10.48550/arXiv.0806.2228},
archivePrefix = {arXiv},
       eprint = {0806.2228},
 primaryClass = {astro-ph},
       adsurl = {https://ui.adsabs.harvard.edu/abs/2008arXiv0806.2228C}
}

@ARTICLE{Ofringa2017MNRAS.471..301O,
       author = {{Offringa}, A.~R. and {Smirnov}, O.},
        title = "{An optimized algorithm for multiscale wideband deconvolution of radio astronomical images}",
      journal = {\mnras},
         year = 2017,
        month = oct,
       volume = {471},
       number = {1},
        pages = {301-316},
          doi = {10.1093/mnras/stx1547},
archivePrefix = {arXiv},
       eprint = {1706.06786},
 primaryClass = {astro-ph.IM},
       adsurl = {https://ui.adsabs.harvard.edu/abs/2017MNRAS.471..301O}
}

@ARTICLE{GunnPeterson1965ApJ...142.1633G,
       author = {{Gunn}, James E. and {Peterson}, Bruce A.},
        title = "{On the Density of Neutral Hydrogen in Intergalactic Space.}",
      journal = {\apj},
         year = 1965,
        month = nov,
       volume = {142},
        pages = {1633-1636},
          doi = {10.1086/148444},
       adsurl = {https://ui.adsabs.harvard.edu/abs/1965ApJ...142.1633G}
}

@ARTICLE{Monsalve2018ApJ...863...11M,
       author = {{Monsalve}, Raul A. and {Greig}, Bradley and {Bowman}, Judd D. and {Mesinger}, Andrei and {Rogers}, Alan E.~E. and {Mozdzen}, Thomas J. and {Kern}, Nicholas S. and {Mahesh}, Nivedita},
        title = "{Results from EDGES High-band. II. Constraints on Parameters of Early Galaxies}",
      journal = {\apj},
         year = 2018,
        month = aug,
       volume = {863},
       number = {1},
          eid = {11},
        pages = {11},
          doi = {10.3847/1538-4357/aace54},
archivePrefix = {arXiv},
       eprint = {1806.07774},
 primaryClass = {astro-ph.CO},
       adsurl = {https://ui.adsabs.harvard.edu/abs/2018ApJ...863...11M}
}

@ARTICLE{Spinelli2019MNRAS.489.4007S,
       author = {{Spinelli}, Marta and {Bernardi}, Gianni and {Santos}, Mario G.},
        title = "{On the contamination of the global 21-cm signal from polarized foregrounds}",
      journal = {\mnras},
         year = 2019,
        month = nov,
       volume = {489},
       number = {3},
        pages = {4007-4015},
          doi = {10.1093/mnras/stz2425},
archivePrefix = {arXiv},
       eprint = {1908.05303},
 primaryClass = {astro-ph.CO},
       adsurl = {https://ui.adsabs.harvard.edu/abs/2019MNRAS.489.4007S}
}

@ARTICLE{HERA2022ApJ...924...51A,
       author = {{Abdurashidova}, Zara and {Aguirre}, James E. and {Alexander}, Paul and {Ali}, Zaki S. and {Balfour}, Yanga and {Barkana}, Rennan and {Beardsley}, Adam P. and {Bernardi}, Gianni and {Billings}, Tashalee S. and {Bowman}, Judd D. and {Bradley}, Richard F. and {Bull}, Philip and {Burba}, Jacob and {Carey}, Steve and {Carilli}, Chris L. and {Cheng}, Carina and {DeBoer}, David R. and {Dexter}, Matt and {de Lera Acedo}, Eloy and {Dillon}, Joshua S. and {Ely}, John and {Ewall-Wice}, Aaron and {Fagnoni}, Nicolas and {Fialkov}, Anastasia and {Fritz}, Randall and {Furlanetto}, Steven R. and {Gale-Sides}, Kingsley and {Glendenning}, Brian and {Gorthi}, Deepthi and {Greig}, Bradley and {Grobbelaar}, Jasper and {Halday}, Ziyaad and {Hazelton}, Bryna J. and {Heimersheim}, Stefan and {Hewitt}, Jacqueline N. and {Hickish}, Jack and {Jacobs}, Daniel C. and {Julius}, Austin and {Kern}, Nicholas S. and {Kerrigan}, Joshua and {Kittiwisit}, Piyanat and {Kohn}, Saul A. and {Kolopanis}, Matthew and {Lanman}, Adam and {La Plante}, Paul and {Lekalake}, Telalo and {Lewis}, David and {Liu}, Adrian and {Ma}, Yin-Zhe and {MacMahon}, David and {Malan}, Lourence and {Malgas}, Cresshim and {Maree}, Matthys and {Martinot}, Zachary E. and {Matsetela}, Eunice and {Mesinger}, Andrei and {Mirocha}, Jordan and {Molewa}, Mathakane and {Morales}, Miguel F. and {Mosiane}, Tshegofalang and {Mu{\~n}oz}, Julian B. and {Murray}, Steven G. and {Neben}, Abraham R. and {Nikolic}, Bojan and {Nunhokee}, Chuneeta D. and {Parsons}, Aaron R. and {Patra}, Nipanjana and {Pieterse}, Samantha and {Pober}, Jonathan C. and {Qin}, Yuxiang and {Razavi-Ghods}, Nima and {Reis}, Itamar and {Ringuette}, Jon and {Robnett}, James and {Rosie}, Kathryn and {Santos}, Mario G. and {Sikder}, Sudipta and {Sims}, Peter and {Smith}, Craig and {Syce}, Angelo and {Thyagarajan}, Nithyanandan and {Williams}, Peter K.~G. and {Zheng}, Haoxuan},
        title = "{HERA Phase I Limits on the Cosmic 21 cm Signal: Constraints on Astrophysics and Cosmology during the Epoch of Reionization}",
      journal = {\apj},
         year = 2022,
        month = jan,
       volume = {924},
       number = {2},
          eid = {51},
        pages = {51},
          doi = {10.3847/1538-4357/ac2ffc},
archivePrefix = {arXiv},
       eprint = {2108.07282},
 primaryClass = {astro-ph.CO},
       adsurl = {https://ui.adsabs.harvard.edu/abs/2022ApJ...924...51A}
}

@ARTICLE{HERA2023ApJ...945..124H,
       author = {{HERA Collaboration} and {Abdurashidova}, Zara and {Adams}, Tyrone and {Aguirre}, James E. and {Alexander}, Paul and {Ali}, Zaki S. and {Baartman}, Rushelle and {Balfour}, Yanga and {Barkana}, Rennan and {Beardsley}, Adam P. and {Bernardi}, Gianni and {Billings}, Tashalee S. and {Bowman}, Judd D. and {Bradley}, Richard F. and {Breitman}, Daniela and {Bull}, Philip and {Burba}, Jacob and {Carey}, Steve and {Carilli}, Chris L. and {Cheng}, Carina and {Choudhuri}, Samir and {DeBoer}, David R. and {de Lera Acedo}, Eloy and {Dexter}, Matt and {Dillon}, Joshua S. and {Ely}, John and {Ewall-Wice}, Aaron and {Fagnoni}, Nicolas and {Fialkov}, Anastasia and {Fritz}, Randall and {Furlanetto}, Steven R. and {Gale-Sides}, Kingsley and {Garsden}, Hugh and {Glendenning}, Brian and {Gorce}, Ad{\'e}lie and {Gorthi}, Deepthi and {Greig}, Bradley and {Grobbelaar}, Jasper and {Halday}, Ziyaad and {Hazelton}, Bryna J. and {Heimersheim}, Stefan and {Hewitt}, Jacqueline N. and {Hickish}, Jack and {Jacobs}, Daniel C. and {Julius}, Austin and {Kern}, Nicholas S. and {Kerrigan}, Joshua and {Kittiwisit}, Piyanat and {Kohn}, Saul A. and {Kolopanis}, Matthew and {Lanman}, Adam and {La Plante}, Paul and {Lewis}, David and {Liu}, Adrian and {Loots}, Anita and {Ma}, Yin-Zhe and {MacMahon}, David H.~E. and {Malan}, Lourence and {Malgas}, Keith and {Malgas}, Cresshim and {Maree}, Matthys and {Marero}, Bradley and {Martinot}, Zachary E. and {McBride}, Lisa and {Mesinger}, Andrei and {Mirocha}, Jordan and {Molewa}, Mathakane and {Morales}, Miguel F. and {Mosiane}, Tshegofalang and {Mu{\~n}oz}, Julian B. and {Murray}, Steven G. and {Nagpal}, Vighnesh and {Neben}, Abraham R. and {Nikolic}, Bojan and {Nunhokee}, Chuneeta D. and {Nuwegeld}, Hans and {Parsons}, Aaron R. and {Pascua}, Robert and {Patra}, Nipanjana and {Pieterse}, Samantha and {Qin}, Yuxiang and {Razavi-Ghods}, Nima and {Robnett}, James and {Rosie}, Kathryn and {Santos}, Mario G. and {Sims}, Peter and {Singh}, Saurabh and {Smith}, Craig and {Swarts}, Hilton and {Tan}, Jianrong and {Thyagarajan}, Nithyanandan and {Wilensky}, Michael J. and {Williams}, Peter K.~G. and {van Wyngaarden}, Pieter and {Zheng}, Haoxuan},
        title = "{Improved Constraints on the 21 cm EoR Power Spectrum and the X-Ray Heating of the IGM with HERA Phase I Observations}",
      journal = {\apj},
         year = 2023,
        month = mar,
       volume = {945},
       number = {2},
          eid = {124},
        pages = {124},
          doi = {10.3847/1538-4357/acaf50},
archivePrefix = {arXiv},
       eprint = {2210.04912},
 primaryClass = {astro-ph.CO},
       adsurl = {https://ui.adsabs.harvard.edu/abs/2023ApJ...945..124H}
}

@ARTICLE{Bosman2022MNRAS.514...55B,
       author = {{Bosman}, Sarah E.~I. and {Davies}, Frederick B. and {Becker}, George D. and {Keating}, Laura C. and {Davies}, Rebecca L. and {Zhu}, Yongda and {Eilers}, Anna-Christina and {D'Odorico}, Valentina and {Bian}, Fuyan and {Bischetti}, Manuela and {Cristiani}, Stefano V. and {Fan}, Xiaohui and {Farina}, Emanuele P. and {Haehnelt}, Martin G. and {Hennawi}, Joseph F. and {Kulkarni}, Girish and {Mesinger}, Andrei and {Meyer}, Romain A. and {Onoue}, Masafusa and {Pallottini}, Andrea and {Qin}, Yuxiang and {Ryan-Weber}, Emma and {Schindler}, Jan-Torge and {Walter}, Fabian and {Wang}, Feige and {Yang}, Jinyi},
        title = "{Hydrogen reionization ends by z = 5.3: Lyman-{\ensuremath{\alpha}} optical depth measured by the XQR-30 sample}",
      journal = {\mnras},
         year = 2022,
        month = jul,
       volume = {514},
       number = {1},
        pages = {55-76},
          doi = {10.1093/mnras/stac1046},
archivePrefix = {arXiv},
       eprint = {2108.03699},
 primaryClass = {astro-ph.CO},
       adsurl = {https://ui.adsabs.harvard.edu/abs/2022MNRAS.514...55B}
}

@ARTICLE{Euclid2026arXiv260703432Y,
       author = {{Yang}, D. and {Hennawi}, J.~F. and {Guarneri}, F. and {Wolf}, J. and {Belladitta}, S. and {Schindler}, J.-T. and {Hughes}, A.~C.~N. and {Ba{\~n}ados}, E. and {Mortlock}, D.~J. and {Yang}, J. and {Wang}, F. and {Fan}, X. and {Jahnke}, K. and {Stern}, D. and {Willott}, C.~J. and {Barth}, A.~J. and {Rottgering}, H.~J.~A. and {Varadaraj}, R.~G. and {Decarli}, R. and {Eilers}, A.-C. and {Ezziati}, M. and {Fu}, Y. and {Huang}, J. and {Jin}, X. and {Kang}, Y. and {Martinez-Ramirez}, L.~N. and {Matsuoka}, Y. and {Onoue}, M. and {Pello}, R. and {Remigio}, R.~P. and {Tee}, W.~L. and {Venemans}, B. and {Vietri}, G. and {Wang}, B. and {Abbo}, L.~J. and {Atek}, H. and {Bisogni}, S. and {Bosman}, S.~E.~I. and {Bowler}, R.~A.~A. and {Conselice}, C.~J. and {Davies}, F.~B. and {Gutierrez}, C.~M. and {Harikane}, Y. and {Rubinur}, K. and {Lovell}, C.~C. and {Magliocchetti}, M. and {Matthee}, J. and {Ricci}, F. and {Scialpi}, M. and {Scott}, D. and {Spinoglio}, L. and {Tarsitano}, F. and {Toba}, Y. and {Walter}, F. and {Weaver}, J.~R. and {Zamorani}, G. and {Altieri}, B. and {Amara}, A. and {Andreon}, S. and {Aussel}, H. and {Baccigalupi}, C. and {Baldi}, M. and {Balestra}, A. and {Bardelli}, S. and {Battaglia}, P. and {Biviano}, A. and {Branchini}, E. and {Brescia}, M. and {Camera}, S. and {Ca{\~n}as-Herrera}, G. and {Capobianco}, V. and {Carbone}, C. and {Carretero}, J. and {Castellano}, M. and {Castignani}, G. and {Cavuoti}, S. and {Chambers}, K.~C. and {Cimatti}, A. and {Colodro-Conde}, C. and {Congedo}, G. and {Conversi}, L. and {Copin}, Y. and {Courbin}, F. and {Courtois}, H.~M. and {Cropper}, M. and {Cuillandre}, J.-C. and {Degaudenzi}, H. and {De Lucia}, G. and {Dolding}, C. and {Dole}, H. and {Douspis}, M. and {Dubath}, F. and {Dupac}, X. and {Dusini}, S. and {Escoffier}, S. and {Farina}, M. and {Farinelli}, R. and {Ferriol}, S. and {Finelli}, F. and {Fourmanoit}, N. and {Frailis}, M. and {Franceschi}, E. and {Fumana}, M. and {Galeotta}, S. and {George}, K. and {Gillis}, B. and {Giocoli}, C. and {G{\'o}mez-Alvarez}, P. and {Gracia-Carpio}, J. and {Grazian}, A. and {Grupp}, F. and {Guzzo}, L. and {Gwyn}, S. and {Haugan}, S.~V.~H. and {Hoekstra}, H. and {Holmes}, W. and {Hook}, I.~M. and {Hormuth}, F. and {Hornstrup}, A. and {Jhabvala}, M. and {Kermiche}, S. and {Kubik}, B. and {Kuijken}, K. and {K{\"u}mmel}, M. and {Kunz}, M. and {Kurki-Suonio}, H. and {Le Brun}, A.~M.~C. and {Ligori}, S. and {Lilje}, P.~B. and {Lindholm}, V. and {Lloro}, I. and {Mainetti}, G. and {Maino}, D. and {Maiorano}, E. and {Mansutti}, O. and {Marggraf}, O. and {Martinelli}, M. and {Martinet}, N. and {Marulli}, F. and {Massey}, R.~J. and {McCracken}, H.~J. and {Medinaceli}, E. and {Mei}, S. and {Mellier}, Y. and {Meneghetti}, M. and {Merlin}, E. and {Meylan}, G. and {Mohr}, J.~J. and {Mora}, A. and {Moresco}, M. and {Moscardini}, L. and {Munari}, E. and {Nakajima}, R. and {Neissner}, C. and {Nichol}, R.~C. and {Niemi}, S.-M. and {Padilla}, C. and {Paltani}, S. and {Pasian}, F. and {Pedersen}, K. and {Percival}, W.~J. and {Pettorino}, V. and {Pires}, S. and {Polenta}, G. and {Poncet}, M. and {Popa}, L.~A. and {Pozzetti}, L. and {Racca}, G.~D. and {Raison}, F. and {Rebolo}, R. and {Renzi}, A. and {Rhodes}, J. and {Riccio}, G. and {Rix}, H.-W. and {Romelli}, E. and {Roncarelli}, M. and {Rosset}, C. and {Rusholme}, B. and {Saglia}, R. and {Sakr}, Z. and {Sapone}, D. and {Sauvage}, M. and {Schirmer}, M. and {Schneider}, P. and {Schrabback}, T. and {Secroun}, A. and {Seidel}, G. and {Serrano}, S. and {Sihvola}, E. and {Simon}, P. and {Sirignano}, C. and {Sirri}, G. and {Stanco}, L. and {Steinwagner}, J. and {Tallada-Cresp{\'\i}}, P. and {Tereno}, I. and {Tessore}, N. and {Toft}, S. and {Toledo-Moreo}, R. and {Torradeflot}, F.},
        title = "{Euclid: Discovery of 31 new quasars at $6.6 < z < 7.8$}",
      journal = {arXiv e-prints},
         year = 2026,
        month = jul,
          eid = {arXiv:2607.03432},
        pages = {arXiv:2607.03432},
archivePrefix = {arXiv},
       eprint = {2607.03432},
 primaryClass = {astro-ph.GA},
       adsurl = {https://ui.adsabs.harvard.edu/abs/2026arXiv260703432Y}
}

@ARTICLE{Matsuoka2026arXiv260610160M,
       author = {{Matsuoka}, Yoshiki and {Decarli}, Roberto and {Farina}, Emanuele Paolo and {Gloudemans}, Anniek J. and {Ba{\~n}ados}, Eduardo and {Arrigoni Battaia}, Fabrizio and {Eilers}, Anna-Christina and {Mazzucchelli}, Chiara and {Strauss}, Michael A. and {Suh}, Hyewon and {Trebitsch}, Maxime and {Walter}, Fabian and {Wang}, Feige and {Aoki}, Kentaro and {Arita}, Junya},
        title = "{Aether-SHELLQs: JWST integral-field spectroscopy of candidate obscured quasars at z \raisebox{-0.5ex}\textasciitilde 6}",
      journal = {arXiv e-prints},
         year = 2026,
        month = jun,
          eid = {arXiv:2606.10160},
        pages = {arXiv:2606.10160},
          doi = {10.48550/arXiv.2606.10160},
archivePrefix = {arXiv},
       eprint = {2606.10160},
 primaryClass = {astro-ph.GA},
       adsurl = {https://ui.adsabs.harvard.edu/abs/2026arXiv260610160M}
}

@ARTICLE{Becker2015PASA...32...45B,
       author = {{Becker}, George D. and {Bolton}, James S. and {Lidz}, Adam},
        title = "{Reionisation and High-Redshift Galaxies: The View from Quasar Absorption Lines}",
      journal = {\pasa},
         year = 2015,
        month = dec,
       volume = {32},
          eid = {e045},
        pages = {e045},
          doi = {10.1017/pasa.2015.45},
archivePrefix = {arXiv},
       eprint = {1510.03368},
 primaryClass = {astro-ph.CO},
       adsurl = {https://ui.adsabs.harvard.edu/abs/2015PASA...32...45B}
}

@ARTICLE{Eilers2018ApJ...867...30E,
       author = {{Eilers}, Anna-Christina and {Hennawi}, Joseph F. and {Davies}, Frederick B.},
        title = "{First Spectroscopic Study of a Young Quasar}",
      journal = {\apj},
         year = 2018,
        month = nov,
       volume = {867},
       number = {1},
          eid = {30},
        pages = {30},
          doi = {10.3847/1538-4357/aae081},
archivePrefix = {arXiv},
       eprint = {1806.05691},
 primaryClass = {astro-ph.GA},
       adsurl = {https://ui.adsabs.harvard.edu/abs/2018ApJ...867...30E}
}

@ARTICLE{Mertens2025A&A...698A.186M,
       author = {{Mertens}, F.~G. and {Mevius}, M. and {Koopmans}, L.~V.~E. and {Offringa}, A.~R. and {Zaroubi}, S. and {Acharya}, A. and {Brackenhoff}, S.~A. and {Ceccotti}, E. and {Chapman}, E. and {Chege}, K. and {Ciardi}, B. and {Ghara}, R. and {Ghosh}, S. and {Giri}, S.~K. and {Hothi}, I. and {H{\"o}fer}, C. and {Iliev}, I.~T. and {Jeli{\'c}}, V. and {Ma}, Q. and {Mellema}, G. and {Munshi}, S. and {Pandey}, V.~N. and {Yatawatta}, S.},
        title = "{Deeper multi-redshift upper limits on the epoch of reionisation 21 cm signal power spectrum from LOFAR between z = 8.3 and z = 10.1}",
      journal = {\aap},
         year = 2025,
        month = jun,
       volume = {698},
          eid = {A186},
        pages = {A186},
          doi = {10.1051/0004-6361/202554158},
archivePrefix = {arXiv},
       eprint = {2503.05576},
 primaryClass = {astro-ph.CO},
       adsurl = {https://ui.adsabs.harvard.edu/abs/2025A&A...698A.186M}
}

@ARTICLE{Nunhokee2025ApJ...989...57N,
       author = {{Nunhokee}, C.~D. and {Null}, D. and {Trott}, C.~M. and {Barry}, N. and {Qin}, Y. and {Wayth}, R.~B. and {Line}, J.~L.~B. and {Jordan}, C.~H. and {Pindor}, B. and {Cook}, J.~H. and {Bowman}, J. and {Chokshi}, A. and {Ducharme}, J. and {Elder}, K. and {Guo}, Q. and {Hazelton}, B. and {Hidayat}, W. and {Ito}, T. and {Jacobs}, D. and {Jong}, E. and {Kolopanis}, M. and {Kunicki}, T. and {Lilleskov}, E. and {Morales}, M.~F. and {Pober}, J.~C. and {Selvaraj}, A. and {Shi}, R. and {Takahashi}, K. and {Tingay}, S.~J. and {Webster}, R.~L. and {Yoshiura}, S. and {Zheng}, Q.},
        title = "{Limits on the 21 cm Power Spectrum at z = 6.5─7.0 from Murchison Widefield Array Observations}",
      journal = {\apj},
         year = 2025,
        month = aug,
       volume = {989},
       number = {1},
          eid = {57},
        pages = {57},
          doi = {10.3847/1538-4357/adda45},
archivePrefix = {arXiv},
       eprint = {2505.09097},
 primaryClass = {astro-ph.CO},
       adsurl = {https://ui.adsabs.harvard.edu/abs/2025ApJ...989...57N}
}

@ARTICLE{Kulkarni2019MNRAS.485L..24K,
       author = {{Kulkarni}, Girish and {Keating}, Laura C. and {Haehnelt}, Martin G. and {Bosman}, Sarah E.~I. and {Puchwein}, Ewald and {Chardin}, Jonathan and {Aubert}, Dominique},
        title = "{Large Ly {\ensuremath{\alpha}} opacity fluctuations and low CMB {\ensuremath{\tau}} in models of late reionization with large islands of neutral hydrogen extending to z < 5.5}",
      journal = {\mnras},
         year = 2019,
        month = may,
       volume = {485},
       number = {1},
        pages = {L24-L28},
          doi = {10.1093/mnrasl/slz025},
archivePrefix = {arXiv},
       eprint = {1809.06374},
 primaryClass = {astro-ph.CO},
       adsurl = {https://ui.adsabs.harvard.edu/abs/2019MNRAS.485L..24K}
}

@article{Furlanetto:2006jb,
    author = "Furlanetto, Steven and Oh, S. Peng and Briggs, Frank",
    title = "{Cosmology at Low Frequencies: The 21 cm Transition and the High-Redshift Universe}",
    eprint = "astro-ph/0608032",
    archivePrefix = "arXiv",
    doi = "10.1016/j.physrep.2006.08.002",
    journal = "Phys. Rept.",
    volume = "433",
    pages = "181--301",
    year = "2006"
}

@ARTICLE{Ghara2024MNRAS.530..191G,
       author = {{Ghara}, Raghunath and {Bag}, Satadru and {Zaroubi}, Saleem and {Majumdar}, Suman},
        title = "{The morphology of the redshifted 21-cm signal from the Cosmic Dawn}",
      journal = {\mnras},
         year = 2024,
        month = may,
       volume = {530},
       number = {1},
        pages = {191-202},
          doi = {10.1093/mnras/stae895},
archivePrefix = {arXiv},
       eprint = {2308.00548},
 primaryClass = {astro-ph.CO},
       adsurl = {https://ui.adsabs.harvard.edu/abs/2024MNRAS.530..191G}
}

@ARTICLE{Ceccotti2025A&A...696A..56C,
       author = {{Ceccotti}, E. and {Offringa}, A.~R. and {Koopmans}, L.~V.~E. and {Mertens}, F.~G. and {Mevius}, M. and {Acharya}, A. and {Brackenhoff}, S.~A. and {Ciardi}, B. and {Gehlot}, B.~K. and {Ghara}, R. and {Chege}, J.~K. and {Ghosh}, S. and {H{\"o}fer}, C. and {Hothi}, I. and {Iliev}, I.~T. and {McKean}, J.~P. and {Munshi}, S. and {Zaroubi}, S.},
        title = "{Spectral modelling of Cygnus A between 110 and 250 MHz: Impact on the LOFAR 21-cm signal power spectrum}",
      journal = {\aap},
         year = 2025,
        month = apr,
       volume = {696},
          eid = {A56},
        pages = {A56},
          doi = {10.1051/0004-6361/202453106},
archivePrefix = {arXiv},
       eprint = {2502.18459},
 primaryClass = {astro-ph.CO},
       adsurl = {https://ui.adsabs.harvard.edu/abs/2025A&A...696A..56C}
}

@ARTICLE{deOliveira-Costa2008MNRAS.388..247D,
       author = {{de Oliveira-Costa}, Ang{\'e}lica and {Tegmark}, Max and {Gaensler}, B.~M. and {Jonas}, Justin and {Landecker}, T.~L. and {Reich}, Patricia},
        title = "{A model of diffuse Galactic radio emission from 10 MHz to 100 GHz}",
      journal = {\mnras},
         year = 2008,
        month = jul,
       volume = {388},
       number = {1},
        pages = {247-260},
          doi = {10.1111/j.1365-2966.2008.13376.x},
archivePrefix = {arXiv},
       eprint = {0802.1525},
 primaryClass = {astro-ph},
       adsurl = {https://ui.adsabs.harvard.edu/abs/2008MNRAS.388..247D}
}

@ARTICLE{bernardi13,
       author = {{Bernardi}, G. and {Greenhill}, L.~J. and {Mitchell}, D.~A. and {Ord}, S.~M. and {Hazelton}, B.~J. and {Gaensler}, B.~M. and {de Oliveira-Costa}, A. and {Morales}, M.~F. and {Udaya Shankar}, N. and {Subrahmanyan}, R. and {Wayth}, R.~B. and {Lenc}, E. and {Williams}, C.~L. and {Arcus}, W. and {Arora}, B.~S. and {Barnes}, D.~G. and {Bowman}, J.~D. and {Briggs}, F.~H. and {Bunton}, J.~D. and {Cappallo}, R.~J. and {Corey}, B.~E. and {Deshpande}, A. and {deSouza}, L. and {Emrich}, D. and {Goeke}, R. and {Herne}, D. and {Hewitt}, J.~N. and {Johnston-Hollitt}, M. and {Kaplan}, D. and {Kasper}, J.~C. and {Kincaid}, B.~B. and {Koenig}, R. and {Kratzenberg}, E. and {Lonsdale}, C.~J. and {Lynch}, M.~J. and {McWhirter}, S.~R. and {Morgan}, E. and {Oberoi}, D. and {Pathikulangara}, J. and {Prabu}, T. and {Remillard}, R.~A. and {Rogers}, A.~E.~E. and {Roshi}, A. and {Salah}, J.~E. and {Sault}, R.~J. and {Srivani}, K.~S. and {Stevens}, J. and {Tingay}, S.~J. and {Waterson}, M. and {Webster}, R.~L. and {Whitney}, A.~R. and {Williams}, A. and {Wyithe}, J.~S.~B.},
        title = "{A 189 MHz, 2400 deg$^{2}$ Polarization Survey with the Murchison Widefield Array 32-element Prototype}",
      journal = {\apj},
         year = 2013,
        month = jul,
       volume = {771},
       number = {2},
          eid = {105},
        pages = {105},
          doi = {10.1088/0004-637X/771/2/105},
archivePrefix = {arXiv},
       eprint = {1305.6047},
 primaryClass = {astro-ph.CO},
       adsurl = {https://ui.adsabs.harvard.edu/abs/2013ApJ...771..105B}
}

\begin{appendix}
    \section{Literature measurements for the high redshift quasar sample}\label{apx:High_redshift_Quasars_Sample_and_their_Continuum_Spectra}

In this Appendix, we report the literature measurements used to fit the continuum spectrum of the high-redshift quasar sample
    
    \begin{table}[H]
    \centering
    \caption{J0410-0139 radio flux density measurements in the literature.}
    \begin{tabular}{lcc}
    \hline\hline
    \textbf{Survey} & \textbf{Frequency (MHz)} & \textbf{Flux density (mJy)} \\
    \hline
    
    \multicolumn{3}{l}{FLASH\tablefootmark{a}}\\
    & 856  & $5.09 \pm 0.30$ \\
    
    \multicolumn{3}{l}{RACS\tablefootmark{a}}\\
    & 888  & $5.64 \pm 0.90$ \\
    
    \multicolumn{3}{l}{VAST\tablefootmark{a}}\\
    & 888  & $5.42 \pm 0.97$ \\
    & 888  & $5.10 \pm 0.80$ \\
    & 1370 & $7.89 \pm 1.62$ \\
    & 1370 & $11.99 \pm 1.84$ \\
    & 1370 & $5.83 \pm 1.50$ \\
    
    \multicolumn{3}{l}{NVSS\tablefootmark{a}}\\
    & 1400 & $4.20 \pm 0.50$ \\
    
    \multicolumn{3}{l}{VLA\tablefootmark{a}}\\
    & 1500 & $8.12 \pm 0.02$ \\
    & 1500 & $9.21 \pm 0.17$ \\
    & 3000 & $14.48 \pm 0.01$ \\
    & 3000 & $10.63 \pm 0.02$ \\
    
    \multicolumn{3}{l}{VLBA\tablefootmark{a}}\\
    & 1500 & $4.46 \pm 0.04$ \\
    
    \multicolumn{3}{l}{VLASS\tablefootmark{a}}\\
    & 3000 & $7.60 \pm 0.20$ \\
    & 3000 & $10.63 \pm 0.02$ \\
    & 3000 & $10.10 \pm 0.15$ \\
    \hline
    \end{tabular}
    \tablefoot{
    \tablefoottext{a}{\citet{Banados2025NatAs...9..293B} and reference therein.}
    }
    \label{tab:J0410_fluxes}
    
    \end{table}

        \begin{table}[H]
    \centering
    \caption{PSO J172+18 radio flux density measurement in the literature.}
    \begin{tabular}{lcc}
    \hline\hline
    \textbf{Survey} & \textbf{Frequency (MHz)} & \textbf{Flux density (mJy)} \\
    \hline
    
    \multicolumn{3}{l}{FIRST\tablefootmark{a}}\\
    & 1400 & $1.02 \pm 0.144$ \\
    
    \multicolumn{3}{l}{VLA-L\tablefootmark{a}}\\
    & 1520 & $0.51 \pm 0.015$ \\
    
    \multicolumn{3}{l}{VLA-S\tablefootmark{a}}\\
    & 2870 & $0.222 \pm 0.009$ \\
    
    \hline
    \end{tabular}
    \tablefoot{
    \tablefoottext{a}{\citet{Banados2021ApJ...909...80B} and reference therein.}
    }
    \label{tab:PSOJ172_fluxes}
    \end{table}
    
    \begin{table}[H]
    \centering
    \caption{J1429+5547 radio flux density measurement in the literature.}
    \begin{tabular}{lcc}
    \hline\hline
    \textbf{Survey} & \textbf{Frequency (MHz)} & \textbf{Flux density (mJy)} \\
    \hline
    \multicolumn{3}{l}{LoTSS \tablefootmark{a}}\\
    & 150  & $10.865 \pm 2.672$ \\
    
    \multicolumn{3}{l}{uGMRT \tablefootmark{b}}\\
    & 323 & $4.91 \pm 0.16$ \\
    
    \multicolumn{3}{l}{VLBI \tablefootmark{c}}\\
    & 1600 & $3.03 \pm 0.05$ \\
    
    \multicolumn{3}{l}{VLA \tablefootmark{d}}\\
    & 1600 & $3.003 \pm 0.039$ \\
    
    \hline
    \end{tabular}
    \tablefoot{
    \tablefoottext{a}{\citet{LoTSS2017AA...598A.104S}}; \tablefoottext{b}{\citet{Shao2020A&A...641A..85S}}; \tablefoottext{c}{\citet{Frey2011AA...531L...5F}; \tablefoottext{d}{\citet{Keller2024MNRAS.528.5692K}}}
    }
    \label{tab:J1429_fluxes}
    \end{table}
    
    \begin{table}[H]
    \centering
    \caption{Radio measurements for quasar J1427+3312.}
    \begin{tabular}{lcc}
    \hline\hline
    \textbf{Survey} & \textbf{Frequency (MHz)} & \textbf{Flux density (mJy)} \\
    \hline
    
    \multicolumn{3}{l}{LOFAR \tablefootmark{a}}\\
    & 150 & $5.43 \pm 0.25$ \\
    
    \multicolumn{3}{l}{uGMRT \tablefootmark{b}}\\
    & 323 & $3.96 \pm 0.10$ \\
    
    \multicolumn{3}{l}{VLA-P \tablefootmark{c}}\\
    & 324.5 & $4.8 \pm 0.8$ \\
    
    \multicolumn{3}{l}{uGMRT \tablefootmark{d}}\\
    & 650 & $2.65 \pm 0.15$ \\
    
    \multicolumn{3}{l}{WSRT \tablefootmark{e}}\\
    & 1380 & $2.12 \pm 0.1$ \\
    
    \multicolumn{3}{l}{VLBA \tablefootmark{f}}\\
    & 1400 & $1.778 \pm 0.109$ \\
    
    \multicolumn{3}{l}{VLA \tablefootmark{g}}\\
    & 1600  & $1.281 \pm 0.035$ \\
    
    \hline
    \end{tabular}
    \tablefoot{
    \tablefoottext{a}{\citet{LOFARBootes2016MNRAS.460.2385W}}; \tablefoottext{a}{\citet{Shao2020A&A...641A..85S}}; \tablefoottext{c}{\citet{Coppejans2015MNRAS.450.1477C}}; \tablefoottext{d}{\citet{Shao2022A&A...659A.159S}}; \tablefoottext{e}{\citet{deVries2002AJ....123.1784D}}; \tablefoottext{f}{\cite{Momjian2008AJ....136..344M}}; \tablefoottext{g}{\cite{Keller2024MNRAS.528.5692K}}
    }
    \label{tab:J1427_fluxes}
    \end{table}
    
    \begin{table}[H]
    \centering
    \caption{Radio measurements for quasar PSO J0309+27 in the literature.}
    \begin{tabular}{lcc}
    \hline\hline
    \textbf{Survey} & \textbf{Frequency (MHz)} & \textbf{Flux density (mJy)} \\
    \hline
    
    \multicolumn{3}{l}{LoTSS \tablefootmark{a}}\\
    & 144 & $66.5 \pm 7.1$ \\
    
    \multicolumn{3}{l}{TGSS \tablefootmark{b}}\\
    & 147.5 & $56.6 \pm 6.6$ \\
    
    \multicolumn{3}{l}{uGMRT \tablefootmark{a}}\\
    & 383 & $40.8 \pm 4.3$ \\
    & 675 & $37.7 \pm 3.9$ \\
    
    \multicolumn{3}{l}{RACS \tablefootmark{a}}\\
    & 887 & $35.6 \pm 4.0$ \\
    
    \multicolumn{3}{l}{VLA FIRST \tablefootmark{a}}\\
    & 1400 & $23.9 \pm 3.3$ \\
    
    \multicolumn{3}{l}{VLASS \tablefootmark{a}}\\
    & 3000 & $13.7 \pm 1.6$ \\
    
    \hline
    \end{tabular}
    \tablefoot{
    \tablefoottext{a}{\citet{Gloudemans2023} and reference therein.};
    \tablefoottext{b}{\citet{TGSS2017A&A...598A..78I}}
    }
    \label{tab:PSOJ0309_fluxes}
    \end{table}
    
    \begin{table}[H]
    \centering
    \caption{Radio measurements for quasar J0856+0224 in the literature.}
    \begin{tabular}{lcc}
    \hline\hline
    \textbf{Survey} & \textbf{Frequency (MHz)} & \textbf{Flux density (mJy)} \\
    \hline

    \multicolumn{3}{l}{TGSS \tablefootmark{a}}\\
    & 147.5 & $870.1 \pm 87.1$ \\

    \multicolumn{3}{l}{GLEAM \tablefootmark{a}}\\
    & 76  & $1400 \pm 81.4$ \\
    & 84  & $1320 \pm 64.9$ \\
    & 92  & $1250 \pm 55.4$ \\
    & 99  & $1170 \pm 49.6$ \\
    & 107 & $1200 \pm 57.1$ \\
    & 115 & $1060 \pm 45.5$ \\
    & 122 & $1000 \pm 42.3$ \\
    & 130 & $903 \pm 40.1$ \\
    & 143 & $871 \pm 31.6$ \\
    & 151 & $894 \pm 27.1$ \\
    & 158 & $778 \pm 26.5$ \\
    & 166 & $816 \pm 25.0$ \\
    & 174 & $732 \pm 27.9$ \\
    & 181 & $705 \pm 23.8$ \\
    & 189 & $685 \pm 26.7$ \\
    & 197 & $583 \pm 26.4$ \\
    & 204 & $585 \pm 23.4$ \\
    & 212 & $583 \pm 21.7$ \\
    & 220 & $521 \pm 20.1$ \\
    & 227 & $532 \pm 19.5$ \\

    \multicolumn{3}{l}{NVSS \tablefootmark{a}}\\
    & 1400 & $86.5 \pm 2.6$ \\
    \hline
    \end{tabular}
    \tablefoot{
    \tablefoottext{a}{\cite{Drouart2020PASA...37...26D} and reference therein.}
    }
    \label{tab:J0856_fluxes}
    \end{table}

    \begin{figure*}[htbp]
        \centering
        \begin{tabular}{ccc}
            \begin{subfigure}{0.3\textwidth}
                \includegraphics[width=\linewidth]{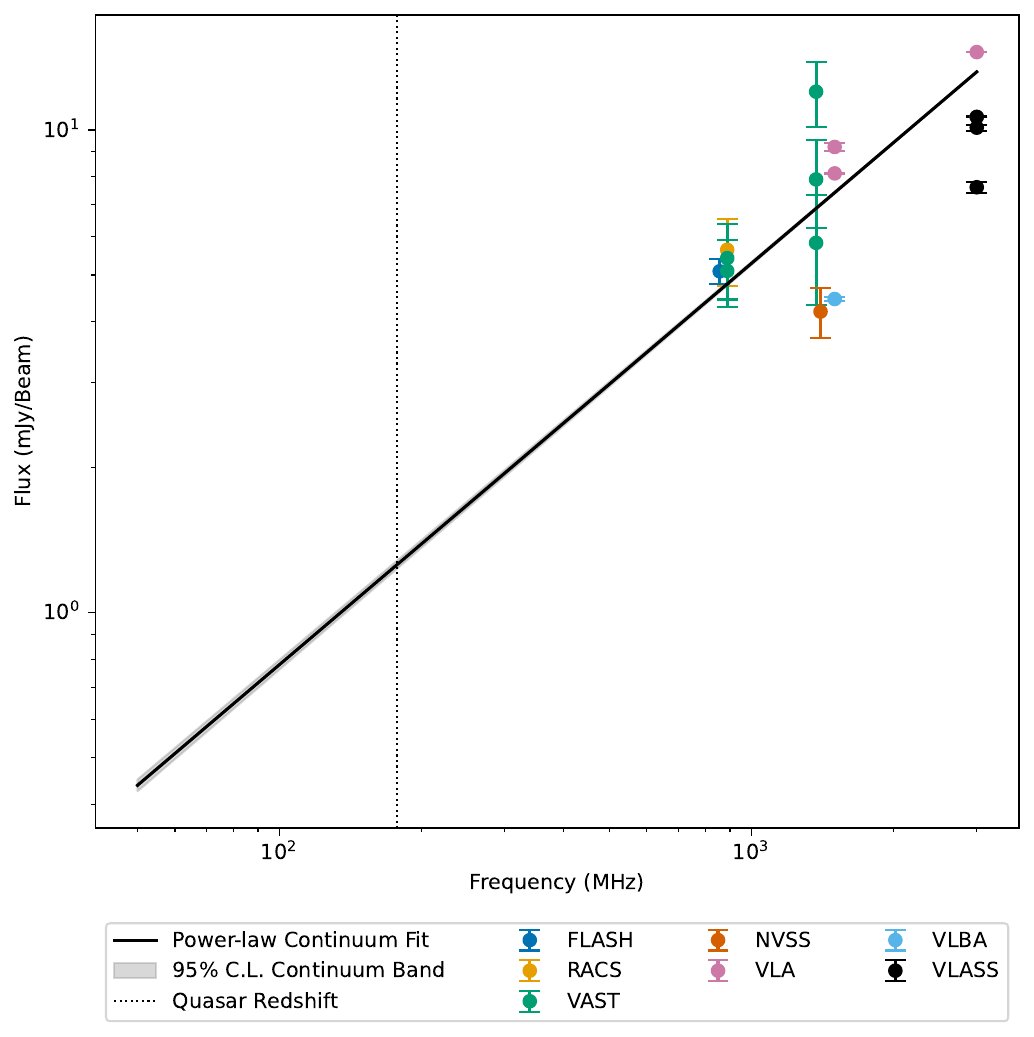}
                \subcaption{J0410-0139}
            \end{subfigure} &
            \begin{subfigure}{0.3\textwidth}
                \includegraphics[width=\linewidth]{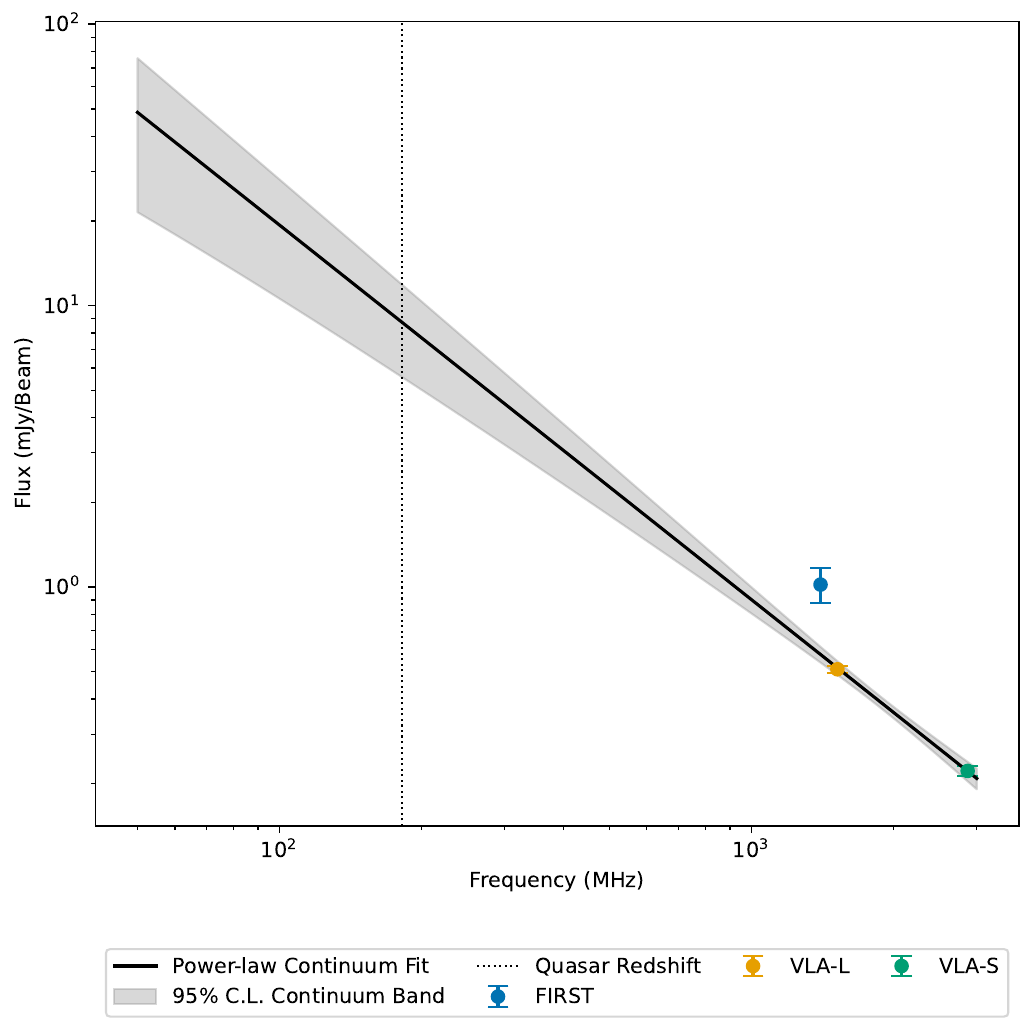}
                \subcaption{PSO J172+18}
            \end{subfigure} &
            \begin{subfigure}{0.3\textwidth}
                \includegraphics[width=\linewidth]{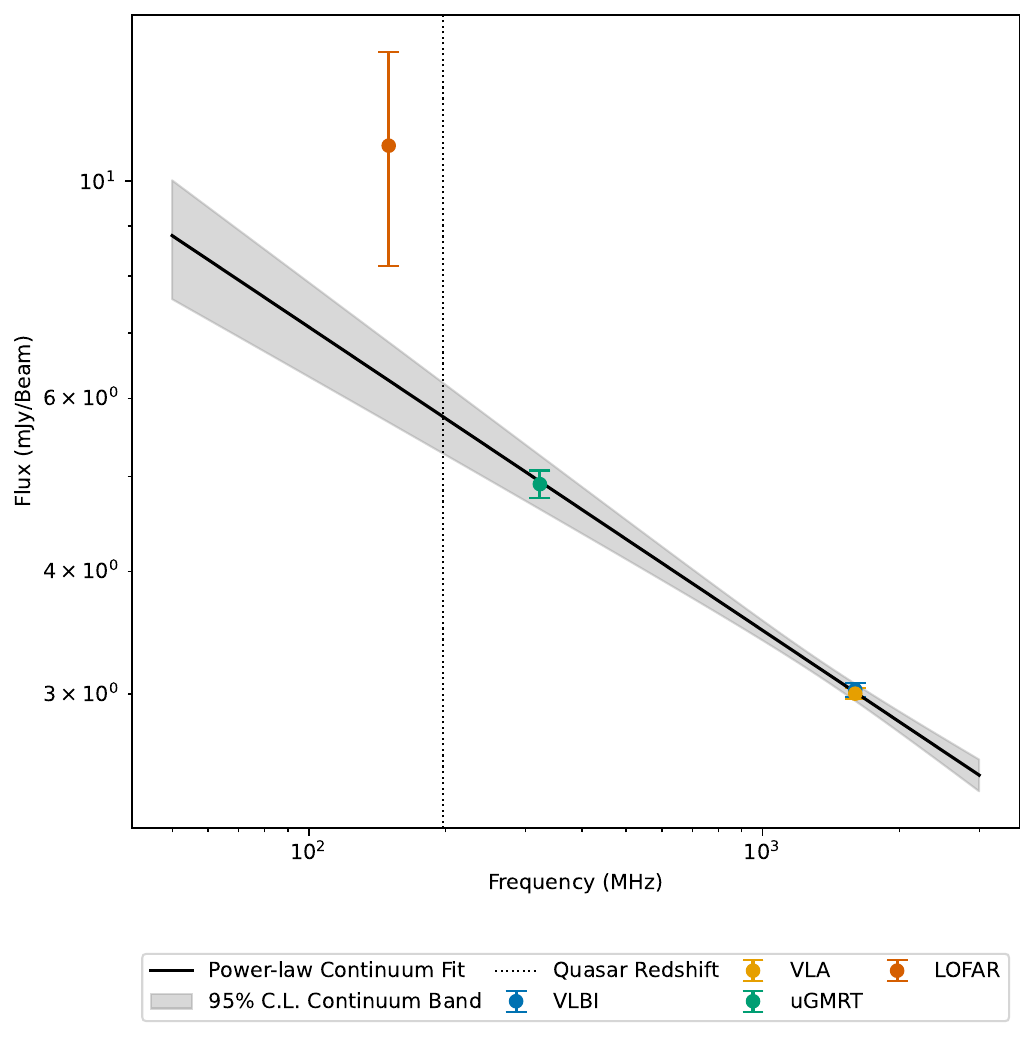}
                \subcaption{J1429+5547}
            \end{subfigure} \\
    
            \begin{subfigure}{0.3\textwidth}
                \includegraphics[width=\linewidth]{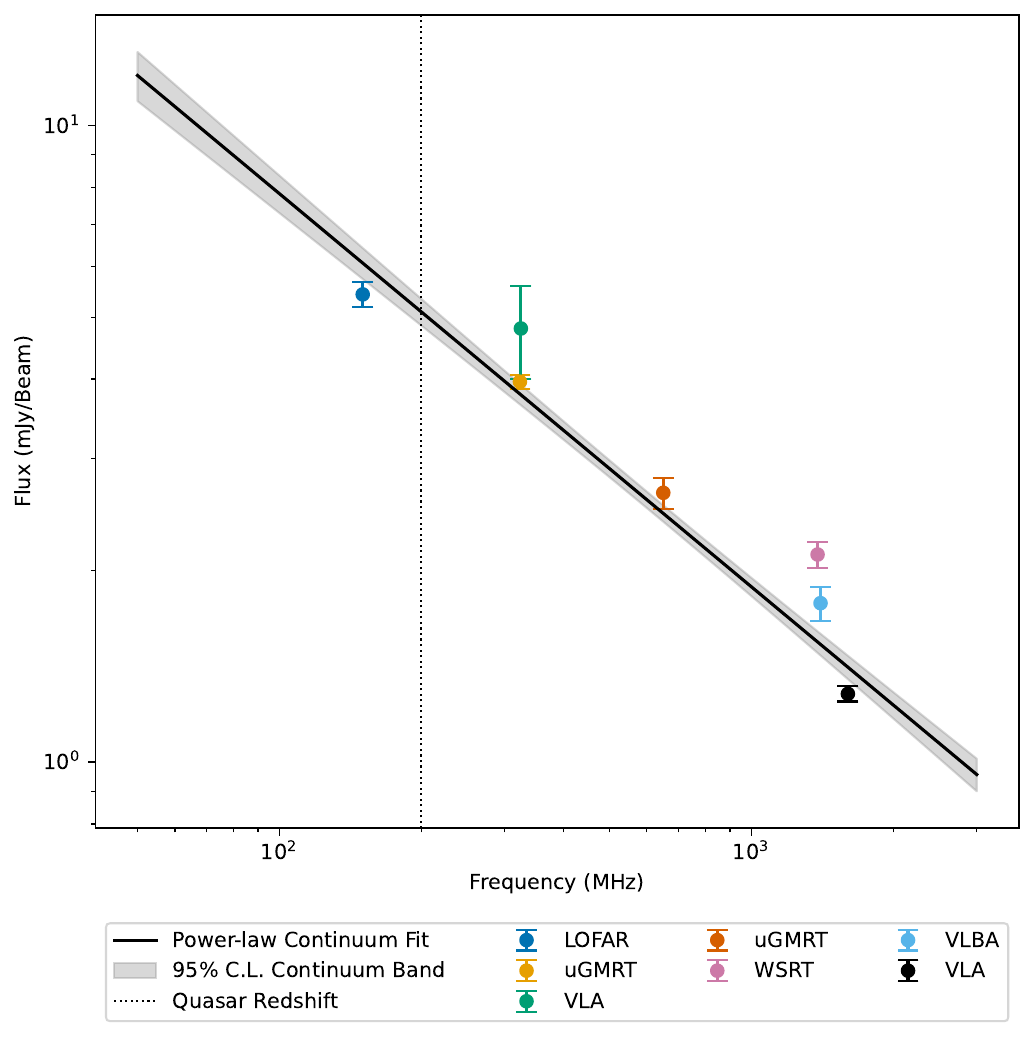}
                \subcaption{J1427+3312}
            \end{subfigure} &
            \begin{subfigure}{0.3\textwidth}
                \includegraphics[width=\linewidth]{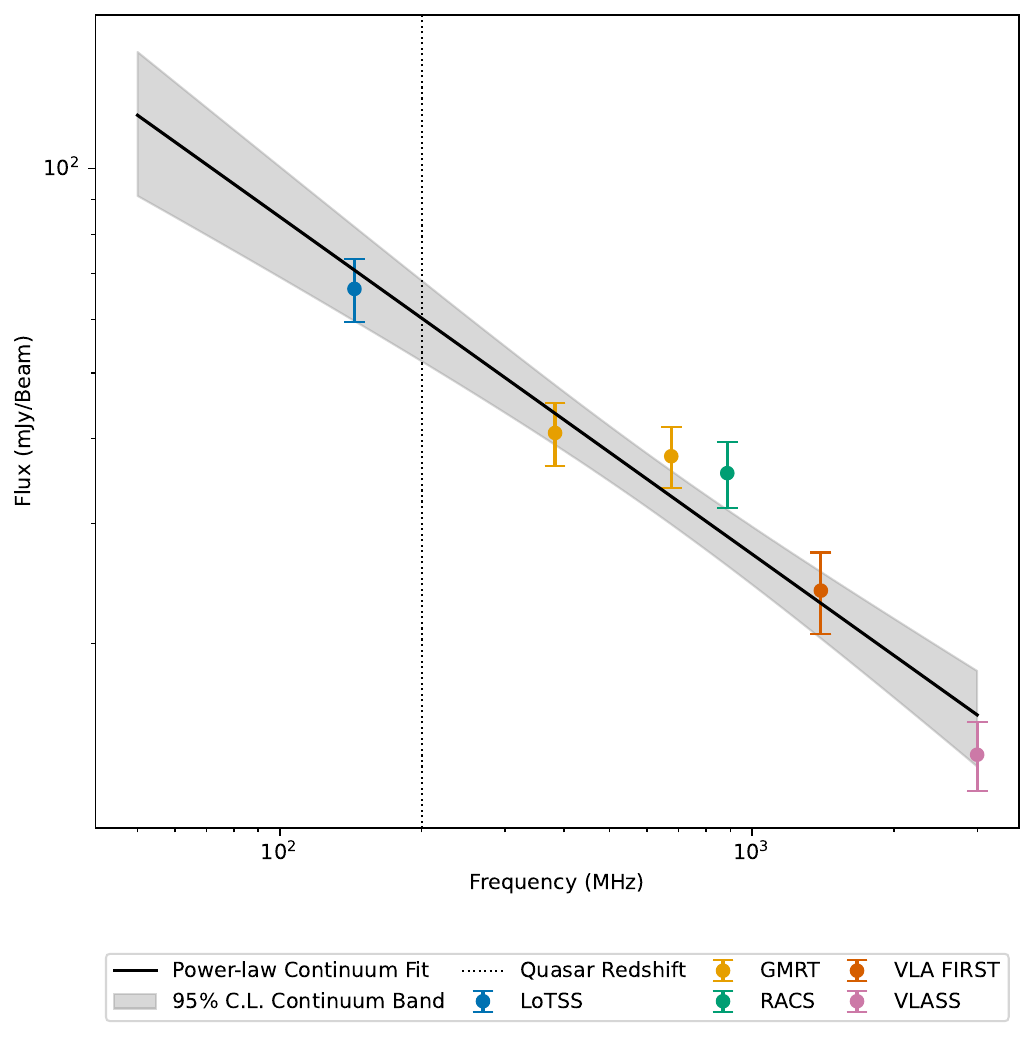}
                \subcaption{PSO J0309+27}
            \end{subfigure} &
            \begin{subfigure}{0.3\textwidth}
                \includegraphics[width=\linewidth]{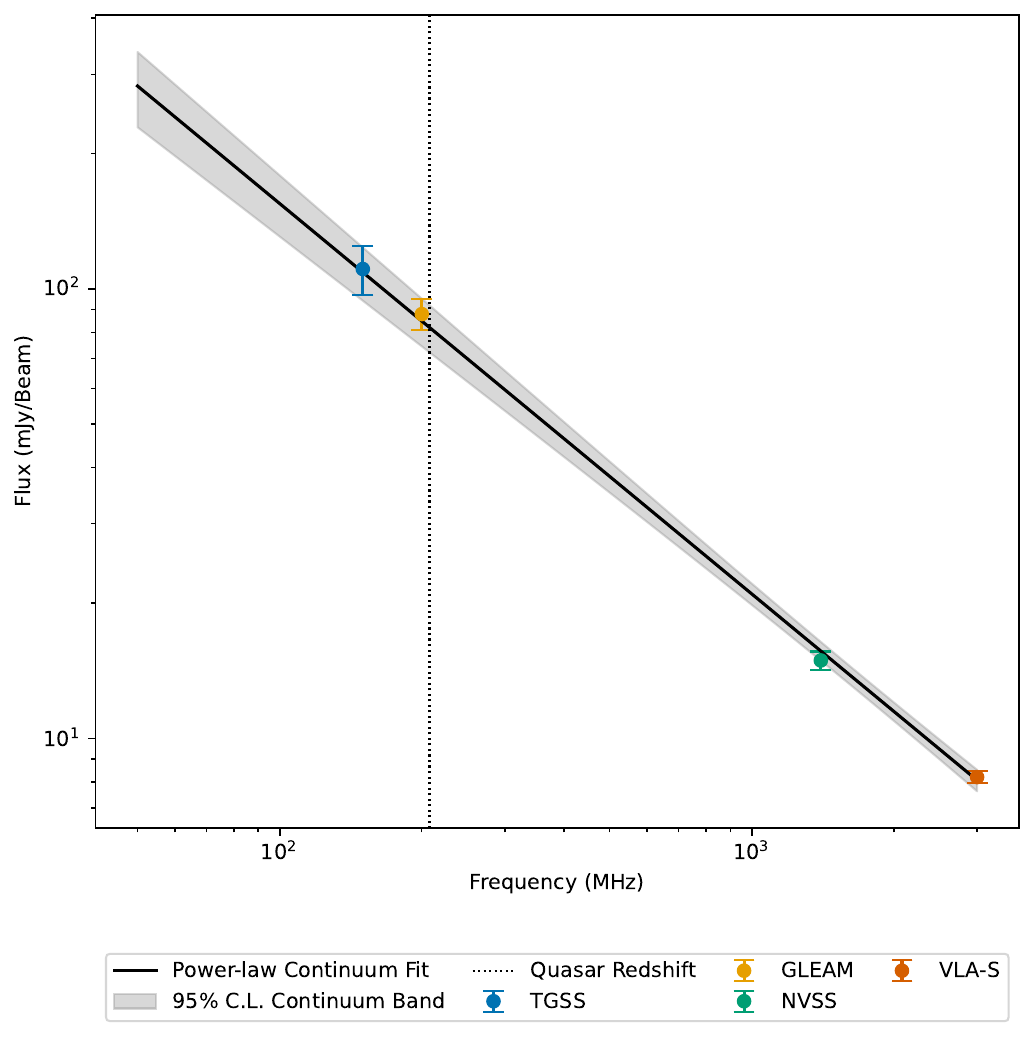}
                \subcaption{J2329-1520}
            \end{subfigure} \\
    
            \begin{subfigure}{0.3\textwidth}
                \includegraphics[width=\linewidth]{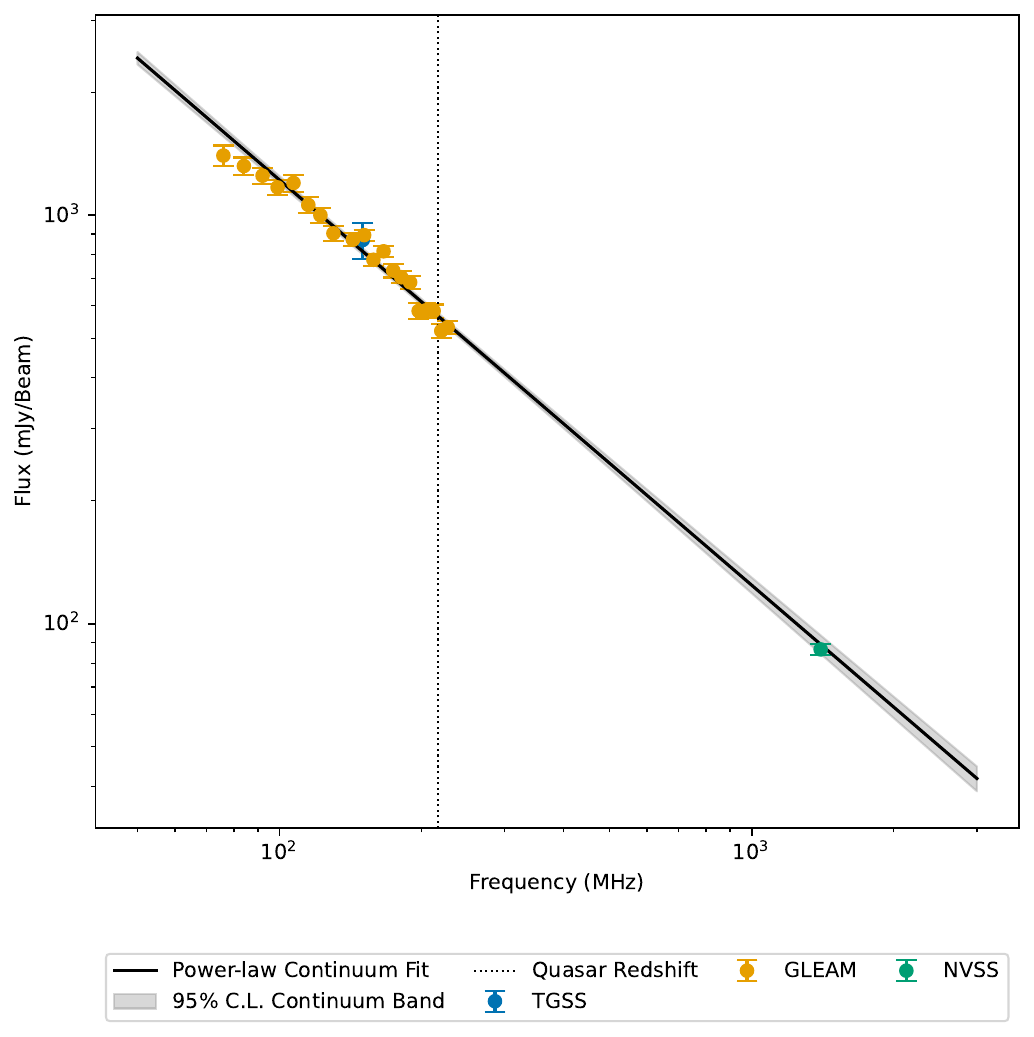}
                \subcaption{J0856+0224}
            \end{subfigure}
        \end{tabular}
        \caption{Continuum spectra fit of radio-loud sources. The scatters are flux density measurements from the literature, and the vertical dotted lines mark the redshifted 21-cm frequency at each source's redshift. The black solid lines are the best-fit power-law model on the spectra, where the grey contours are the 68\%~C.L. of the reconstructed spectra.}
        \label{fig:sim2_continuum_fit}
    \end{figure*}
    
    \section{Stokes-V noise distribution}
    \begin{figure}[H]
    \centering
    \includegraphics[width=\linewidth]{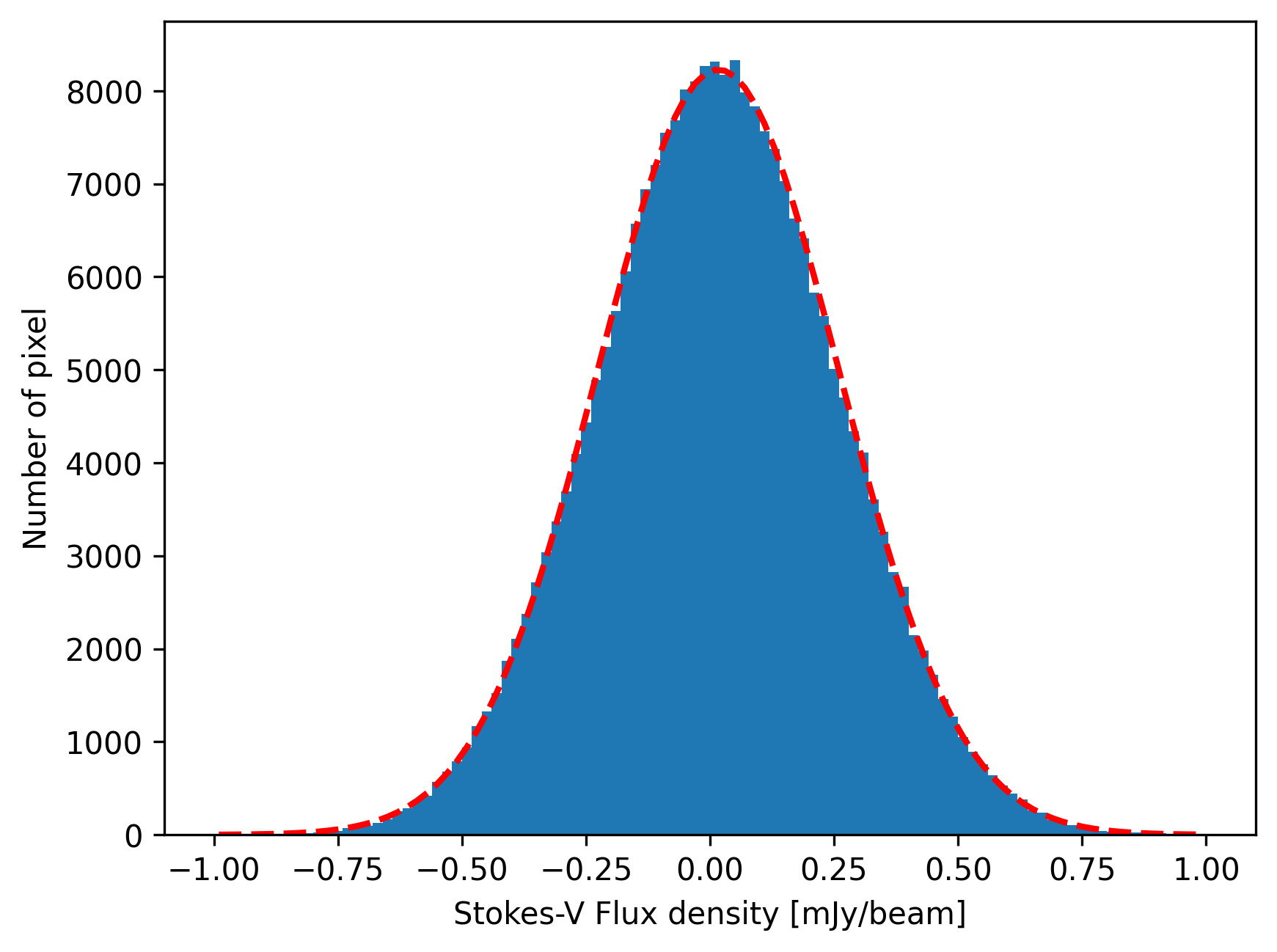}
    \caption{Distribution of pixel values from the continuum, frequency-averaged Stokes V image of J2329-1520. We note how closely the distribution follows a Gaussian shape, as expected for thermal noise, which, in turn, is expected to be the only contribution to the Stokes V image. The dashed red line represents the Gaussian best fit to the distribution with a $\sigma$ of $0.3$~mJy~beam$^{-1}$.}
    \label{fig:StokesV_histogram}
    \end{figure}
\end{appendix}

\end{document}